\documentclass[11pt]{article}
\usepackage{setspace,graphicx,epstopdf,amsmath,amsfonts,subcaption,amssymb,amsthm}
\usepackage{marginnote,datetime,enumitem,rotating,fancyvrb}
\usepackage{float}
\usepackage{natbib}
\usdate
\usepackage{setspace}

\usepackage[english]{babel}
\usepackage{tabulary}
\usepackage{longtable}
\usepackage[left=1.0in,right=1.0in,top=1.0in,bottom=1.0in]{geometry}
\usepackage[symbol]{footmisc}
\renewcommand*{\thefootnote}{\fnsymbol{footnote}}
\usepackage{booktabs}
\usepackage{multirow}
\usepackage{graphicx}
\usepackage{setspace}
\usepackage{amssymb}
\usepackage{amsthm}
\usepackage{bm}
\usepackage{tabulary}
\usepackage{longtable}
\usepackage{booktabs}
\mathchardef\mhyphen="2D 
\usepackage{changepage}
\usepackage{ragged2e}
\usepackage{bbm}
\usepackage{dsfont}
\usepackage{float}
\usepackage{nicefrac}
\usepackage{setspace}
\usepackage[titletoc,title]{appendix}
\usepackage{rotating}

\newcolumntype{L}[1]{>{\raggedright\let\newline\\\arraybackslash\hspace{0pt}}m{#1}}
\newcolumntype{C}[1]{>{\centering\let\newline\\\arraybackslash\hspace{0pt}}m{#1}}
\newcolumntype{R}[1]{>{\raggedleft\let\newline\\\arraybackslash\hspace{0pt}}m{#1}}

\usepackage{cellspace}

\usepackage{colortbl}
\usepackage{lscape}

\usepackage{amsmath}
\newcommand\numberthis{\addtocounter{equation}{1}\tag{\theequation}}

 \definecolor{miewblue}{RGB}{217,239,251}

 \definecolor{rsgreen}{RGB}{0,255,77}

\makeatletter
\renewcommand*{\@fnsymbol}[1]{\ensuremath{\ifcase#1\or *\or \dagger\or \ddagger\or
    \mathsection\or \| \or \|\or **\or \dagger\dagger
    \or \ddagger\ddagger \else\@ctrerr\fi}}
\makeatother

\theoremstyle{plain}

\makeatother
\providecommand{\lemmaname}{Lemma}
\providecommand{\propositionname}{Proposition}

\makeatletter
\patchcmd{\NAT@test}{\else \NAT@nm}{\else \NAT@nmfmt{\NAT@nm}}{}{}
\DeclareRobustCommand\citepos
  {\begingroup
  \let\NAT@nmfmt\NAT@posfmt
  \NAT@swafalse\let\NAT@ctype\z@\NAT@partrue
  \@ifstar{\NAT@fulltrue\NAT@citetp}{\NAT@fullfalse\NAT@citetp}}
\let\NAT@orig@nmfmt\NAT@nmfmt
\def\NAT@posfmt#1{\NAT@orig@nmfmt{#1's}}
\makeatother

\usepackage[labelfont = bf, labelsep = period, font = small]{caption}

\usepackage[T1]{fontenc}

\usepackage{amsmath}

\usepackage[sc]{mathpazo}
\usepackage{times}

\DeclareMathOperator*{\argmax}{arg\,max}

\usepackage[table,xcdraw,svgnames,dvipsnames]{xcolor}
\usepackage{natbib}  
\usepackage[colorlinks]{hyperref}

\bibpunct{(}{)}{;}{a}{,}{,}

\hypersetup{
  colorlinks   = true, 
  urlcolor     = blue, 
  linkcolor    = blue, 
  citecolor   = blue 
}

\usepackage{tikz}
 
\usetikzlibrary{calc}
\usepackage{pgfplots}
\pgfplotsset{compat=1.5}

\usepackage{tikz}

\usepackage{tikz}

\tikzset{
  LabelStyle/.style = { rectangle, rounded corners, draw,
                        minimum width = 2em, 
                        text = red}, 
  VertexStyle/.append style = { inner sep=5pt,
                                font = \Large\bfseries},
  EdgeStyle/.append style = {->, bend left} }

\usetikzlibrary{arrows,positioning} 
\tikzset{
    >=stealth',
    punkt/.style={
           rectangle,
           rounded corners,
           draw=black, very thick,
           text width=6.5em,
           minimum height=2em,
           text centered},
    pil/.style={
           ->,
           thick,
           shorten <=2pt,
           shorten >=2pt,}
}

\usepackage{pgfplots}
\usetikzlibrary{pgfplots.dateplot}
\pgfplotsset{compat=1.10}
\usepgfplotslibrary{fillbetween}
\usetikzlibrary{patterns}

\usepackage{tikz}

\usepackage[nameinlink,noabbrev]{cleveref}


\hypersetup{colorlinks = true}
\AtBeginDocument{%
\hypersetup{%
  	linkcolor =     blue!85!black,
   	anchorcolor=    blue!85!black,
   	citecolor =     blue!85!black,
   	filecolor =     blue!85!black,
   	urlcolor =      blue!40!black,
   	menucolor =     blue!85!black,
	}
	}

\defcitealias{data4science}{MIT Election Data And Science Lab (2018)}

\date{\normalsize\mydate\today}

\renewcommand*{\thefootnote}{\arabic{footnote}}
\usepackage{dcolumn}

\begin{document}

\renewcommand*{\thefootnote}{\arabic{footnote}}
\setcounter{footnote}{0}
\title{\Large \textbf{Government Guarantees and Banks' Income Smoothing}\thanks{We thank Warren Bailey, Matt Baron, Riddha Basu (discussant), Sanjeev Bhojraj, Murillo Campello, Gustavo Cortes, Steve Crawford,
Richard Frankel, George Gao, Yadav Gopalan, Bob Jarrow, Gaurav Kankanhalli, Inder Khurana, Alan Kwan, Bob Libby, Josh Madsen, Xiumin
Martin, Roni Michaely, Maureen O'Hara, Guilherme Pimentel, Manju Puri, Cathy Schrand (discussant),  Haluk {\"U}nal (the editor), Biqin Xie (discussant), one anonymous reviewer, as well as workshop
participants at Cornell University, University of Minnesota, Singapore Management University, Washington University in St. Louis,
University of Missouri, the 18th FDIC/JFSR Annual Bank Research Conference, the 2017 FARS Midyear Meeting, the 2017 Trans-
Atlantic Doctoral Conference, and the 2017 AAA Annual Meeting for their helpful comments and support. An earlier version of
this paper has been circulated under the title ``Government Guarantees and Banks' Earnings Management." Errors are our own.}} \vspace{3.0cm}

\author{
\begin{tabular}{ccccc}
\\
\textsc{\normalsize \href{https://sites.google.com/site/dantaswebsite/}{Manuela M. Dantas}}\thanks{David Nazarian College of Business and Economics, California State University, Northridge. Email: $\textbf{\href{manuela.dantas@csun.edu}{manuela.dantas@csun.edu}}$.} & \hspace{0.75cm} & \textsc{\normalsize \href{https://kelley.iu.edu/faculty-research/faculty-directory/profile.cshtml?id=KENMERK}{Kenneth J. Merkley}}\thanks{Kelley School of Business, Indiana University. Email: $\textbf{\href{mailto:kenmerk@iu.edu}{kenmerk@iu.edu}}$.} & \hspace{0.75cm} & \textsc{\normalsize \href{https://business.missouri.edu/departments-faculty/people-directory/felipe-bastos-gurgel-silva}{Felipe B. G. Silva}}\thanks{Trulaske College of Business, University of Missouri. Email: $\textbf{\href{mailto:bastosgurgelsilvf@missouri.edu}{bastosgurgelsilvf@missouri.edu}}$} \\
\textit{\normalsize California State University} & \hspace{0.75cm} & \textit{\normalsize Indiana University} & \hspace{0.75cm} & \textit{\normalsize University of Missouri}\\
\textit{\normalsize Northridge} & \hspace{0.75cm} &  & \hspace{0.75cm} & 
\vspace{0.25cm}
\\
\end{tabular}
}

\date{\normalsize\today \\ \vspace{1cm} \large\emph{Journal of Financial Services Research} \\ \normalsize(forthcoming) }

\maketitle

\begin{abstract}
\noindent We propose four channels through which government guarantees  affect banks’ incentives to smooth income. Empirically, we exploit two complementary settings that represent plausible exogenous changes in government guarantees: the \emph{increase} in \emph{implicit} guarantees following the creation of the Eurozone  and the \emph{removal} of \emph{explicit} guarantees granted to the Landesbanken. We show that increases (decreases) in government guarantees are associated with significant decreases (increases) in banks' income smoothing. Taken together, our results  largely corroborate the predominance of a tail-risk channel, wherein government guarantees reduce banks’ tail risk, thereby reducing managers’ incentives to engage in income smoothing.
\end{abstract}

\thispagestyle{empty}   
\pagenumbering{arabic}  

\newpage
\maketitle
\vspace{2cm}
\begin{center}

\title{\Large \textbf{Government Guarantees and Banks' Income Smoothing}} \vspace{4.0cm}    

 \large\emph{Journal of Financial Services Research} \\ \normalsize(forthcoming)
 
\end{center}

\begin{abstract}
\noindent We propose four channels through which government guarantees  affect banks’ incentives to smooth income. Empirically, we exploit two complementary settings that represent plausible exogenous changes in government guarantees: the \emph{increase} in \emph{implicit} guarantees following the creation of the Eurozone  and the \emph{removal} of \emph{explicit} guarantees granted to the Landesbanken. We show that increases (decreases) in government guarantees are associated with significant decreases (increases) in banks' income smoothing. Taken together, our results  largely corroborate the predominance of a tail-risk channel, wherein government guarantees reduce banks’ tail risk, thereby reducing managers’ incentives to engage in income smoothing.
\end{abstract}

\thispagestyle{empty}   
\pagenumbering{arabic}  

\def\llp {$llp_{i,t}$}
\def\LLR {$llr_{i,t-1}$}
\def\ebllp {$ebllp_{i,t}$}
\def\PostNinetyNine  {$Post1999_{t}$}
\def\ebllpPostNinetyNine  {$ebllp_{i,t} \times Post1999_{t}$}
\def\lagonellp {$llp_{i,t-1}$}
\def\lagtwollp {$llp_{i,t-2}$}
\def\lagonesize {$Size_{i,t-1}$}
\def\lagonecap {$CAP_{i,t-1}$}
\def\changeloans {$\Delta Loans_{i,t}$}
\def\IInc {$IInc_{i,t}$}
\def\IExp {$IExp_{i,t}$}
\def\BigBath {$I^{BigBath}_{i,t}$}
\def\SmallP {$I^{SmallProfit}_{i,t}$}

\def\ebllpDNinetyeight {$ebllp_{i,t} \times D1998_{t}$}

\def\PercGDPPC{$\Delta \% GDP.PC_{c,t}$}

\def\ebllpDNinetyeight {$ebllp_{i,t} \times D1998_{t}$}

\def\FEA {$FEA_{c}$}
\def\FEAPostNinetyNine {$FEA_{c}\times Post1999_{t}$}

\def\FEADNinetyEight {$FEA_{c}\times Post1998_{t}$}

\def\ebllpFEA {$ebllp_{i,t} \times FEA_{c}$}
\def\ebllpFEAPostNinetyNine {$ebllp_{i,t} \times FEA_{c} \times Post1999_{t} $}

\def\ebllpFEADNinetyeight {$ebllp_{i,t} \times FEA_{c} \times D1998_{t}$}

\def\ebllpchangeloans {$ebllp_{i,t} \times \Delta Loans_{i,t}$}

\def\meanLibor {$Libor_{t}$}
\def\ebllpmeanLibor {$ebllp_{i,t} \times Libor_{t}$}

\def\VolLibor {$\sigma^{Libor}_{t}$}
\def\ebllpVolLibor {$ebllp_{i,t} \times \sigma^{Libor}_{t}$}

\def\RRtussandfour {$MP^{Shock}_{t}$}
\def\ebllpRRtussandfour {$ebllp_{i,t} \times MP^{Shock}_{t}$}

\def\RetEURUSD {$ Ret^{EUR.USD}_{t}$}
\def\ebllpRetEURUSD {$ebllp_{i,t} \times Ret^{EUR.USD}_{t}$}

\def\VolEURUSD {$ \sigma^{EUR.USD}_{t}$}
\def\ebllpVolEURUSD {$ebllp_{i,t} \times \sigma^{EUR.USD}_{t}$}

\def\Isubs {$I^{Subs}_{i,t}$}
\def\ebllpIsubs {$ebllp_{i,t} \times I^{Subs}_{i,t}$}

\def\EPUEuro {$EPU^{Euro}_{t}$}
\def\ebllpEPUEuro {$ebllp_{i,t} \times EPU^{Euro}_{t}$}

\def\Landes {$Landes_{i}$}

\def\PostTussandFive {$Post2005_{t} $}

\def\ebllpPostTussandFive {$ebllp_{i,t} \times Post2005_{t} $}

\def\LandesPostTussandFive {$Landes_{i} \times Post2005_{t}$}

\def\ebllpLandes {$ebllp_{i,t} \times Landes_{i}$}

\def\ebllpLandesPostTussandFive {$ebllp_{i,t} \times Landes_{i} \times Post2005_{t}$}

\def\ebllpDTussandFour {$ebllp_{i,t} \times D2004_{t}$}

\def\LandesDTussandFour {$Landes_{i} \times D2004_{t}$}

\def\ebllpLandesDTussandFour {$ebllp_{i,t} \times Landes_{i} \times D2004_{t}$}

\newpage \onehalfspacing

\section{Introduction} \label{sec:intro}
 \linespread{1.5}
\onehalfspacing
\setcounter{page}{1}

Government guarantees play a central role in the stability and functioning of the banking sector, and are associated with a plethora of asset pricing and real effects \citep{o1990deposit,flannery1996evidence,acharya2007too,gandhi2015size}. Notwithstanding their importance, the extent to which these guarantees affect the information banks report to capital markets and regulators remains largely unclear. This question is particularly relevant given that the pro-cyclicality of banks’ provisions and reported income is a matter of utmost importance to regulators and investors.  In this paper, we attempt to fill this void in the literature by investigating how government guarantees relate to banks’ income smoothing in two complementary empirical settings. 

The empirical challenge of testing the relationship between government guarantees and banks’ income smoothing is twofold. First, the direct measurement of such guarantees is problematic for several reasons. Economically speaking, government guarantees are expected cash infusions in extreme left-tail scenarios. Because such guarantees are often implicitly granted rather than contractually written, measurement errors arise directly from the uncertainty regarding how much financial aid (if any) banks can get from the government. Even when guarantees are explicitly granted, their measurement entails the probability assessment of extreme left-tail events. 

The second challenge underlying this investigation is that government guarantees are endogenous to a variety of factors that reflect characteristics of individual banks and the entire financial sector. As for their etiology, government guarantees are not determined in an arbitrary or whimsical manner, and the factors that explain the differences in such guarantees are likely to be directly or indirectly correlated with banks’ income smoothing. As for their economic consequences, such guarantees are known to be associated with a gamut of real effects, suggesting that the link between government guarantees and banks’ income smoothing operates concomitantly through several channels. The fundamentally endogenous nature of the problem is such that even if the econometrician can perfectly exogenize levels of government guarantees and isolate the factors intrinsically explaining different levels of guarantees across banks, the real effects (consequences) of such guarantees would still involve a myriad of channels.

 Figure 1 illustrates of the four possible channels identified in this paper. First, government guarantees directly curtail the tail-risk of banks, thereby reducing managers' incentives to engage in income smoothing (a channel which we refer to as the \emph{tail-risk channel}). Importantly, the same shift in risk from bank investors to the government may elicit an endogenous response by banks, which internalize the government's safeguarding power and take on more risks \citep{duchin2014safer,gropp2014impact,fischer2014government}, potentially leading to higher earnings volatility and increasing incentives to smooth earnings (\emph{risk-taking channel}). The censoring effect of government guarantees on tail risk may also induce less monitoring from bank investors, which in turn allows managers to increase their income smoothing (\emph{investor-monitoring channel}). Lastly, insofar as these guarantees ultimately represent fiscal costs to the government, increased guarantees may elicit stronger monitoring actions by regulators, which in turn reduces managers’ ability to smooth earnings (\emph{government-monitoring channel}).\footnote{In \Cref{app:sec:theorymodel} we present a stylized theoretical framework showing the micro-foundations of the channels proposed in this paper.}

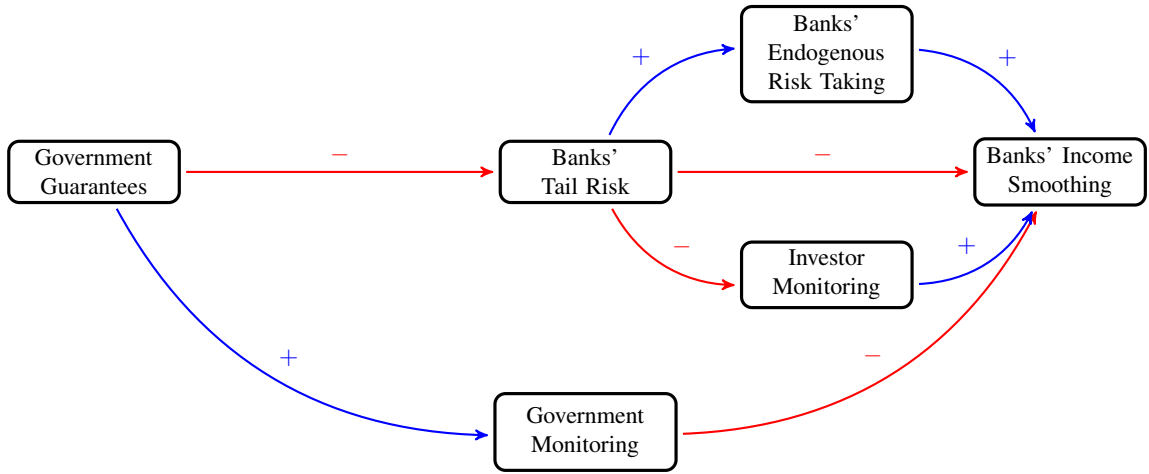
\begin{figure}[h]

\begin{center}
\begin{tikzpicture}[node distance=0.8cm, auto,]
\footnotesize
 \node (market) {};
 \node[above=of market] (dummy) {};
  \node[right=6.25cm of dummy] (dummy2) {};

\node[punkt,left=2.0cm of dummy] (g) {Government Guarantees};

\node[punkt,right=2.0cm of dummy] (btr) {Banks' Tail Risk};

 \node[punkt,right=4.0cm of btr] (smth) {Banks' Income Smoothing};

\node[punkt,below= of dummy2] (invmo) {Investor Monitoring};
\node[punkt,above= of dummy2] (brt) {Banks' Endogenous Risk Taking};

 \node[punkt, inner sep=5pt,below=2.5cm of btr] (regmo) {Government Monitoring};

\path (btr) edge [pil,->,red] node{\textcolor{red}{$-$}} (smth);
\path (g) edge [pil,->,red] node{\textcolor{red}{$-$}} (btr);
\path (g) edge [pil,->, bend right=30,blue] node{\textcolor{blue}{$+$}} (regmo);
\path (regmo) edge [pil,->, bend right=30,red] node{\textcolor{red}{$-$}} (smth);
\path (btr) edge [pil,->, bend left=30,blue] node{\textcolor{blue}{$+$}} (brt);
\path (btr) edge [pil,->, bend right=30,red] node{\textcolor{red}{$-$}} (invmo);
\path (brt) edge [pil,->, bend left=30,blue] node{\textcolor{blue}{$+$}} (smth);
\path (invmo) edge [pil,->, bend right=30,blue] node{\textcolor{blue}{$+$}} (smth);

\end{tikzpicture}

\end{center}

\normalsize
\vspace{-0.5cm}
     \begin{minipage}{0.95\linewidth}\scriptsize
   \singlespacing
   This figure illustrates the four channels proposed through which government guarantees affect banks' incentives to engage in income smoothing: the tail-risk channel, the risk-taking channel, the investor-monitoring channel, and the government-monitoring channel. Paths in blue (red) indicate increasing (decreasing) effects.  \\
  \end{minipage}
\caption{ \normalsize Channels Through Which Government Guarantees Affect Banks' Income Smoothing}
\label{fig:channels}
\end{figure}

Empirically, we attempt to circumvent the endogenous nature of government guarantees by exploiting the complementary nature of our research settings. Notably, these settings complement each other along several dimensions: (i) internal versus external validity, (ii) explicit versus implicit guarantees, and (iii) positive versus negative shifts in government guarantees. We acknowledge that the individual magnitudes of four (or more) channels operating in conjunction cannot be assessed ex-ante based merely on the literature. Thus, instead of making a directional prediction, we opt to take an agnostic stance regarding the relative magnitude of each channel. Since two of the channels (the \emph{tail-risk channel} and the \emph{government-monitoring channel}) imply a negative association between government guarantees and banks’ income smoothing, whereas a positive association is expected for the other two, the problem calls for empirical scrutiny to investigate the overall effect and shed light on which channel (or channels) is more prevalent.

In both settings, we measure income smoothing by examining banks’ discretionary use of loan loss provisions in connection with pre-provision earnings. Representing perhaps the most important bank accrual, loan loss provisions are highly correlated with income and capital ratios \citep{ahmed1999bank,beatty2014financial}. Managers exercise their discretion and judgment when setting up provisions in order to engage in income smoothing, aiming to positively influence the risk perceptions of investors and regulators \citep{greenawalt1988bank,moyer1990capital,scholes1990tax,bushman2010pros}.

Our first setting is the creation of the Eurozone in 1999. We use this setting as a proxy for an increase in implicit government guarantees. It serves as an appropriate setting for several reasons. First, the monetary union itself resulted from years of discussions between potential member countries, with substantial uncertainty regarding its viability even a few months prior to its enaction.\footnote{Heightened uncertainty is possibly owing to the fact that failed attempts to establish a monetary union occurred in the past (e.g., the European Monetary System (EMS)). Our time series analysis suggests that uncertainty peaked around the third or fourth quarter of 1998 (see \Cref{fig:macroseries}).} Importantly, despite the unquestionable economic consequences of the project, many scholars argue that the main reasons for its creation were geopolitical rather than economic \citep{stiglitz2016euro}.\footnote{In an interview at the Stigler Center at the University of Chicago, Joseph Stiglitz argued: \emph{``In economics, we don’t have very many experiments, and this is a natural experiment.
Nobody in their right mind would have done it, but they did it. (...) It was a political project, not an economic one.''}}

The creation of the Eurozone represents a positive shift in implicit government guarantees for various reasons. First, it improved the credit worthiness of member countries’ sovereign bonds (through the convergence of interest rates and the establishment of the monetary union). This improvement consequently increased  government guarantees to the banking sector \citep{acharya2012tale}. The credit risk of sovereign bonds is theoretically and empirically related to the strength of implicit government guarantees \citep{gerlach2010banking,acharya2014pyrrhic,correa2014sovereign,gandhi2020equity,silva2020fiscal}. Moreover, the creation of the Eurozone increased the importance of the banking sectors of each member country because the costs of potential bank failures could jeopardize the future of the project \citep{gerlach2010banking,chinn2012eurozone}. The creation of the Eurozone established the ECB as a lender of last resort that increased the likelihood of banks receiving assistance either through lower interest rates or capital infusions.\footnote{Ex-post behavior corroborates this assertion as Eurozone banks received lower interest rates and a significant amount of capital during the recent financial crisis \citep{hannon2016}.}

We start our Eurozone analysis by defining treatment banks as banks headquartered in a group of 10 countries comprising the first wave of the Eurozone in 1999 (first Euro adopters, or \emph{FEA} hereafter). Using a difference estimation, we find that banks significantly reduced their income smoothing after the creation of the Eurozone, which supports either (both) the tail-risk channel or (and) the government-monitoring channel. Lending weight to our inferences, we show that our difference estimation is robust to a variety of tests. Probing further, we expand our analysis by considering a difference-in-differences design where our group of control banks correspond to those headquartered in a large set of developed countries that never joined the Eurozone (never Euro adopters or \emph{NEA} hereafter). This analysis shows that \emph{FEA} banks reduced their income smoothing as compared to NEA banks after the formation of the Eurozone. Notably, our difference-in-differences results are robust to a battery of tests, including the estimation with propensity-score-matched groups and Nordic banks alone and the construction of bounding values for our estimates following \citet{oster2019unobservable}.

The advantage of the Eurozone setting is that it represents a broad and economically important shift in implicit government guarantees, prioritizing the external validity of the findings. The disadvantage is that it is still plausible that its creation affected other bank-specific or macroeconomic variables that could also affect banks’ discretionary use of provisions to smooth earnings.\footnote{In our \Cref{app:sec:eurozoneadd}, we perform a battery of robustness tests to investigate whether changes in bank performance stemming from the creation of the Eurozone alternatively explain our findings.} To address the limitations of the Eurozone in terms of internal validity, we move to our second setting, which naturally trades off generalizability for a cleaner identification of economic channels. 

In our second setting, we examine the removal of explicit government guarantees in 2005 from a group of state-owned German banks known as the Landesbanken. Prior to this event, these banks were granted specific government guarantees that included an explicit guarantee of all their liabilities and a maintenance obligation requiring the injection of additional equity when necessary. An investigation by the European Commission  determined that these guarantees represented potentially unfair government aid and recommended their subsequent removal.

We use both differences and difference-in-differences models to estimate how the removal of government guarantees to the Landesbanken affects their income smoothing. We find evidence that the Landesbanken initiate income smoothing after the government guarantees are removed. In our difference-in-differences analyses, we find that the Landesbanken increased their income smoothing after controlling for changes in the reporting behavior of two control groups: (i) German commercial banks that serve as our within-country different-type counterfactuals, and (ii) French government-owned banks that represent the different-country same-type counterfactuals. As in our Eurozone analysis, we conduct additional tests to substantiate our findings.

The directional consistency of the findings of the two settings indicates that the \emph{tail-risk channel} and the \emph{government-monitoring channel} are more economically meaningful than the other two. Between these two channels, we argue that the former has a more prominent role. With respect to the Eurozone setting, the creation of the monetary union represented an increase in government guarantees at the supranational level. The ECB was established with the mandate of conducting monetary policy and, as ex-post evidence shows, works as a lender of last resort. However, the creation of the Eurozone maintained individual countries’ central banks that continued to be responsible for regulatory activities. Regarding the Landesbanken, we emphasize that despite the removal of government guarantees, it is plausible that the \emph{government-monitoring channel} is not substantially triggered by the event since the banks remained government-owned. Lastly, it is possible that regulators still prefer smoothing even when government guarantees increase as it allows for banks to build reserves when their performance is good enough to offset subsequent poor performance, mitigating the pro-cyclicality of their capital and performance \citep{Dugan2009} and potentially greater regulatory forbearance \citep{gallemore2020bank}. Expressly, given that each of the settings have strengths and limitations, we invite the reader to interpret our results as a collective body of evidence rather than isolated sets of analysis.

The findings of our paper speak to several strands of literature. First, we contribute to the broader literature on the determinants of banks’ managerial incentives to smooth earnings \citep[e.g.,][]{greenawalt1988bank,laeven2003loan,bikker2005bank,fonseca2008cross,perez2008earnings} by demonstrating that government guarantees are
related to bank managers’ decisions to smooth income. Second, our evidence speaks to the literature on the effects of government guarantees on capital market outcomes \citep[e.g.,][]{o1990deposit,flannery1996evidence,gandhi2015size,kelly2016too,baron2017credit,silva2020fiscal}.Lastly, our findings should be of particular interest to regulators in that we provide evidence regarding the macroeconomic implications of government guarantees. Specifically, our results suggest that these guarantees also have spillover effects for the information environment of the banking sector. Whereas a reduction in income smoothing might be construed as an increase in financial transparency that enhances financial stability \citep{acharya2016banks}, its effect may be associated with undesirable consequences, insofar as counter-cyclical capital buffers can be effective in mitigating the pro-cyclicality of the financial system \citep{dewatripont2012macroeconomic,bouvatier2012provisioning,agenor2015loan}.

The remainder of the paper proceeds as follows: In \Cref{sec:limodhyp}, we  summarize the literature, develop our conceptual framework, and discuss our empirical strategy. \Cref{sec:euro} and \Cref{sec:landes} respectively present our analyses of the creation of the Eurozone and the removal of guarantees to the Landesbanken. \Cref{sec:conc} concludes the paper.

\section{Literature Review, Conceptual Development, and Empirical Strategy}
\label{sec:limodhyp}

\subsection{Government Guarantees and the Banking Sector}

Government guarantees play a particularly important role in the banking sector because of this sector's inherent leverage and the duration mismatch of assets and liabilities that makes it vulnerable to financial crises \citep{calomiris2014fragile}. Governments and quasi-governmental entities sometimes provide banks both explicit and implicit guarantees to limit the size and scope of  financial disasters and to reduce the instability of the banking sector as a whole.

Explicit guarantees are publicly recognized promises made by governments, ideally,
to serve some societal purpose (e.g., stable banking sector). As these guarantees are by definition contractual, they provide reasonable assurance to stakeholders  that their investments will be protected according to specific terms and conditions.\footnote{In this paper, we employ the term ``bank investors’’ as an abuse of terminology to refer to any bank capital provider who relies on financial performance metrics provided by banks and can benefit from potential government bailouts in case the bank itself is unable to fulfill its financial obligations. These include equity holders, debt holders, and depositors.} Implicit government guarantees are less formal and are the most common. While implicit guarantees involve greater uncertainty
due to the lack of a formal contract, they are generally recognized as having an
important effect on the banking sector.

The research on banking finds that government guarantees have important  implications for the asset pricing of banks' equity and debt. \citet*{o1990deposit} investigate the effect of the Comptroller of Currency's announcement that some banks were ``too big to fail", showing that this categorization  was associated with positive wealth effects for equity shareholders of banks included in the ``total insurance policy" and with negative effects for actively traded banks not
included in the Comptroller of the Currency's statement (i.e., control banks). \citet{flannery1996evidence} provide evidence that debt investors impound the value of implicit guarantees into bond prices as the U.S. government's willingness to absorb losses from private debt changes over time. The value of implicit guarantees is also observed in the prices of put options written in the whole banking sector vis-\`{a}-vis a weighted basket of put options written to insure individual banks \citep{kelly2016too}. Other studies provide evidence that government guarantees have pricing implications by showing that banks with greater guarantees have lower adjusted stock returns \citep[e.g.,][]{gandhi2015size,gandhi2020equity}. 

Beyond these positive asset pricing effects from the standpoint of banks' stakeholders, there is also evidence that government guarantees relate to banks' risk-taking  and the competitiveness of the banking sector \citep{stern2004too}. \citet{acharya2007too,acharya2008information} find that government guarantees can lead banks to herd and acquire common risks. \citet*{gropp2011competition} provide evidence that bailout perceptions are associated with an increase in risk taking by the competitors of banks with government guarantees. \citet{duchin2014safer} find that bailed-out banks ex post initiate riskier loans and shift assets toward riskier securities. \citet{baron2020countercyclical} shows that the counter-cyclicality of bank-equity issuance is modulated by implicit guarantees, and \citet{silva2020fiscal} finds that governments' fiscal deficits convey the strength of their guarantees, thereby affecting banks' credit risk and overall financial stability. 

\subsection{Banks’ Incentives for Financial Reporting}

Banks' financial and regulatory reports play a significant role in helping external parties evaluate their financial performance \citep{beatty2014financial,acharya2016banks}. Investors use these reports to evaluate bank performance and identify investment opportunities, while regulators use them in connection with other information to guide their efforts to limit banks’ excessive risk taking and to promote the stability of the financial sector through macroprudential policies. Banks' financial reports are inherently opaque and there is considerable information asymmetry regarding the quality of their loan portfolios \citep*{diamond1984financial,boyd1986financial,morgan2002rating,flannery2004market,flannery20132007}.

Consistent with the importance of banks’ financial reports to investors and regulators, a large body of research has shown that banks have the discretion to use financial reporting  to achieve favorable regulatory and earnings outcomes. Although the results vary in some cases based on the time period, setting, and methodology, some studies have found a variety of possible channels for banks to exercise discretion in reporting in order to achieve (i) higher capital ratios, (ii) higher earnings, and (iii) smoother earnings. These channels include loan loss provisions, the realization of gains and losses on investment securities, pension settlement transactions, taking a big bath (i.e., excess expenses), and other gains and losses \citep*[e.g.,][]{moyer1990capital,scholes1990tax,beatty1995managing,collins1995bank,kim1998impact,ahmed1999bank,beatty2002earnings,barth2017bank}. 

While each of these channels is potentially significant, loan loss provisions have received the most attention in the literature \citep{beatty2014financial}, perhaps in part because they are the most economically significant bank accrual. In this study we focus on banks’ incentives to smooth income as we expect these incentives are the most likely to be affected by government guarantees.

\subsection{The Loan Loss Provisions and Income Smoothing}

Banks’ financial statements are generally opaque, not only because of the inherent complexity of their business models, but also because of managerial incentives to withhold information from capital providers and regulators. In fact, banks' ability to generate money-like safe and liquid securities is contingent on their ability to keep detailed information about their loan portfolios hidden from their capital providers \citep{dang2017banks}. Bank capital providers rely on reported earnings to assess bank performance because they cannot perfectly observe the composition and risk characteristics of bank loans. Consequently, bank managers are encouraged to manage reported earnings due to short-termism and career concerns \citep{rajan1994bank}.

To the extent that the assessment of loan loss provisions is subjective,  allowing a range of possible provisions, even under the scrutiny of auditors and regulators, bank managers may take advantage of such flexibility to smooth earnings \citep{greenawalt1988bank}. Income smoothing can be construed as a way of increasing bank opacity because it reduces investors’ (and possibly regulators’) risk perceptions, leading to positive valuation effects \citep{trueman1988explanation}. Corroborating this argument, some studies have shown that banks use discretion in their reporting in order to manage earnings and circumvent capital requirements \citep*{ahmed1999bank,huizinga2012bank,beatty2014financial,jiang2016competition} which in turn can interfere with bank regulation and capital allocation \citep{jayaratne1996finance,cohen2014bank}. Consistent with the idea that banks’ income smoothing is an instance of reporting distortion, \citet{kilic2013impact} provide evidence that the use of loan loss provisions for income smoothing can impair the informativeness of those provisions for future loan defaults and future stock performance. However, it is important to underscore that in some circumstances investors and regulators might prefer smoother earnings, even if they result in an underlying performance distortion, either because the predictability of accounting earnings is a feature of earnings quality \citep{ewert2015economic,ewert2016more} or because bank opacity enables regulatory forbearance \citep{gallemore2020bank}.

\subsection{Income Smoothing Incentives and Government Guarantees}

In this subsection, we conceptually develop four channels through which government guarantees affect banks’ income smoothing.

Theoretically speaking, government guarantees can be construed as cash infusions from the government to banks in high-marginal-utility states. Hence, the simple introduction of government guarantees has a censoring effect on tail risk that directly benefits banks’ stakeholders. Studies have found evidence of this effect for equity securities \citep{o1990deposit,gandhi2015size}, debt securities  \citep{flannery1996evidence}, and option prices \citep{kelly2016too}. Consequently, the most direct channel through which this effect would operate is by \emph{reducing} the amount of income smoothing needed by bank managers to mitigate stakeholders’ risk perceptions. We formally refer to this channel as the \emph{tail-risk channel}.  

At this juncture, it is important to note that bailouts are not without fiscal costs, since financial aid ultimately would have to be backed by government resources (i.e., fiscal deficits to be financed by either taxes, the issuance of sovereign debt, or the creation money). Thus, it stands to reason that governments may increase regulatory scrutiny of banks when the scope of government guarantees increases. To the extent that  income smoothing can be construed as a form of misreporting, an increase in regulatory monitoring may curtail banks’ ability to smooth earnings. This \emph{government-monitoring channel} could also mean that government guarantees \emph{reduce} banks’ income smoothing. 

Whereas the two channels stated above would similarly indicate that government guarantees are associated with lower levels of income smoothing, the multitude of real (endogenous) effects that such guarantees have on banks may lead to other channels that could predict an opposite association. For example, as government guarantees practically transfer away risks from banks to the government, banks may internalize this exogenous reduction in risk due to such guarantees and endogenously increase their risk-taking behavior, which in turn would increase the volatility of banks' actual earnings and, consequently, lead to an \emph{increase} in income smoothing.  The existence of this third channel, which we refer to as the \emph{risk-taking channel}, is supported by a large body of literature \citep{duchin2014safer,gropp2014impact,fischer2014government}.

Another endogenous response to increased government guarantees is that bank investors can naturally reduce their monitoring in accordance with the direct censoring effect of government guarantees on tail risk (or in anticipation of stronger monitoring by the government). In fact, some forms of income smoothing might inhibit outside monitoring \citep[e.g.,][]{bushman2012accounting}. Overall, the reduction in investors' monitoring would allow bank managers to \emph{increase} income smoothing. We refer to this fourth channel as the \emph{investor-monitoring channel}.  

Critically, we emphasize that while the existence of the four channels is well grounded theoretically, assessing their individual magnitudes ex-ante is a daunting task. Their individual manifestation can occur in parallel or sequentially, and any attempt to ascribe economic magnitudes to the four would entail going beyond traditional reduced-form designs and structurally estimating a full equilibrium model of government guarantees, endogenous risk taking and monitoring by investors and regulators. Hence, for the purpose of this paper, we take an agnostic approach to the direction of their overall effect, acknowledging that their prevalence is a question that cannot be answered without subjecting our settings to an empirical investigation. Furthermore, we are aware that the magnitude of each channel and overall directional effect may be contingent on the specificities of the events studied herein.

\subsection{Empirical Strategy}
\label{subsec:empest}

A significant number of studies investigate banks’ income smoothing by considering their discretionary use of loan loss provisions to distort actual earnings \citep[e.g.,][]{greenawalt1988bank,laeven2003loan,bikker2005bank,fonseca2008cross,perez2008earnings,gebhardt2011mandatory,kilic2013impact}. The goal is to capture a bank’s income smoothing by estimating the association between its loan loss provisions and its earnings before such provisions and taxes \citep[see][]{beatty2014financial}. 

The reasoning behind this approach is that earnings before loan loss provisions should not be related to such provisions  after controlling for other determinants of loan losses. In other words, provisions are simply intended to serve as a buffer for future credit losses---meaning that their levels should largely reflect the risk characteristics of banks' credit portfolios.

However, since setting up provisions requires significant degrees of subjectivity and professional judgment, managers can exert their discretion and record higher (lower) levels of loan loss provisions when earnings before provisions are higher (lower), and this in turn enables them to defer (accelerate) the recognition of net income to (from) a future period. That is, the income smoothing in question would be manifested in the data as a positive association between earnings before the loan loss provisions and the loan loss provisions that managers report. This empirical approach does not simply reflect general volatility in earnings which likely depends on variation in banks’ risk taking, but rather captures smoothing based on the  decision to vary the amount of loan loss provisions in connection with earnings prior to the provision expense.

The aforementioned arguments provide the conceptual framework for the reduced-form approach to modeling the determinants of banks’ loan loss provisions illustrated below, without loss of generality:
\begin{equation}\label{eq:llpgeneral}
\begin{split}
llp_{i,t} = \beta_0 + \underbrace{\beta_1 \times ebllp_{i,t}}_{smoothing\;component} + \Gamma \cdot X_{i,t} + \epsilon_{i,t}
\end{split} \numberthis
\end{equation}
where \llp{} is bank $i$'s loan loss provision for year $t$ scaled by lagged total loans, \ebllp{} is earnings before loan loss provisions and taxes scaled by lagged total loans, and $X_{i,t}$ is a vector of control variables. Along these lines, the term $\beta_1 \times ebllp_{i,t}$ represents the fraction of the reported loan loss provisions that is due to discretionary income smoothing. The vector of control variables $X_{i,t}$, intended to represent the ``non-smoothing'' component of banks' provisions, epitomize bank-level and macro variables that represent banks' credit risks, and business cycles, among others. The coefficient $\beta_1$ is referred to as the income smoothing coefficient.

It is important to emphasize that the scope of our study is to investigate how banks' income smoothing incentives vary with changes in government guarantees. Differently stated, the outcome variable of interest in this article is not the dependent variable of a moodel of loan loss provisions (\llp{} in \Cref{eq:llpgeneral}) but the income smoothing coefficient itself ($\beta_1$). 

In both research settings, we are interested in measuring how banks' income smoothing ($\beta_1$) varies for a specific set of banks affected by a change in government guarantees that occurs in a given year, either unconditionally (differences estimation) or relative to another set of banks (difference-in-differences estimation). This approach entails augmenting the model by adding the interaction terms of \ebllp{} with dummy variables that represent (1) the 
 period after the change in government guarantees, (2) the treatment group, and (3) the post-treatment interaction. In a differences estimation, the coefficient of the interaction term of \ebllp{} with the post-event dummy is our key estimate of interest. In a difference-in-differences design, we are primarily interested in the triple interaction (\ebllp $\times$ treatment-group indicator $\times$ post-event indicator). 

The literature on banks’ income smoothing via loan loss provisions suggests that the set of bank-level controls $X_{i,t}$  should generally reflect the ``non-smoothing’’ determinants of banks’ provisions (such as differences in banks’ risk taking and regulatory scrutiny). Thus, our starting reference is the set of controls considered by \citet{bushman2012accounting,bushman2015delayed} that comprise bank size (\lagonesize{},
 the natural logarithm of the bank's dollar-nominated total assets in millions of USD)
to account for variation in regulatory scrutiny and monitoring, variation in banks' capital structure (\lagonecap{}, measured as total equity divided by total assets), and serial (lagged, contemporaneous, and forward-looking) terms that represent changes in non-performing loans (which proxy for banks’ credit risk taking). However, an important limitation of the two samples is that most banks do not report changes in non-performing loans for the periods in question.\footnote{Specifically, requiring valid observations on changes in non-performing loans would reduce our Eurozone adoption sample from 4,425 to 1,123 bank-year observations, of which 1,041 would come from a single FEA country (Italy) and four other FEA countries (Austria, Belgium, Germany, and the Netherlands) are not even represented (i.e., no valid bank-year observations). As for the Landesbanken sample, none of the bank-year observations report data on non-performing loans.} Insofar as the purpose of these changes is to proxy for banks’ overall credit risk, we overcome this limitation by using additional controls also used in other studies \citep[e.g.,][]{fonseca2008cross}. We account for the creditworthiness of banks’ lending assets by adding the lagged values of the prior two years of the dependent variable (\lagonellp{} and \lagtwollp{}), interest income per interest-earning assets (\IInc{}), and interest expense per interest-bearing liabilities (\IExp{}). We also control for changes in loans (\changeloans{}), inasmuch as credit expansions lead to increases in crash risk and expectations of banking crises  \citep{schularick2012credit,baron2017credit}.
 We obtain all the fundamental data on banks from Bankscope and the macroeconomic data from multiple sources.\footnote{Detailed variable descriptions are given in \Cref{app:sec:variable_definitions}.} 

\section{The Eurozone Creation}
\label{sec:euro}

\subsection{Institutional Background and Data}

In 1999, 11 European countries formally created the Eurozone by adopting the Euro as
their common currency and creating the European Central Bank (ECB). They conceived the Eurozone
 to achieve financial stability within the region, as well as to
enhance economic integration and trade among the member countries. The ECB administers
monetary policy for the Eurozone, holds and manages foreign reserves for member states,
and oils the operation of payment systems.\footnote{Some studies have considered the economic benefits of the Euro through increased capital market integration
and increased growth opportunities \citep*{micco2003currency,zingales2003banks,bekaert2013european,jayaraman2014reporting}. For a timeline of relevant events related to the creation of the Eurozone, see \Cref{tab:eurotimeline} in \Cref{app:sec:eurozoneadd}.}

Academic and anecdotal evidence suggests that the creation of the Eurozone led to the formation of stronger implicit guarantees for banks headquartered in member countries. Scholars have argued theoretically and demonstrated empirically that the implicit guarantees granted to sovereign countries spill over into the banking sector  \citep{acharya2014pyrrhic} for different samples of banks spanning a variety of countries  \citep{correa2014sovereign,gandhi2020equity,silva2020fiscal}. The creation of the Eurozone---through
the adoption of a sole currency and convergence of sovereign interest rates to German rates---represents a clear example of implicit guarantees granted to sovereign bonds of the member countries. In fact, the magnitude and geopolitical relevance of the event represents a shift in the relative importance of individual banking systems for the monetary
union and for the global economy \citep{chinn2012eurozone}.\footnote{The later Eurozone crisis also provides anecdotal examples of how the deterioration of the fiscal stability of peripheral countries affected the credit risk of their major banks \citep{acharya2012tale}.}

The  creation of the Eurozone is a particularly advantageous setting to analyze the income smoothing of a large set of banks. We argue that this setting is plausibly exogenous to banks' decision to smooth income at the time of its implementation because, as emphasized by \citet{stiglitz2016euro}, the creation of the Eurozone was a mostly geopolitical event determined by factors in place long before its official establishment. The idea of a common European currency followed years of discussions among countries and jurisdictions, and the fact that previous attempts to establish a monetary union in the region ultimately failed---notably the European Monetary System (EMS)---suggests that substantial uncertainty still remained even shortly before the monetary union went into force. We empirically validate this premise by analyzing the time series of the Economic Policy Uncertainty (EPU) index for Eurozone jurisdictions, as well as the spreads between 10-year sovereign bonds of Eurozone members and the 10-year US Treasury bond, which serve as a benchmark of the sovereign credit risk and yields of a developed market. The EPU indices collectively reached their peaks in September 1998 and, in line with markets deeming the Eurozone project credible, we observe decreasing sovereign yield spreads from September and October 1998 up to the first months of 1999.\footnote{The time-series dynamics of the macro series are shown in \Cref{fig:macroseries}. Anecdotally, news media articles dating a few days before the onset of the new monetary regime underscored important dimensions of uncertainty that the single currency regime would bring to markets, businesses, and consumers alike \citep{dahlburg1998}.} To avoid overlapping with other major economic events that plausibly affect banks’ performance, we define our sample period as the three years before (1996-1998) and the three years after (1999--2001) the creation of the Eurozone.\footnote{Specifically, it is crucial to avoid overlaps with the burst of the ``dot-com bubble''  and the 9/11 terrorist attacks. The latter is particularly relevant as it represented a major spike in policy uncertainty \citep[]{baker2016measuring} and an increase in military spending. Military spending shocks are problematic for our research purposes because they represent  major fiscal shocks \citep[see][]{ramey2011identifying,auerbach2012measuring,owyang2013government,ramey2018government} and widening deficits affect the strength and credibility of implicit guarantees \citep{acharya2014pyrrhic}. Keeping a balanced pre-post sample also allows us to avoid overlaps with major banking crises \citep[see][]{baron2021banking}, for example, the Swedish/Scandinavian crisis of 1990--1994 \citep{englund1999swedish}. While these reasons corroborate our choice of sample period, it is important to acknowledge that a panel of bank-years spanning over six years is shorter than those typically used in the literature of income smoothing \citep[see][]{liu2006income,bushman2012accounting}. That is, using short panels to estimate income smoothing comes with the caveat that smoothing coefficients may be subject to measurement errors, insofar as smoothing is a dynamic (intertemporal) bank choice.}

We start our sample selection by focusing on 10 out of the 11 countries that joined the Eurozone in 1999---namely, Austria, Belgium, Finland, France, Germany, Ireland, Italy, Luxembourg, the Netherlands, and Portugal---yielding group of treatment countries (``First Euro Adopters'' or FEA hereafter). We exclude Spain because of the contemporaneous implementation of dynamic provisioning mandates by the Spanish Central Bank that compelled banks to smooth provisions over the business cycle \citep{bouvatier2012provisioning,agenor2015loan,jimenez2017macroprudential}.\footnote{In short, dynamic provisioning is a special macro-prudential tool that forces banks to set up provisions incorporating  a general provision that is proportional to the amount of
the increase in the loan portfolio and a general countercyclical provision element.}
Given that our overarching goal is to capture earnings smoothing stemming from managers' discretion and professional judgment in setting loan loss provisions, Spanish banks are excluded from our sample insofar as their provisions reflect a regulatory mandate, not managers' choices. We define our set of control banks  in our difference-in-differences analysis as banks headquartered in countries that never joined the monetary union (``Never Euro Adopters'' or NEA). These countries include European nations that never joined the Eurozone and other non-European developed nations (Australia, Canada, Japan, and the United States).

We gather our bank-level financial data  from Bankscope and our macroeconomic data from multiple sources. We restrict our treatment (FEA) and control (NEA) samples to financial institutions primarily engaged in lending, which includes Commercial Banks, Bank Holding Companies, Savings Banks, and Cooperative Banks. Following the literature, we require banks to have valid financial data that cover at least three years and spans both the pre- and post- event periods, as well as total assets greater than 100 million USD.\footnote{Using more restrictive sampling criteria requiring banks to have at least four (five) years of valid data to be included in our sample reduces the number of bank-year observations from 4,425 to 4,315 (3,537). Our main results are robust to the requirement of at least four (five) years of valid bank-level data.} We also trim the bank-specific variables of \Cref{eq:llpgeneral} at  1\% and 99\%.  

\Cref{tab.euro.sample} shows how our sample of FEA and NEA banks are distributed across different countries (Panel A) and presents the summary statistics of our bank fundamentals data used to estimate empirical specifications of loan loss provisions.
\begin{center}
 \
    [ \textsc{Insert \Cref{tab.euro.sample} About Here} ]
\end{center}

\subsection{Differences Estimation}\label{subsec:eurodiff}

We start our investigation by focusing on the effect of the creation of the Eurozone on the income smoothing of banks headquartered in FEA countries. As mentioned in \Cref{subsec:empest}, we augment \Cref{eq:llpgeneral} specifically with the inclusion of a post-1999 term (i.e., $Post1999_t=1_{\{t\ge 1999\}}$, a dummy variable equal to 1 for the year 1999 onwards, and zero otherwise). \Cref{eq:llpeurodiff} represents the general reduced-form expression to be estimated in our differences estimation analysis, with the coefficient estimate of the $ebllp_{i,t} \times Post1999_{t}$ as our main outcome variable of interest, in that it represents how much income smoothing changed after the shift in implicit guarantees.
\begin{equation}\label{eq:llpeurodiff}
\begin{split}
llp_{i,t} =& \beta_0 + \beta_1 \times ebllp_{i,t} + \beta_2 \times Post1999_{t} + \beta_3 \times ebllp_{i,t} \times Post1999_{t} + \Gamma \cdot X_{i,t} + \epsilon_{i,t}\\
\end{split} \numberthis
\end{equation}

The set of control variables $X_{i,t}$ typically encompasses different bank-level and macroeconomic factors that explain the differences in banks' provisioning choices other than their discretionary income smoothing \citep{beatty2014financial}. That is, abstracting from incentives for smoothing income, banks' provisions should generally reflect the main purpose of this expense account as a buffer for future credit losses. In line with the variety of bank-level controls considered in other studies, we start our analysis by estimating \Cref{eq:llpeurodiff} with different sets of bank-level fundamental variables. Specifically, we consider the model proposed by \citet{bushman2012accounting,bushman2015delayed}  as our reference. Data on changes in non-performing loans for the period surrounding the creation of the Eurozone is limited for the vast majority of banks. We compensate for this limitation by adding other bank-level proxies that capture the overall credit risk incurred by financial institutions. We attempt to account for differences in banks' risk taking by augmenting the model with the interest-rate income (\IInc{}, interest income normalized by the average earning assets) and interest-rate expense (\IExp{}, interest expense normalized by the average interest-bearing liabilities), in addition to changes in loans (\changeloans{}) as well as two lagged terms of the dependent variable (\lagonellp{} and \lagtwollp{}) that summarize the recent history of credit risk expectations by the bank in question \citep{fonseca2008cross}.

\Cref{tab:euro.d} reports the coefficient estimates of our pre-post differences design. In this analysis, as throughout the paper, we dually cluster standard errors at the bank and year levels. They are reported in parenthesis in \Cref{tab:euro.d}. We also include the percentage change in GDP per capita (\PercGDPPC{}) as a macroeconomic control and country and bank-type fixed effects in our differences estimation of the Eurozone creation. Columns 1 to 3 consider different combinations of bank-level controls to attest to the robustness of the main findings. Column 3 represents our baseline model.\footnote{As shown in column 2, requiring loan loss reserves to be included in the model considerably reduces the number of observations of FEA countries, from 4,425 to 1,813. Critically, this condition also reduces the cross-country representation of our sample, the observations of which are distributed as follows: Austria (0), Belgium (1), Finland (5), France (319), Germany (0), Ireland (20), Italy (1,339), Luxembourg (16), the Netherlands (10), and Portugal (103). Hence, we opt to perform the subsequent estimations in the paper with a model that does not consider $llr_{i,t}$ as a control in order to preserve the generalizability of our findings to a larger set of banks. Because the reduced sample is disproportionately comprised of banks headquartered in fewer countries, the coefficient estimate of \PercGDPPC{} should be interpreted with caution.} In column 4 we hold constant the controls of our baseline model but consider a fully-interacted (fully-saturated) model---that is, interacting $Post1999_{t}$ with every control variable in the model.\footnote{Throughout the paper, the specific estimations of fully-interacted models include the interaction terms with the ``post'' indicator (differences and DID) and the interaction terms with ``treatment'' and ``post-treatment'' indicators (DID). We do not tabulate the coefficient estimates of the interaction terms for clarity of exposition.} In column 5 we create a dummy variable representing systemically important financial institutions $I_{I,t}^{LargeBank}$ that equals one if the bank is within the top quintile of assets for a given country and year, and zero otherwise. We then interact this indicator variable with \PostNinetyNine{}, \ebllp{}, and \ebllpPostNinetyNine{} to shed light on heterogeneous effects for large banks (which presumably were granted higher levels of implicit guarantees even before the creation of the Eurozone) vis-\`{a}-vis other banks. In column 6 we re-estimate the baseline model excluding German banks.\footnote{This is done to address concerns that changes in provisioning choices due to the early adoption of some IFRS credit risk standards by German banks alternatively explain our main findings.} In column 7, we add dummy variables indicating whether a given bank year experienced a ``small'' (positive) profit (\SmallP{}) and whether provisions reported are considerably higher than in the previous three years  (\BigBath{}, potentially indicating ``big bath accounting'').

In our last two columns we hold constant the bank-level controls of the baseline specification and augment the model with  other macroeconomic variables. In column 8 we consider two macroeconomic components that vary for different years and countries which are plausibly affected by the establishment of the Eurozone---namely the unemployment rate ($Unemp_{c,t}$) that moves with the business cycle, and foreign trade flows ($Trade_{c,t}$) that could have changed due to the enhanced economic integration between member countries. Last, to understand how non-country-specific macro variables drive banks’ provisions over our sample period, in column 9 we include proxies for economic policy uncertainty and monetary policy shocks. The variation in policy uncertainty happens 
as a matter of course when the setting involves the formation of a monetary union, so we add the \EPUEuro{} index by \citet{baker2016measuring} as a control. Regarding monetary policy, interest rates may affect banks' funding costs and overall profitability. Moreover, monetary easing can be associated with credit booms, asset price bubbles, and risk premia \citep{gali2014monetary,paul2019time,dong2020asset,cieslak2019non,cieslak2019stock}. Given the major role played by US monetary policy decisions in other markets \citep{brusa2020one,dedola2020does,cortes2022unconventional}, we also aggregate the US monetary policy surprises estimated by \citet{romer2004new} and augmented by \citet{breitenlechner2018update}. This aggregation yields our monetary policy proxy (\RRtussandfour).

\begin{center}
 \
    [ \textsc{Insert \Cref{tab:euro.d} About Here} ]
\end{center}

The results shown in \Cref{tab:euro.d} suggest that banks reduced earnings smoothing following the adoption of the Euro, consistent with increases in implicit government guarantees reducing banks' smoothing incentives, as evidenced by negative estimates of $ebllp_{i,t}\times Post1999_t$ (all significant at $p<0.01$). In terms of economic significance, our main result (column 3) suggests a reduction of approximately 49\% ($0.060/0.122$) in banks' earnings smoothing following the creation of the Eurozone. Besides the main effect of interest, we underscore the significant estimates of the risk-taking proxies (i.e., \lagonellp{} and \lagtwollp{}). Moreover, column 5 displays significant interaction effects, consistent with the notion that the large, systemically important banks within a given country and year are naturally granted higher levels of implicit guarantees even before the formation of the Eurozone, as evidenced by the negative coefficient estimate of $I_{I,t}^{LargeBank}\times ebllp_{i,t}$ (i.e., these banks engage in less earnings smoothing than their smaller counterparts). Whereas the main effect for the entire sample of FEA banks is a reduction in income smoothing after the creation of the Eurozone (represented by $ebllp_{i,t}\times Post1999_t$), the positive estimate of the triple interaction $I_{I,t}^{LargeBank}\times ebllp_{i,t}\times Post1999_t$ means that this reduction is attenuated for large banks. Surprisingly, the coefficient estimates of the two global macroeconomic variables in column 9 (i.e., \RRtussandfour{} and \EPUEuro{}) are both statistically insignificant,  mitigating concerns that unusual changes in monetary policy and economic policy in general could have a measurable effect on banks' loan loss provisions. Overall, these results are consistent with the notion that the creation of the Eurozone brought about a positive shift in implicit guarantees at the supranational level, consequently increasing the importance of the banking systems of the member countries and the role of the ECB as a lender of last resort.

While the creation of the Eurozone  may have occurred mostly for geopolitical forces unrelated to banks' incentives to smooth income, the relevance of the event is such that subsequent changes in economic conditions can operate as omitted confounding  factors driving our main results. Differently stated, the event may have triggered a credit expansion, leading to increases in cash risk and expectations of banking crises \citep{schularick2012credit,baron2017credit}. Moreover, the adoption of the Euro may have reduced foreign-exchange risks \citep{bartram2006impact,hutson2010openness}.\footnote{The new currency adoption followed a
convergence process wherein national currency conversions had to be carried out by a triangulation via the Euro. The definite rates were determined by the Council of the
European Union as a function of market rates on December 31, 1998.} Regardless of the confounding macroeconomic variable that is contemporaneously affected by the event, we start with the premise that for any macroeconomic factor to alternatively explain changes in banks' reporting decisions, it should be economically meaningful enough to have real effects on banks' operating performance. Because our main outcome variable is not the levels of banks' loan loss provisions but rather the \emph{association} between provisions and banks' profitability before provisions, we then augment the model of \Cref{{eq:llpeurodiff}} with the levels ($z_{i,t}$) and the interaction terms ($ebllp_{i,t} \times z_{i,t}$), where the variables $z_{i,t}$ correspond to different bank-level performance metrics---namely, changes in loans (\changeloans{}), net-interest margin ($NIM_{i,t}$), interest-income rate (\IInc{}), and interest-expense rate (\IExp{}). The results are reported in \Cref{tab.euro.interacperf} of \Cref{app:sec:eurozoneadd}.

Before proceeding to our difference-in-differences analysis, it is important to note that the validity of our inferences hinges on the idea that the fiscal year of 1999 indeed represented the first reporting period in which managers could fully set up loan loss provisions that incorporated greater expectations of government guarantees.  To better understand the evolution of earnings smoothing during the six years covered by our analysis, we modify the baseline model of \Cref{eq:llpeurodiff} by interacting $ebllp_{i,t}$ with dummies for each of the years, as shown in \Cref{eq:llpeurodiffserial}. 
\begin{equation}\label{eq:llpeurodiffserial}
\begin{split}
llp_{i,t} =& \beta_0 + \displaystyle\sum_{y=1996}^{2001} \beta_1^{(y)} \times 1_{y} \times ebllp_{i,t}  +  \Gamma \cdot X_{i,t} +  \displaystyle\sum_{y=1996}^{2001} \alpha_y \times 1_{y} + \epsilon_{i,t}
\end{split} \numberthis
\end{equation}

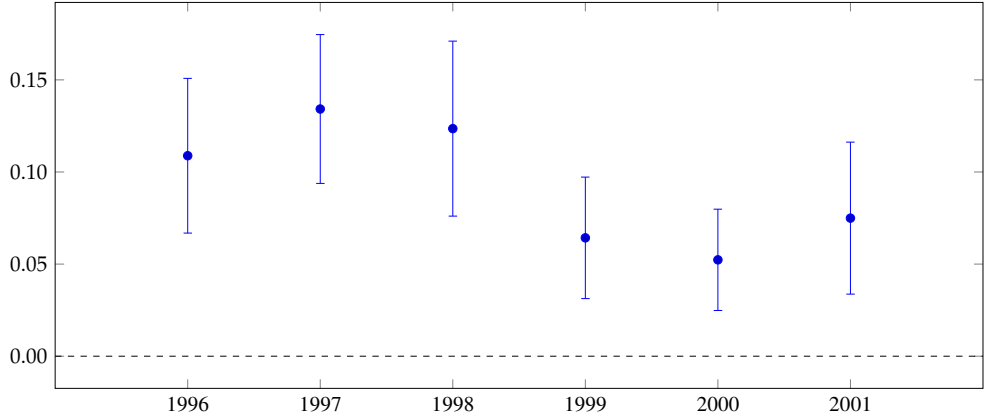
\begin{figure}[t!]
    \centering
\begin{center}    
\begin{tikzpicture}[scale=0.8,trim axis left, trim axis right],
\centering
\begin{axis}[
height=8cm,
width=17cm,  
  xmin=1995,
  xmax=2002,
yticklabel style={
            /pgf/number format/fixed,
            /pgf/number format/precision=2,
            /pgf/number format/fixed zerofill,
	/pgf/number format/1000 sep={,},
        },
xticklabels={1996, 1997, 1998, 1999, 2000, 2001},
ticklabel style = {font=\small},
xtick={1996, 1997, 1998, 1999, 2000, 2001}
]
    \addplot +[only marks,  error bars/.cd, y dir=both, y explicit,
]
    table [x=x, y=y, y error=y-err]{%
      x y y-err
1996  0.1088240834 0.04199108
1997  0.1341875578 0.04039405
1998  0.1235390988 0.04749362
1999 0.0642482296 0.03297597
2000  0.0523102865 0.02749315
2001  0.0749532178 0.04125327
    };
\addplot[dashed] coordinates {(1995,0) (2002,0)};
\end{axis}
\end{tikzpicture}%
\end{center}
\vspace{-0.5cm}
     \begin{minipage}{0.95\linewidth}\scriptsize
   \singlespacing
This figure reports the $\beta_1^{(y)}$ coefficient estimates of \Cref{eq:llpeurodiffserial} considering the baseline specification (controls of column 3 of \Cref{tab:euro.d}) and computed for the sample of FEA banks. Vertical bars represent 95\% confidence intervals. \\
  \end{minipage}
  \caption{Eurozone Creation and Income Smoothing---Evolution of Smoothing Coefficients}
\label{fig:evoluton}
\end{figure}

The series of coefficients $\beta_1^{(y)}$ depicted in \Cref{fig:evoluton} trace out the evolution of loan loss provision smoothing, indicating a noticeable reduction starting in the fiscal year of 1999.

\subsection{Difference-in-Differences Estimation}

The results of \Cref{tab:euro.d} suggest an economically and statistically significant reduction in  banks' income smoothing coinciding with the creation of the Eurozone. Nevertheless, the broad nature of the event still poses additional challenges with respect to the internal validity of our results. While our findings are robust to different specifications, reduced-form models bear important limitations in isolating potential economic factors that vary in time simultaneously with the event in question. We address this issue with a difference-in-differences (DID) approach using bank-year observations from NEA countries as the control group. This augments \Cref{eq:llpeurodiff} as follows:
\begin{equation}\label{eq:llpeurodid}
\begin{split}
llp_{i,t} =& \beta_0 +  \beta_1 \times ebllp_{i,t} + \beta_2 \times Post1999_{t} + \beta_3 \times ebllp_{i,t} \times Post1999_{t} + \beta_4 \times FEA_{i} + \\ 
& +  \beta_5 \times FEA_{i} \times Post1999_{t} + \beta_6 \times ebllp_{i,t} \times FEA_{i} + \beta_7 \times ebllp_{i,t} \times FEA_{i} \times Post1999_{t} + \\
& +\Gamma \cdot X_{i,t} + \epsilon_{i,t}
\end{split} \numberthis
\end{equation}

While we impose similar filters to our treatment and control sets,  concerns may still exist regarding the comparability of the two groups of banks. We address these limitations considering two subsamples of our difference-in-differences observations. First, to address the concerns regarding the cross-country comparability of the main difference-in-differences sample, we restrict our analysis to only Nordic countries. The main advantage of this approach is that it yields one treatment country (Finland) and four control countries (Denmark, Iceland, Norway, and Sweden) whose economies and banking sectors are relatively more comparable than our broader set. This approach, whereas mechanically enhancing cross-country comparability, comes at the expense of a substantial reduction in generalizability. To counterbalance this limitation, in our second approach we obtain propensity-score-matched (PSM) subsamples of treatment and control banks, in which the matching is based on four independent variables in \Cref{eq:llpeurodid} which are relevant in explaining differences in banks' loan loss provisioning: \ebllp{}, the two lagged terms (\lagonellp{} and \lagtwollp{}), and \IInc{}. We  consider a caliper of 0.25, require exact matching of fiscal years, and perform the matching one to one.\footnote{In many situations, propensity score matching algorithms are performed with the underlying objective of controlling for factors that determine the ``selection into treatment'' of the observations. For our purposes, banks cannot endogenously self-select into a treatment or control condition because, by design, the treatment condition is solely a function of the country the bank is domiciled in. Thus, in our analysis, the PSM subsamples simply serve to semi-parametrically account for individual differences between the treatment and control groups.} We then obtain two different PSM samples: one with replacement (\emph{PSM 1}) and one without replacement (\emph{PSM 2}).\footnote{The cross-country distribution of the PSM samples with (without) replacement is as follows: Australia 97 (47), Austria 20 (20), Belgium 37 (36), Canada 200 (58), Denmark 415 (83), Finland 4 (4), France 254 (185), Germany 949 (305), Iceland 2 (2), Ireland 27 (21), Italy 501 (200), Luxembourg 51 (31), the Netherlands 45 (23), Norway 111 (53), Portugal 38 (21), Sweden 53 (17),  Switzerland 53 (42), the United Kingdom 82 (35), and the United States 913 (509). }

\Cref{tab:euro.did} depicts our difference-in-differences results. Panel A presents the OLS coefficient estimates of the main sample (column 1), the main sample with a fully interacted model (column 2), the propensity score-matched samples (columns 3 and 4), and the subsample of Nordic countries (columns 5 and 6). Panel B contrasts the sample properties of the matching variables by contrasting the treatment and control samples before and after matching. Specifically, we present (i) the t-statistics of the tests of differences in means of the treatment and control samples before and after matching, and (ii) the percentage improvement in comparability between  treatment and control banks as a result of the matching process, as per \citet*{ho2007matching}.\footnote{As noted by \citet{ho2007matching}, this is a superior method for comparing the balance between the covariates of treatment and control groups  before versus after matching relative to simply contrasting differences in means.} 

Overall, negative and statistically significant coefficients of the triple interaction term $ebllp_{i,t} \times FEA_{i} \times Post1999_{t}$ in all specifications lend credence to the argument that banks headquartered in FEA countries alone experienced a reduction in earnings smoothing following the creation of the Eurozone. 
\begin{center}
 \
    [ \textsc{Insert \Cref{tab:euro.did} About Here} ]
\end{center}

While the difference-in-differences results provide evidence in line with the existence of direct channels stemming from government guarantees, unobservable time-variant fundamentals that deferentially affect the treatment and control groups other than changes in government guarantees could still be the drivers of our results. To shed more light on this possibility, we conduct two additional analyses. 

First, we perform a standard (linear) difference-in-differences estimation for a variety of outcome variables that represent banks' operating performance:
\begin{equation}\label{eq:yeurodid}
\begin{split}
y_{i,t} &= \beta_0 + \beta_1 \times Post1999_{t} + \beta_2 \times FEA_{i} +  \beta_3 \times FEA_{i} \times Post1999_{t} + \epsilon_{i,t}
\end{split} \numberthis
\end{equation}
where $y_{i,t}$ is (i) banks' earnings before loan loss provisions and taxes (\ebllp{}), (ii) banks' net interest margin ($NIM_{i,t}$), (iii) banks' return on assets ($ROA_{i,t}$), (iv) banks' return on equity ($ROE_{i,t}$); and (v) banks' interbank ratio ($IBank_{i,t}$, defined as the ratio between interbank assets and interbank liabilities). We also consider the linear effects on the indicator variables representing the instances of the two other forms of earnings management considered in \Cref{tab:euro.d}---namely, \BigBath{} and \SmallP{}. The results are reported in \Cref{tab:euro.adddid} of \Cref{app:sec:eurozoneadd}. Notably, the coefficient estimates of $FEA_{i} \times Post1999_{t}$ are all statistically insignificant, failing to provide evidence that banks headquartered in FEA countries experienced significant changes in profitability and operating performance after the creation of the Eurozone.  

Critically, we neither directly assess every possible change in economic conditions and regulatory environment following the formation of the monetary union, nor contradict the literature documenting sizeable real economic effects \citep{rose2001national,frankel2002estimate,bris2006real,lane2006real}. Instead, what we can glean from \Cref{tab:euro.adddid} is that the structural changes associated with the event, whether directly observable or not, are insufficient in the short run to significantly alter the real performance (profitability) of treatment banks vis-\`{a}-vis control banks. 

In our second analysis, we econometrically assess the concerns that confounding factors drive our results by constructing parameter bounds to assess the robustness to omitted variable bias, based on \citet{oster2019unobservable}. The procedure is akin to the one used by \citet{verner2020household}, predicated on the work by \citet{oster2019unobservable} and \citet*{altonji2005selection}. Assuming that the selection based on unobservable factors is proportional to the selection based on observable  variables included in our models of loan loss provisions, we estimate the bounding values ($\beta^{*}$) of our difference-in-differences estimate.  As shown in \Cref{tab:ostereuro}, even in the most conservative case our bounding value is still negative (i.e., indicating a reduction in earnings smoothing).\footnote{As in \citet{verner2020household}, we consider two different choices of baseline (simplified) models as our starting points to implement the \citet{oster2019unobservable} test: (i) a model only with the seven terms \PostNinetyNine{}, \FEA{}, \FEAPostNinetyNine{}, \ebllp{}, \ebllpPostNinetyNine{}, \ebllpFEA{}, and \ebllpFEAPostNinetyNine{} of a standard DID model (i.e., without bank-level and macro controls) and (ii) a model with the seven terms \PostNinetyNine{}, \FEA{}, \FEAPostNinetyNine{}, \ebllp{}, \ebllpPostNinetyNine{}, \ebllpFEA{}, and \ebllpFEAPostNinetyNine{}, augmented with our proxy for business cycles (\PercGDPPC{}) but excluding bank-level controls.} The baseline models are used to estimate $\dot{\beta}$ and $\dot{R}^2$. In both cases, the augmented model (which is used to compute $\tilde{\beta}$ and $\tilde{R}^2$) is the full difference-in-differences model in column 1 of \Cref{tab:euro.did}. These results suggest that our difference-in-differences inferences are robust to potentially omitted correlated variables.

\section{The Removal of Guarantees to the Landesbanken}
\label{sec:landes}

\subsection{Institutional Background and Data}

The German banking system is essentially comprised of private-sector commercial banks, state-owned banks (Landesbanken and savings banks)
and cooperative banks.\footnote{For additional information regarding the banking sector in Germany, see \citet{gropp2014impact}, \citet{fischer2014government} and \citet{baron2020countercyclical}.} The Landesbanken are a group of international wholesale banks, each of which is affiliated with one or more German federal states. They were
established in the nineteenth century with the objective of promoting regional development.

For most of their history, the government has granted the Landesbanken  two layers of  guarantees: an explicit guarantee of all liabilities (``Gew\"{a}hrtr\"{a}gerhaftung") and a maintenance
obligation that requires owners to inject additional equity capital when necessary (``Anstaltslast"). However, in 2001, the German government together with the European
Commission agreed to stop guaranteeing any debt issued after July  2005.\footnote{For a detailed description of the event, see the ``Brussels Agreement" of July 17, 2001. The economic and political facts that led to this decision relate to past complaints from German commercial
banks that such guarantees gave the Landesbanken a competitive advantage. Commercial banks argued that guarantees represented state aid, and therefore violated Article 47 of the European Union Treaty.} Accordingly, we use this plausibly exogenous reduction in government guarantees as an empirical setting to test the effects of this protection on banks' income smoothing.

Using the Landesbanken setting has some important empirical advantages. First, this setting represents a clear shock to government guarantees. Prior to 2005, unlike other German commercial banks, the Landesbanken had the ability to issue debt explicitly guaranteed by the German government. However, the Landesbanken lost these guarantees in 2005, equating them with other German banks. Second, this setting allows us to restrict our attention to banks in a single country, eliminating confounding  cross-country factors.

However, this setting has some limitations. Explicit guarantees are less common than implicit guarantees and governments may want to avoid them for political reasons. In addition, the Landesbanken setting involves state-owned banks that may respond to changes in government guarantees to a different degree than other banks. Thus, while the removal of explicit government guarantees to the Landesbanken likely provides an empirical setting with strong internal validity, the results may have less external validity and, as such, may not generalize to other banks.

We start our sample selection for this analysis by identifying the Landesbanken operating at the end of the fiscal year of 2001. This is accomplished by using the sample of Landesbanken from \citet{fischer2014government} and then checking corporate news articles for subsequent mergers or acquisition activities affecting the sample.  To the extent that our empirical strategy relies on differences and difference-in-differences estimations, we restrict our sample to the Landesbanken for which we have valid observations both before and after the removal of the liability guarantees. We define our sample period starting in 2002 and ending in 2007 for two reasons. First, a six-year period (balancing 3 pre-event and 3 post-event years) allows us to avoid the onset of the Global Financial Crisis. Second, the time frame is consistent with our first empirical setting (i.e., the creation of the Eurozone). Because the ability to issue liabilities with explicit guarantees officially ceased in July 2005, the first fiscal year in which the Landesbanken could have reported provisions under the new regime without government guarantees would be the fiscal year of 2005. Thus, our final sample contains the eight unique Landesbanken depicted in \Cref{tab.landes.sample} (Panel A).

For our difference-in-differences analysis, we consider two groups of control banks. First, we consider a sample of German commercial banks (\emph{control group 1}, hereafter). To control for potential differences in regulation, monitoring, and operating activities, we limit the control sample to banks with a minimum of five billion dollars (USD) in assets.  This requirement eliminates from the control sample small banks which would likely be very different from the Landesbanken. Second, we consider a sample of French government-owned banks (\emph{control group 2}). The two control groups represent the counterfactuals along different components. In other words,  \emph{control group 1}  represents same-country-different-type counterfactuals while \emph{control group 2} can be construed as same-type-different-country counterfactuals. Panel B of \Cref{tab.landes.sample} reports the summary statistics of the bank fundamentals of the treatment sample and the two control samples.

\begin{center}
 \
    [ \textsc{Insert \Cref{tab.landes.sample} About Here} ]
\end{center}

\subsection{Differences Estimation}

To investigate whether the removal of government guarantees relates to changes in the income smoothing  of the Landesbanken, we first use a difference estimation approach. Specifically, we estimate \Cref{eq:llplandesdiff} as follows:
\begin{equation}\label{eq:llplandesdiff}
\begin{split}
llp_{i,t} =& \beta_0 + \beta_1 \times ebllp_{i,t} + \beta_2 \times Post2005_{t} + \beta_3 \times ebllp_{i,t} \times Post2005_{t} +  \Gamma \cdot X_{i,t} + \epsilon_{i,t} 
\end{split} \numberthis
\end{equation}
where \PostTussandFive{} equals one for fiscal years from 2005 onwards (i.e., $Post2005_t=1_{\{t\ge 2005\}}$), when government guarantees were removed. We predict that the coefficient for $ebllp_{i,t} \times Post2005_{t}$ will be positive if the removal of government guarantees increases banks' incentives to smooth earnings via their loan loss provisions.  \Cref{tab.landes.d} reports the results of estimating \Cref{eq:llplandesdiff}, where column 1 presents a standard differences estimation model and column 2 is the fully interacted version thereof. For robustness, we re-estimate the standard differences model eight times, with each estimation excluding one Landesbanken from the sample following the order in Panel A of  \Cref{tab.landes.sample}. The results are shown in columns 3 to 10.

Consistent with the prevalence of either the tail-risk channel or the government-monitoring channel proposed in this paper, we find that the coefficient for $Ebllp_{i,t} \times Post2005_{t}$ is positive and significant, indicating the Landesbanken started engaging in income smoothing after 2005 when government guarantees
were removed. Interestingly, the coefficients  for $Ebllp_{i,t}$ are generally inconsistent with the idea that the Landesbanken used the loan loss provisions to smooth earnings before 2005. These findings are consistent with government guarantees reducing banks' incentives to smooth earnings. 

It is worth noting that, although explicit guarantees were removed, implicit guarantees from German savings banks still provided an additional layer of support to Landesbanken.\footnote{German savings banks were granted the same explicit guarantees before 2001 as the Landesbanken and were similarly affected by the withdrawal of explicit guarantees. We focus on the Landesbanken for two reasons. First, German savings banks were primarily deposit-financed rather than bond-financed. Hence, they were less affected by the removal as depositors remained covered by deposit insurance. Second, German savings banks have no shareholders other than the government and  therefore have weaker incentives to engage in earnings smoothing even after 2005.} As documented in \citet*{puri2011global}, Moody's explicitly considered such support mechanisms as a wider mandate, thereby protecting all creditors rather than only depositors.  
In fact, the main justification provided by Fitch Ratings to downgrade the liabilities of all Landesbanken was that these banks predominantly relied on capital market liability financing. The removal of explicit guarantees would thereby elicit a response from creditors, demanding higher risk premia \citep{korner2013abolishing}. New credit ratings were issued in 2004 but they did not come into effect until July 18, 2005 (i.e., after the loss of explicit guarantees). 

\begin{center}
 \
    [ \textsc{Insert \Cref{tab.landes.d} About Here} ]
\end{center}

\subsection{Difference-in-Differences Estimation}

The results of the difference estimation  in \Cref{tab.landes.d} show an economically and statistically significant change in the income smoothing of Landesbanken after the removal of government guarantees. Despite the cleaner identification when compared to the Eurozone analysis, one possible concern with this approach is that it does not account for potentially correlated omitted variables that vary in time simultaneously with the removal of the guarantees. To alleviate these concerns, we also conduct a difference-in-differences analysis, strengthening the internal validity of our inferences. Finding the ideal set of counterfactuals for our treatment group  of the Landesbanken (i.e., a hypothetical group of German government-owned banks not subject to the removal of guarantees of the Landesbanken) is an impractical task. We attempt to overcome this limitation by using two alternatives of controls in our difference-in-differences estimation: \emph{control group 1} comprised of large German commercial banks (same-country, different-type counterfactuals) and \emph{control group 2} represented by French government-owned banks (different-country, same-type counterfactuals).
\begin{equation}\label{eq:llplandesdid}
\begin{split}
llp_{i,t} =& \beta_0 +  \beta_1 \times ebllp_{i,t} + \beta_2 \times Post2005_{t} + \beta_3 \times ebllp_{i,t} \times Post2005_{t}+ \\ 
&+ \beta_4 \times Landes_{i} +  \beta_5 \times Landes_{i} \times Post2005_{t} + \beta_6 \times ebllp_{i,t} \times Landes_{i} + \\
&+ \beta_7 \times ebllp_{i,t} \times Landes_{i} \times Post2005_{t} +\Gamma \cdot X_{i,t} + \epsilon_{i,t} \\ 
\end{split} \numberthis
\end{equation}

\Cref{tab.landes.did} reports the difference-in-differences estimates of \Cref{eq:llplandesdid}, with columns 1 and 2 giving the estimations for  \emph{control-group 1} and columns 3 and 4 giving the estimations for  \emph{control-group 2}. Overall, the positive and significant estimates of the difference-in-differences effect ($\beta_7$) suggest that the removal of explicit guarantees to the Landesbanken had different effects on their incentives to smooth income relative to the two groups of control banks.

\begin{center}
 \
    [ \textsc{Insert \Cref{tab.landes.did} About Here} ]
\end{center}

A noteworthy event contemporaneous to the removal of government guarantees to the  Landesbanken is the mandatory adoption of IFRS by European Union jurisdictions (including Germany) in 2005. We discuss its potential implications for the inferences of our results in several ways. First, to the extent that we rely on a difference-in-differences estimation, for the IFRS mandate to posit a threat to our inferences it must not only affect the treatment group (Landesbanken) differently than our \emph{control group 1} (German commercial banks)---the direction of the effect should be that the IFRS country mandate increased the smoothing incentives of the Landesbanken vis-\`{a}-vis German commercial banks. On this matter, it is important to note that differences in loan loss provisions cannot stem from the comprehensive adoption of IFRS itself, but specifically from the  incurred loss approach of IAS 39. As explained by \citet{gebhardt2011mandatory}, critically relevant to our setting is the fact that many German (and Austrian) banks were early adopters of specific IFRS rules before the country-level mandate in 2005. In fact, the existence of early IAS 39 adopters, the step-by-step transition to IFRS, and the fact that under German GAAP some banks were allowed to set up hidden reserves are the reasons why \citet{gebhardt2011mandatory} exclude German and Austrian banks when assessing the effect of IFRS adoption on banks’ accounting quality. Overall, \citet{gebhardt2011mandatory} argue that the adoption of IAS 39 is associated with decreases in earnings smoothing, so the potential early adoption could bias our smoothing estimates in the pre-period.\footnote{With respect to the broad adoption of IFRS, the literature highlights mixed evidence as to whether such adoption increase \citep{ahmed2013does} or decrease \citep{barth2008international} earnings management. \citet*{capkun2016effect} investigate the reasons for such conflicting results, emphasizing that the increase in earnings management for countries adopting IFRS after 2005 can be attributed to IFRS standards that went into effect in 2005 providing greater flexibility of accounting choices. However, in our case, the key standard that would directly affect banks' provisioning choices is IAS 39. As pointed out by \citet{gebhardt2011mandatory}, the incurred loss approach of IAS 39 requires banks to provide only for losses incurred as of the balance sheet date, in contrast with the principles-based rules of local (German) GAAP which gave managers considerable leeway  to use discretion (i.e., to smooth income).}

To the extent that early IAS 39 adoption can be an important confounding factor, we provide additional arguments and empirical tests to mitigate such concerns. First, while  \emph{control group 1}  may include early adopters, to the best of our knowledge \emph{control group 2} does not. We underscore that our difference-in-differences results are inferentially consistent for the two groups (and consistent with our main differences estimation result). Second, we perform the test proposed by \citet{oster2019unobservable} to investigate whether our main difference-in-differences parameter is robust to omitted correlated variables. As in our Eurozone creation analysis, we follow the calibration proposed by \citet{oster2019unobservable} and estimate parameter bounds for the difference-in-differences coefficient considering both  control groups (see \Cref{tab:osterlandes}).

These arguments also help alleviate concerns that the difference-in-differences estimation for \emph{control-group 1} can be biased due to a violation of the stable unit treatment value assumption (SUTVA). Specifically, it is possible that the removal of guarantees to the Landesbanken allowed the German government to provide stronger government guarantees to other German banks, including those from \emph{control-group 1}. This argument is theoretically justifiable if the government funds used to bail out the Landesbanken capital providers could have been used to bail out other German banks. While we cannot perfectly rule out this possibility, the consistency of the results for the two control groups and their robustness to omitted variable biases  attenuate such concerns.

\section{Concluding Remarks}
\label{sec:conc}

This article provides the first empirical evidence of the effect of government guarantees on banks' income smoothing. Specifically, we propose four channels through which government guarantees change bank managers' incentives to smooth reported income and use two different settings that constitute changes in government guarantees to empirically investigate this question.

In our first setting we consider the creation of the Eurozone in 1999 as a positive change in implicit government guarantees of banks headquartered in countries comprising the monetary union. We find that \emph{stronger} government guarantees lead to a statistically and economically significant \emph{reduction} in banks' income smoothing. Our results show that stronger implicit government guarantees in connection with the creation of the Eurozone are associated with a large and economically significant effect on banks' incentives to smooth income. In our second setting we examine the direct reduction of explicit government guarantees previously granted to a specific class of German banks (Landesbanken). We find
that following the \emph{removal} of the explicit government guarantees these banks \emph{increased} their income smoothing. These results are also robust to a difference-in-differences estimation with two sets of control banks. In both settings, we provide several arguments and pieces of evidence to substantiate the channels or dispel potential explanations beyond the changes in government guarantees.

This negative association is consistent with the predominance of two channels that predict a negative association between government guarantees and banks' income smoothing---namely, the \emph{tail-risk channel} and the \emph{government-monitoring channel}. Since separating these two directionally equivalent channels is a challenging task when using reduced-form methods, we provide additional arguments that give credence to the \emph{tail-risk channel} as our dominating force. First, the Eurozone  represents stronger government guarantees that stem from the nature of this multilateral project (rather than from individual nation governments). Differently put, the creation of the Eurozone grants banks of the member countries an extra layer of implicit guarantees due to the importance of their financial sectors to the well-functioning of the project. The supranational nature of implicit guarantees makes it less likely that individual government monitoring changed due to the creation of the Eurozone. Second, the Landesbanken setting is characterized by the removal of explicit guarantees given to state-owned banks. Whereas the removal of explicit guarantees could lead to weaker government monitoring, it is possible that for government-owned banks changes in monitoring would be less economically meaningful. 

We underscore the important limitations of our research settings, in particular the existence of potential confounding factors for the Eurozone analysis and the external validity for the Landesbanken analysis. Furthermore, the fact that both events were  announced years before their actual implementation could lead to anticipatory responses by banks. While we conduct a variety of tests to alleviate such concerns, we acknowledge the inability to matter-of-factly rule out alternative channels and rely on the directional consistency of our findings and the complementary nature of the setting to substantiate our conclusions.  We also emphasize that evidence generally consistent with the major role of the \emph{tail-risk channel} in our two settings should not be interpreted as conclusive evidence that this channel will always override the other three channels. Critically, depending on other institutional banking sector variables (regulatory environment, development of capital markets, etc.) and the corresponding nature of government guarantees, the other channels may display stronger effects. 

On the whole, our findings highlight the role of government guarantees as a significant and economically important determinant of banks' financial reporting decisions. In particular, understanding banks' income smoothing through loan loss provisions is a matter of utmost importance. On the one hand, income smoothing can be portrayed as a potential distortion of banks' underlying economic fundamentals. On the other hand, smoothing loan loss provisions over the banks' cycle mitigates the procyclicality of credit conditions and enhances financial stability.

Amid important policy debates on the trade-offs between bail-ins and bailouts which emphasize the role of government guarantees and how to curb systemic risk in the banking sector, our results can inform bank regulators and policymakers on a heretofore unexplored effect of government guarantees.

\section*{Declarations}

All authors certify that they have no affiliations with or involvement in any organization or entity with any financial interest or non-financial interest in the subject matter or materials discussed in this manuscript.

\newpage

\singlespacing
\linespread{1.0}

\bibliographystyle{chicago}
{
\begin{small}
\singlespacing
\linespread{1.0}
\bibliography{DMS_main_arxiv}
\end{small}
}

\singlespacing
\linespread{1.0}

\begin{landscape}
\singlespacing

\begin{table}[H]
\centering
\caption{Eurozone Creation and Income Smoothing---Sample Description and Summary Statistics}

    \vspace{-0.3cm}
   \begin{minipage}{0.95\linewidth}\scriptsize
   \singlespacing
This table reports the description of the Eurozone creation sample per country (Panel A) and summary statistics of bank-level fundamental variables of the First Euro Adopter (FEA, treatment group) and Never Euro Adopter banks (NEA, control group) both separately for the ``pre'' and ``post'' periods (Panel B). Correlation coefficients significant at $p<0.05$ are highlighted. \\

    \end{minipage}
\subcaption*{Panel A: Sample Description}

  \label{tab.euro.sample}
\scriptsize
\begin{tabular}{L{2.3cm}rL{2.3cm}rL{2.0cm}L{2.3cm}rL{2.3cm}rL{2.3cm}r}
  \toprule

  FEA Country & Obs. & FEA Country & Obs. &&NEA Country & Obs. &NEA Country & Obs. &NEA Country & Obs.\\ 
  \hline
Austria &  49 & Ireland &  28 &&  Australia & 127 & Iceland &  18 &Switzerland & 210  \\ 
  Belgium &  82 & Italy & 1359 &&  Canada &  89 & Japan &  30 &United Kingdom & 125\\ 
  Finland &  15 & Luxembourg & 210 &&  Croatia &  17 & Norway & 165 &United States & 2657 \\ 
  France & 526 & Netherlands &  71 &&  Denmark & 166 &   Poland &  50 & \\ 
  Germany & 1980 & Portugal & 105 &&   Hungary &  13 & Sweden &  18 &\\ 

   \bottomrule \\

\end{tabular}

\subcaption*{Panel B: Summary Statistics---FEA and NEA Banks}

\begin{tabular}{@{\extracolsep{-2pt}}L{1.3cm}D{.}{.}{2,2}D{.}{.}{2,2}D{.}{.}{2,2}D{.}{.}{2,2}D{.}{.}{2,2}D{.}{.}{2,2}D{.}{.}{2,2}D{.}{.}{2,2}D{.}{.}{2,2}D{.}{.}{2,2}D{.}{.}{2,2}D{.}{.}{2,2}D{.}{.}{2,2}D{.}{.}{2,2}D{.}{.}{2,2}D{.}{.}{2,2}D{.}{.}{2,2}D{.}{.}{2,2}D{.}{.}{2,2}D{.}{.}{2,2}D{.}{.}{2,2}D{.}{.}{2,2}}
  \toprule
  &\multicolumn{5}{c}{Treatment Banks (1996--1998)} &\multicolumn{5}{c}{Treatment Banks (1999--2001)} &\multicolumn{5}{c}{Control Banks (1996--1998)} &\multicolumn{5}{c}{Control Banks (1999--2001)} \\

  \hline
 & \multicolumn{1}{c}{N} & \multicolumn{1}{c}{Mean} & \multicolumn{1}{c}{P25} & \multicolumn{1}{c}{P50} & \multicolumn{1}{c}{P75} & \multicolumn{1}{c}{N} & \multicolumn{1}{c}{Mean} & \multicolumn{1}{c}{P25} & \multicolumn{1}{c}{P50} & \multicolumn{1}{c}{P75} & \multicolumn{1}{c}{N} & \multicolumn{1}{c}{Mean} & \multicolumn{1}{c}{P25} & \multicolumn{1}{c}{P50} & \multicolumn{1}{c}{P75} & \multicolumn{1}{c}{N} & \multicolumn{1}{c}{Mean} & \multicolumn{1}{c}{P25} & \multicolumn{1}{c}{P50} & \multicolumn{1}{c}{P75}\\ 
  \hline

  \llp & \multicolumn{1}{c}{2194}	&0.009&	0.004	&0.007&	0.012&		\multicolumn{1}{c}{2231}&	0.008&	0.003&	0.007	&0.010	&	\multicolumn{1}{c}{1788}&	0.006	&0.001	&0.004	&0.007	&	\multicolumn{1}{c}{1897}&	0.006&	0.001&	0.004&	0.007 \\ 
  \ebllp & \multicolumn{1}{c}{2194}	&0.029&	0.016&	0.023&	0.036	&	\multicolumn{1}{c}{2231}&	0.025&	0.013&	0.020&	0.031&		\multicolumn{1}{c}{1788}&	0.038	&0.021&	0.032&	0.044&		\multicolumn{1}{c}{1897}&	0.037&	0.020&	0.030&	0.041 \\ 
  \lagonellp & \multicolumn{1}{c}{2194}	&0.010&	0.004&	0.008&	0.012&		\multicolumn{1}{c}{2231}&	0.008&	0.003&	0.007&	0.011&		\multicolumn{1}{c}{1788}&	0.006	&0.000	&0.003	&0.007	&	\multicolumn{1}{c}{1897}&	0.006&	0.001	&0.004	&0.007 \\ 
  \lagtwollp & \multicolumn{1}{c}{2194}	&0.010	&0.004	&0.008	&0.013		&\multicolumn{1}{c}{2231}	&0.009	&0.003	&0.007	&0.012		&\multicolumn{1}{c}{1788}	&0.006	&0.000	&0.003	&0.006		&\multicolumn{1}{c}{1897}	&0.006	&0.001	&0.004	&0.007\\ 
  \lagonesize & \multicolumn{1}{c}{2194}	&6.897	&5.740	&6.637	&7.726		&\multicolumn{1}{c}{2231}	&6.910	&5.712	&6.623	&7.763		&\multicolumn{1}{c}{1788}	&7.342	&6.041	&7.132	&8.316		&\multicolumn{1}{c}{1897}	&7.664	&6.368	&7.526	&8.719\\ 
  \lagonecap & \multicolumn{1}{c}{2194}	&0.064	&0.040	&0.051	&0.079		&\multicolumn{1}{c}{2231}	&0.069	&0.043	&0.054	&0.086		&\multicolumn{1}{c}{1788}	&0.089	&0.068	&0.081	&0.101		&\multicolumn{1}{c}{1897}	&0.090	&0.067	&0.081	&0.101
 \\ 
  \changeloans & \multicolumn{1}{c}{2194}	&0.097	&0.027&	0.072&	0.131&		\multicolumn{1}{c}{2231}&	0.102&	0.015&	0.073&	0.149&		\multicolumn{1}{c}{1788}&	0.188&	0.046&	0.116&	0.227&		\multicolumn{1}{c}{1897}&	0.151&	0.022&	0.099&	0.198 \\ 
  \IInc & \multicolumn{1}{c}{2194}	&0.074&	0.065&	0.070&	0.080	&	\multicolumn{1}{c}{2231}	&0.061&	0.056&	0.061&	0.064	&	\multicolumn{1}{c}{1788}	&0.080	&0.073	&0.080	&0.086	&	\multicolumn{1}{c}{1897}	&0.079&	0.072&	0.077	&0.084 \\ 
  \IExp & \multicolumn{1}{c}{2194}	&0.046	&0.037	&0.041&	0.051	&	\multicolumn{1}{c}{2231}	&0.035	&0.028&	0.034	&0.039	&	\multicolumn{1}{c}{1788}	&0.044&	0.036&	0.042&	0.049	&	\multicolumn{1}{c}{1897}&	0.043&	0.035&	0.042&	0.050  \\ 
  
   \hline

\end{tabular}

\begin{tabular}{@{\extracolsep{-2pt}}L{3cm}D{.}{.}{2,2}D{.}{.}{2,2}D{.}{.}{2,2}D{.}{.}{2,2}D{.}{.}{2,2}D{.}{.}{2,2}D{.}{.}{2,2}D{.}{.}{2,2}D{.}{.}{2,2}D{.}{.}{2,2}D{.}{.}{2,2}D{.}{.}{2,2}D{.}{.}{2,2}D{.}{.}{2,2}D{.}{.}{2,2}D{.}{.}{2,2}D{.}{.}{2,2}D{.}{.}{2,2}D{.}{.}{2,2}D{.}{.}{2,2}}
\hline
&\multicolumn{9}{c}{Correlation Matrix: Treatment Sample (1996--1998)}  &\multicolumn{9}{c}{Correlation Matrix: Treatment Sample (1999--2001)} \\

  \hline
\llp &1.000	&\cellcolor{miewblue}0.291	&\cellcolor{miewblue}0.488	&\cellcolor{miewblue}0.384	&0.014	&\cellcolor{miewblue}0.124	&0.026	&\cellcolor{miewblue}0.141	&0.011 & 1.000	&\cellcolor{miewblue}0.194&	\cellcolor{miewblue}0.326	&\cellcolor{miewblue}0.327	&\cellcolor{miewblue}-0.078	&0.033	&-0.011	&\cellcolor{miewblue}0.132&	-0.026

 \\ 
\ebllp & \cellcolor{miewblue}0.291	&1.000&	\cellcolor{miewblue}0.128	&\cellcolor{miewblue}0.078	&\cellcolor{miewblue}0.054	&\cellcolor{miewblue}0.328	&\cellcolor{miewblue}0.122	&\cellcolor{miewblue}0.259	&\cellcolor{miewblue}0.226 &\cellcolor{miewblue}0.194	&1.000	&\cellcolor{miewblue}0.094	&\cellcolor{miewblue}0.103	&\cellcolor{miewblue}0.084	&\cellcolor{miewblue}0.169	&\cellcolor{miewblue}0.198	&0.038	&\cellcolor{miewblue}0.047
 \\ 
 \lagonellp & \cellcolor{miewblue}0.488&	\cellcolor{miewblue}0.128	&1.000	&\cellcolor{miewblue}0.501	&\cellcolor{miewblue}0.043	&\cellcolor{miewblue}0.093	&\cellcolor{miewblue}-0.054	&\cellcolor{miewblue}0.129&	0.013 & \cellcolor{miewblue}0.326	&\cellcolor{miewblue}0.094	&1.000	&\cellcolor{miewblue}0.390	&-0.034	&0.031	&\cellcolor{miewblue}-0.050	&\cellcolor{miewblue}0.127	&-0.024

 \\ 
\lagtwollp &  \cellcolor{miewblue}0.384&	\cellcolor{miewblue}0.078&	\cellcolor{miewblue}0.501&	1.000&	\cellcolor{miewblue}0.061&	\cellcolor{miewblue}0.047&	-0.025&	\cellcolor{miewblue}0.112&	-0.017
& \cellcolor{miewblue}0.327&	\cellcolor{miewblue}0.103&	\cellcolor{miewblue}0.390&	1.000&	0.008&	\cellcolor{miewblue}0.054&	0.012&	\cellcolor{miewblue}0.077&	\cellcolor{miewblue}-0.062\\

 \lagonesize &0.014&	\cellcolor{miewblue}0.054&	\cellcolor{miewblue}0.043&	\cellcolor{miewblue}0.061&	1.000&	\cellcolor{miewblue}-0.162&	-0.001&	\cellcolor{miewblue}0.075&	\cellcolor{miewblue}0.337
 & \cellcolor{miewblue}-0.078&	\cellcolor{miewblue}0.084&	-0.034&	0.008&	1.000&	\cellcolor{miewblue}-0.186&	\cellcolor{miewblue}0.106&	0.029&	\cellcolor{miewblue}0.287  \\ 
 \lagonecap & \cellcolor{miewblue}0.124	&\cellcolor{miewblue}0.328	&\cellcolor{miewblue}0.093&	\cellcolor{miewblue}0.047	&\cellcolor{miewblue}-0.162	&1.000	&\cellcolor{miewblue}0.134	&\cellcolor{miewblue}0.312	&\cellcolor{miewblue}0.084
&0.033	&\cellcolor{miewblue}0.169	&0.031	&\cellcolor{miewblue}0.054	&\cellcolor{miewblue}-0.186	&1.000	&\cellcolor{miewblue}0.205	&-0.034	&\cellcolor{miewblue}-0.306
 \\ 
 \changeloans & 0.026&	\cellcolor{miewblue}0.122	&\cellcolor{miewblue}-0.054&	-0.025&	-0.001&	\cellcolor{miewblue}0.134&	1.000&	\cellcolor{miewblue}0.049	&0.027 & -0.011&	\cellcolor{miewblue}0.198&	\cellcolor{miewblue}-0.050&	0.012	&\cellcolor{miewblue}0.106 &\cellcolor{miewblue}	0.205&	1.000	&\cellcolor{miewblue}0.104&\cellcolor{miewblue}	0.042

 \\ 
  \IInc & \cellcolor{miewblue}0.141&	\cellcolor{miewblue}0.259&	\cellcolor{miewblue}0.129&	\cellcolor{miewblue}0.112&	\cellcolor{miewblue}0.075&	\cellcolor{miewblue}0.312&	\cellcolor{miewblue}0.049&	1.000&	\cellcolor{miewblue}0.708
& \cellcolor{miewblue}0.132&	0.038&	\cellcolor{miewblue}0.127&	\cellcolor{miewblue}0.077&	0.029&	-0.034&	\cellcolor{miewblue}0.104&	1.000&	\cellcolor{miewblue}0.654
 \\ 
  \IExp & 0.011	&\cellcolor{miewblue}0.226	&0.013	&-0.017	&\cellcolor{miewblue}0.337	&\cellcolor{miewblue}0.084	&0.027&	\cellcolor{miewblue}0.708&	1.000
&  -0.026&	\cellcolor{miewblue}0.047&	-0.024&	\cellcolor{miewblue}-0.062&	\cellcolor{miewblue}0.287&	\cellcolor{miewblue}-0.306&	\cellcolor{miewblue}0.042&	\cellcolor{miewblue}0.654&	1.000
\\

\hline
&\multicolumn{9}{c}{Correlation Matrix: Control Sample (1996--1998)}&\multicolumn{9}{c}{Correlation Matrix: Control Sample (1999--2001)} \\

  \hline

\llp & 1.000&\cellcolor{miewblue}	0.209&\cellcolor{miewblue}	0.586&\cellcolor{miewblue}	0.455&	0.023&	\cellcolor{miewblue}0.063&	\cellcolor{miewblue}0.102&	\cellcolor{miewblue}0.310&	\cellcolor{miewblue}0.103 & 1.000&	\cellcolor{miewblue}0.223&\cellcolor{miewblue}	0.628\cellcolor{miewblue}	&0.589\cellcolor{miewblue}	&0.048\cellcolor{miewblue}	&0.177\cellcolor{miewblue}	&0.104\cellcolor{miewblue}	&0.463\cellcolor{miewblue}	&0.186 \\ 
  \ebllp &  \cellcolor{miewblue}0.209	&1.000	&\cellcolor{miewblue}0.204	&\cellcolor{miewblue}0.204	&0.021	&\cellcolor{miewblue}0.287	&\cellcolor{miewblue}0.125	&\cellcolor{miewblue}0.186	&\cellcolor{miewblue}0.056 &\cellcolor{miewblue}0.223	&1.000	&\cellcolor{miewblue}0.163	&\cellcolor{miewblue}0.207	&\cellcolor{miewblue}0.064	&\cellcolor{miewblue}0.233	&\cellcolor{miewblue}0.214	&\cellcolor{miewblue}0.160	&-0.035 \\ 
 \lagonellp &  \cellcolor{miewblue}0.586 &	\cellcolor{miewblue}0.204	&1.000&	\cellcolor{miewblue}0.639	&0.019	&\cellcolor{miewblue}0.092	&0.009	&\cellcolor{miewblue}0.262&	\cellcolor{miewblue}0.060
& \cellcolor{miewblue}0.628	&\cellcolor{miewblue}0.163	&1.000&\cellcolor{miewblue}	0.604&	0.023&\cellcolor{miewblue}	0.149&	-0.017&	\cellcolor{miewblue}0.396&\cellcolor{miewblue}	0.145 \\ 
 \lagtwollp & \cellcolor{miewblue}0.455	&\cellcolor{miewblue}0.204&	\cellcolor{miewblue}0.639	&1.000&	-0.006&	\cellcolor{miewblue}0.156&	0.013	&\cellcolor{miewblue}0.230	&\cellcolor{miewblue}0.089
& \cellcolor{miewblue}0.589	&\cellcolor{miewblue}0.207&	\cellcolor{miewblue}0.604	&1.000	&0.014&	\cellcolor{miewblue}0.181	&0.010	&\cellcolor{miewblue}0.357	&\cellcolor{miewblue}0.117 \\ 
  \lagonesize &  0.023	&0.021	&0.019&	-0.006&	1.000&	\cellcolor{miewblue}-0.307&	\cellcolor{miewblue}-0.059&	0.027&	\cellcolor{miewblue}0.075
& \cellcolor{miewblue}0.048&\cellcolor{miewblue}	0.064&	0.023&	0.014&	1.000&	\cellcolor{miewblue}-0.291&	-0.031&	-0.006&	\cellcolor{miewblue}0.067
\\ 
  \lagonecap & \cellcolor{miewblue}0.063&	\cellcolor{miewblue}0.287&	\cellcolor{miewblue}0.092&	\cellcolor{miewblue}0.156&	\cellcolor{miewblue}-0.307&	1.000&	\cellcolor{miewblue}0.051&	0.043&	\cellcolor{miewblue}-0.068
& \cellcolor{miewblue}0.177	&\cellcolor{miewblue}0.233&	\cellcolor{miewblue}0.149&	\cellcolor{miewblue}0.181&	\cellcolor{miewblue}-0.291&	1.000&	\cellcolor{miewblue}0.050&	\cellcolor{miewblue}0.150&	\cellcolor{miewblue}-0.122
\\ 
 \changeloans &\cellcolor{miewblue}0.102&\cellcolor{miewblue}	0.125&	0.009&	0.013&	\cellcolor{miewblue}-0.059&	\cellcolor{miewblue}0.051&	1.000&	\cellcolor{miewblue}0.229&	\cellcolor{miewblue}0.185
& \cellcolor{miewblue}0.104&	\cellcolor{miewblue}0.214&	-0.017&	0.010&	-0.031&	\cellcolor{miewblue}0.050&	1.000&	\cellcolor{miewblue}0.239&	\cellcolor{miewblue}0.176
 \\ 
 \IInc & \cellcolor{miewblue}0.310	&\cellcolor{miewblue}0.186	&\cellcolor{miewblue}0.262&	\cellcolor{miewblue}0.230&	0.027&	0.043	&\cellcolor{miewblue}0.229&	1.000&	\cellcolor{miewblue}0.636
& \cellcolor{miewblue}0.463&	\cellcolor{miewblue}0.160	&\cellcolor{miewblue}0.396&\cellcolor{miewblue}	0.357	&-0.006&	\cellcolor{miewblue}0.150&	\cellcolor{miewblue}0.239&	1.000&	\cellcolor{miewblue}0.534
 \\ 
 \IExp & \cellcolor{miewblue}0.103&	\cellcolor{miewblue}0.056&	\cellcolor{miewblue}0.060&	\cellcolor{miewblue}0.089&	\cellcolor{miewblue}0.075&	\cellcolor{miewblue}-0.068&	\cellcolor{miewblue}0.185&	\cellcolor{miewblue}0.636&	1.000
 & \cellcolor{miewblue}0.186&	-0.035&	\cellcolor{miewblue}0.145&\cellcolor{miewblue}	\cellcolor{miewblue}0.117&\cellcolor{miewblue}	0.067&	\cellcolor{miewblue}-0.122&	\cellcolor{miewblue}0.176&\cellcolor{miewblue}	0.534&	1.000
 \\ 
 
\bottomrule
\end{tabular}

\end{table}

\end{landscape}

\begin{table}[H] \centering\singlespacing 
  \caption{Eurozone Creation and Income Smoothing---Differences Estimation} 
  
    \vspace{-0.5cm}
     \begin{minipage}{1\linewidth}\scriptsize
   \singlespacing
  This table reports OLS coefficient estimates of the differences model  computed for the Eurozone creation (\Cref{eq:llpeurodiff}). Columns 1--3 depict the main effect with different choices of bank-level controls $X_{i,t}$. Specifications 4--7 hold constant the bank-level controls of the baseline model (column 3) to consider (i) a fully interacted model (column 4), (ii) a model augmented to include the interactions with an indicator variable for large banks ($I_{i,t}^{LargeBank}$) (column 5), (iii) an estimation excluding German banks (column 6), and (iv) the inclusion of variables indicating two other forms of earnings management (column 7). Column 8 augments the model to include two country-level macroeconomic variables: unemployment rate ($Unemp_{c,t}$) and foreign-trade flows ($Unemp_{c,t}$) Column 9 augments the model to include two global macroeconomic variables: US monetary policy shocks following \citet{romer2004new} (\RRtussandfour) and the European variant of the policy uncertainty index (\EPUEuro) following \citet{baker2016measuring}. Country and bank-type fixed effects are included in all specifications.
Robust standard errors clustered at the bank and year level are reported within parentheses.  *, **, and *** indicate significance at the 10\%, 5\%, and 1\% levels, respectively. \\
  \end{minipage}

  \label{tab:euro.d} \scriptsize
\scriptsize 
\begin{tabular}{@{\extracolsep{-2pt}}lD{.}{.}{2,5}D{.}{.}{2,5}D{.}{.}{2,5}D{.}{.}{2,5}D{.}{.}{2,5}D{.}{.}{2,5}D{.}{.}{2,5}D{.}{.}{2,5}D{.}{.}{2,5}D{.}{.}{2,5} D{.}{.}{2,5}} 
\toprule 
 & \multicolumn{9}{c}{\textit{Dependent variable:} \llp{}} \\ 
\cline{2-10} 

\\[-1.8ex] & \multicolumn{1}{c}{(1)} & \multicolumn{1}{c}{(2)} & \multicolumn{1}{c}{(3)} & \multicolumn{1}{c}{(4)} & \multicolumn{1}{c}{(5)} & \multicolumn{1}{c}{(6)} & \multicolumn{1}{c}{(7)} & \multicolumn{1}{c}{(8)} & \multicolumn{1}{c}{(9)} \\ 
\hline \\[-1.8ex]

  \PostNinetyNine{} & 0.001^{**} & 0.0002 & 0.002^{**} & 0.002 & 0.002^{***} & 0.001 & 0.002^{***} & 0.001 & 0.001^{***} \\ 
  & (0.001) & (0.001) & (0.001) & (0.004) & (0.001) & (0.001) & (0.001) & (0.001) & (0.0005) \\ 
  \ebllp{} & 0.128^{***} & 0.071^{**} & 0.122^{***} & 0.118^{***} & 0.136^{***} & 0.086^{***} & 0.134^{***} & 0.126^{***} & 0.127^{***} \\ 
  & (0.021) & (0.031) & (0.021) & (0.020) & (0.023) & (0.019) & (0.024) & (0.022) & (0.022) \\ 
 \rowcolor{miewblue} \ebllpPostNinetyNine{} & -0.063^{***} & -0.052^{***} & -0.060^{***} & -0.051^{***} & -0.074^{***} & -0.059^{***} & -0.059^{***} & -0.068^{***} & -0.067^{***} \\ 
  & (0.009) & (0.015) & (0.011) & (0.006) & (0.010) & (0.011) & (0.012) & (0.013) & (0.014) \\ 
  \lagonellp{} & 0.287^{***} & 0.376^{***} & 0.279^{***} & 0.346^{***} & 0.276^{***} & 0.289^{***} & 0.267^{***} & 0.276^{***} & 0.280^{***} \\ 
  & (0.056) & (0.039) & (0.058) & (0.085) & (0.059) & (0.057) & (0.058) & (0.057) & (0.056) \\ 
  \lagtwollp{} & 0.195^{***} & 0.155^{***} & 0.187^{***} & 0.160^{***} & 0.184^{***} & 0.212^{***} & 0.175^{***} & 0.189^{***} & 0.192^{***} \\ 
  & (0.031) & (0.033) & (0.029) & (0.055) & (0.030) & (0.029) & (0.028) & (0.029) & (0.029) \\ 
  
    \lagonesize{} &  & -0.0001 & -0.0001 & 0.0001 & 0.0002 & -0.0001 & -0.0001 & -0.0001 & -0.0001 \\ 
  &  & (0.0002) & (0.0002) & (0.0003) & (0.0002) & (0.0001) & (0.0002) & (0.0002) & (0.0002) \\ 
  
  \lagonecap{} & 0.005 & 0.007 & -0.001 & 0.002 & -0.001 & 0.007 & -0.007 & -0.001 & -0.002 \\ 
  & (0.005) & (0.009) & (0.006) & (0.006) & (0.007) & (0.009) & (0.007) & (0.006) & (0.006) \\ 
  \changeloans{} & 0.0003 & 0.001 & 0.00001 & 0.001 & 0.0002 & 0.0002 & 0.00004 & 0.0001 & 0.0001 \\ 
  & (0.001) & (0.001) & (0.001) & (0.002) & (0.001) & (0.001) & (0.001) & (0.001) & (0.001) \\ 

 \LLR{} &  & 0.008^{*} &  &  &  &  &  &  &  \\ 
  &  & (0.005) &  &  &  &  &  &  &  \\ 
  \IInc{} &  &  & 0.082^{***} & 0.059^{**} & 0.086^{***} & 0.062^{**} & 0.099^{***} & 0.081^{***} & 0.082^{***} \\ 
  &  &  & (0.030) & (0.027) & (0.030) & (0.029) & (0.032) & (0.029) & (0.027) \\ 
  \IExp{} &  &  & -0.071^{***} & -0.067^{***} & -0.074^{***} & -0.059^{***} & -0.093^{***} & -0.075^{***} & -0.067^{***} \\ 
  &  &  & (0.024) & (0.019) & (0.023) & (0.022) & (0.025) & (0.023) & (0.021) \\ 

  \SmallP{} &  &  &  &  &  &  & 0.004^{***} &  &  \\ 
  &  &  &  &  &  &  & (0.002) &  &  \\ 
  \BigBath{} &  &  &  &  &  &  & 0.037^{***} &  &  \\ 
  &  &  &  &  &  &  & (0.010) &  &  \\ 
  \PercGDPPC{} & -0.003 & -0.003 & -0.003 & -0.003 & -0.004 & 0.001 & -0.002 & -0.003 & -0.008^{***} \\ 
  & (0.003) & (0.002) & (0.003) & (0.004) & (0.003) & (0.002) & (0.003) & (0.003) & (0.003) \\ 
  $I_{I,t}^{LargeBank}$  &  &  &  &  & 0.001^{*} &  &  &  &  \\ 
  &  &  &  &  & (0.001) &  &  &  &  \\ 
  $I_{I,t}^{LargeBank}\times Post1999_t$ &  &  &  &  & -0.003^{***} &  &  &  &  \\ 
  &  &  &  &  & (0.001) &  &  &  &  \\ 
  $I_{I,t}^{LargeBank}\times ebllp_{i,t}$  &  &  &  &  & -0.080^{***} &  &  &  &  \\ 
  &  &  &  &  & (0.024) &  &  &  &  \\ 
  $I_{I,t}^{LargeBank}\times ebllp_{i,t} \times Post1999_t$  &  &  &  &  & 0.077^{***} &  &  &  &  \\ 
  &  &  &  &  & (0.027) &  &  &  &  \\ 
  $Unemp_{c,t}$ &  &  &  &  &  &  &  & 0.0002 & 0.001 \\ 
  &  &  &  &  &  &  &  & (0.0003) & (0.0004) \\ 
  $Trade_{c,t}$ &  &  &  &  &  &  &  & 0.0001^{**} & 0.0001^{**} \\ 
  &  &  &  &  &  &  &  & (0.00004) & (0.00004) \\ 
  \EPUEuro{} &  &  &  &  &  &  &  &  & -0.00004 \\ 
  &  &  &  &  &  &  &  &  & (0.0001) \\ 
  \RRtussandfour{} &  &  &  &  &  &  &  &  & -0.001 \\ 
  &  &  &  &  &  &  &  &  & (0.001) \\ 
 \hline \\[-1.8ex] 
Observations & \multicolumn{1}{c}{4,425} & \multicolumn{1}{c}{1,813} & \multicolumn{1}{c}{4,425} & \multicolumn{1}{c}{4,425} & \multicolumn{1}{c}{4,425} & \multicolumn{1}{c}{2,445} & \multicolumn{1}{c}{4,425} & \multicolumn{1}{c}{4,425} & \multicolumn{1}{c}{4,425} \\ 

Adjusted R$^{2}$ & \multicolumn{1}{c}{0.260} & \multicolumn{1}{c}{0.343} & \multicolumn{1}{c}{0.266} & \multicolumn{1}{c}{0.271} & \multicolumn{1}{c}{0.269} & \multicolumn{1}{c}{0.279} & \multicolumn{1}{c}{0.302} & \multicolumn{1}{c}{0.268} & \multicolumn{1}{c}{0.266} \\

\bottomrule

\end{tabular}
\end{table}

\newpage
\begin{table}[H] \centering\singlespacing\scriptsize  
  \caption{Eurozone Creation and Income Smoothing---DID Estimation} 
  \label{tab:euro.did} 
    \vspace{-0.25cm}
     \begin{minipage}{1\linewidth}\scriptsize
   \singlespacing
  This table reports the results of the DID analysis of the Eurozone creation. Panel A depicts the OLS coefficient estimates of the DID model (\Cref{eq:llpeurodid}), considering (i) the estimation with the full sample of treatment (FEA) and control (NEA) banks (specification 1), (ii) the full sample estimated with a fully interacted model (specification 2), (iii) an analysis with propensity-score-matched subsamples of treatment (FEA) and control (NEA) banks (specifications 3 and 4, respectively  for PSM samples with and without replacement), and (iv) an estimation with a restricted sample of Nordic countries, which yields one FEA (Finland) and three NEA (Denmark, Iceland, Norway, and Sweden) groups (specification 5), also including the pre-event terms as in \citet{bertrand2004much} (specification 6). Bank-type fixed effects are included in all specifications. Robust standard errors clustered at the bank and year level are reported within parentheses.  *, **, and *** indicate significance at the 10\%, 5\%, and 1\% levels, respectively. Panel B presents the effect of the propensity score matching algorithms on the covariate balancing of the baseline loan-loss provisions model. The first two columns represents the t statistics of the differences in means of treatment and control groups pre- and post-matching. The middle six columns report the percent improvement in covariates balance as a result of matching, defined as $(|b|-|a|)/|b|$, where $b$ ($a$) is the statistic for foreign``minus'' the statistic for domestic banks, measured before (after) matching \citep{ho2007matching}. The last two columns represent the Kolmogorov's D statistics comparing the empirical distributions of treatment and control groups pre- and post-matching.   \\
  \end{minipage}
    \subcaption*{Panel A: OLS Coefficient Estimates}
\scriptsize 
\begin{tabular}{@{\extracolsep{-2pt}}lD{.}{.}{2,5} D{.}{.}{2,5} D{.}{.}{2,5} D{.}{.}{2,5} D{.}{.}{2,5} D{.}{.}{2,5} } 
\toprule
 & \multicolumn{5}{c}{\textit{Dependent variable:} \llp} \\ 
\cline{2-7} 
\\[-1.8ex] & \multicolumn{2}{c}{Full Sample} & \multicolumn{1}{c}{PSM 1} & \multicolumn{1}{c}{PSM 2} & \multicolumn{2}{c}{Nordic} \\ 
\\[-1.8ex] & \multicolumn{1}{c}{(1)} & \multicolumn{1}{c}{(2)} & \multicolumn{1}{c}{(3)} & \multicolumn{1}{c}{(4)} & \multicolumn{1}{c}{(5)} & \multicolumn{1}{c}{(6)}\\ 
\hline \\[-1.8ex] 

 \PostNinetyNine{} & 0.0004 & -0.006^{**} & -0.003 & -0.001 & -0.002^{*} & -0.001 \\ 
  & (0.001) & (0.003) & (0.003) & (0.002) & (0.001) & (0.001) \\ 
  \FEA{} & -0.001 & 0.003 & -0.005^{***} & -0.004^{***} & -0.004^{***} & -0.003^{*} \\ 
  & (0.001) & (0.002) & (0.001) & (0.001) & (0.001) & (0.002) \\ 
  \FEAPostNinetyNine{} & 0.001^{***} & 0.009^{*} & 0.005^{***} & 0.004^{***} & 0.007^{***} & 0.006^{***} \\ 
  & (0.001) & (0.005) & (0.002) & (0.001) & (0.002) & (0.002) \\ 
  \ebllp & 0.013 & 0.015 & -0.074 & -0.008 & -0.011 & -0.017 \\ 
  & (0.011) & (0.014) & (0.079) & (0.040) & (0.039) & (0.026) \\ 
  \ebllpPostNinetyNine{} & 0.009 & -0.003 & 0.105 & 0.025 & 0.130^{***} & 0.144^{***} \\ 
  & (0.014) & (0.015) & (0.082) & (0.046) & (0.032) & (0.036) \\ 
  \ebllpFEA{} & 0.088^{***} & 0.092^{***} & 0.221^{***} & 0.160^{***} & 0.093 & 0.123^{*} \\ 
  & (0.013) & (0.017) & (0.043) & (0.014) & (0.066) & (0.064) \\ 
  \rowcolor{miewblue} \ebllpFEAPostNinetyNine{} & -0.060^{***} & -0.044^{***} & -0.179^{***} & -0.134^{***} & -0.317^{***} & -0.362^{***} \\ 
  & (0.012) & (0.012) & (0.060) & (0.030) & (0.040) & (0.048) \\ 
  \lagonellp{} & 0.329^{***} & 0.408^{***} & 0.456^{***} & 0.420^{***} & 0.475^{***} & 0.506^{***} \\ 
  & (0.033) & (0.046) & (0.030) & (0.038) & (0.079) & (0.081) \\ 
  \lagtwollp{} & 0.186^{***} & 0.087^{**} & 0.062 & 0.131^{***} & 0.084^{*} & 0.062 \\ 
  & (0.031) & (0.038) & (0.047) & (0.036) & (0.045) & (0.061) \\ 
  \lagonesize{} & -0.00002 & 0.00004 & -0.0004^{**} & -0.0003^{***} & -0.0004 & -0.0004 \\ 
  & (0.0001) & (0.0002) & (0.0002) & (0.0001) & (0.0003) & (0.0003) \\ 
  \lagonecap{} & -0.005 & -0.007 & -0.024^{***} & -0.018^{***} & -0.020^{*} & -0.026^{**} \\ 
  & (0.003) & (0.005) & (0.008) & (0.006) & (0.012) & (0.011) \\ 
  \changeloans{}& 0.001^{*} & 0.002 & 0.001 & 0.002 & 0.008 & 0.008 \\ 
  & (0.001) & (0.002) & (0.001) & (0.001) & (0.005) & (0.005) \\ 
  \IInc{} & 0.117^{***} & 0.099^{***} & 0.037 & 0.033 & 0.143^{***} & 0.139^{***} \\ 
  & (0.017) & (0.023) & (0.030) & (0.028) & (0.043) & (0.039) \\ 
  \IExp{} & -0.072^{***} & -0.048 & -0.043^{**} & -0.031 & -0.046 & -0.054^{**} \\ 
  & (0.015) & (0.041) & (0.022) & (0.025) & (0.028) & (0.027) \\ 
  \PercGDPPC{} & -0.012^{***} & -0.010^{***} & -0.005 & -0.007^{**} & 0.006 & 0.004 \\ 
  & (0.002) & (0.002) & (0.005) & (0.003) & (0.013) & (0.008) \\ 

  D1998 &  &  &  &  &  & 0.001 \\ 
  &  &  &  &  &  & (0.001) \\ 
  ebllpD1998 &  &  &  &  &  & 0.092^{**} \\ 
  &  &  &  &  &  & (0.044) \\ 
  FEAD1998 &  &  &  &  &  & -0.0005 \\ 
  &  &  &  &  &  & (0.001) \\ 
  ebllpFEAD1998 &  &  &  &  &  & -0.229^{*} \\ 
  &  &  &  &  &  & (0.132) \\ 
 \hline \\[-1.8ex] 
Observations & \multicolumn{1}{c}{8,110} & \multicolumn{1}{c}{8,110} & \multicolumn{1}{c}{3,852} & \multicolumn{1}{c}{1,692} & \multicolumn{1}{c}{382} & \multicolumn{1}{c}{382} \\ 

Adjusted R$^{2}$ & \multicolumn{1}{c}{0.350} & \multicolumn{1}{c}{0.365} & \multicolumn{1}{c}{0.190} & \multicolumn{1}{c}{0.255} & \multicolumn{1}{c}{0.680} & \multicolumn{1}{c}{0.700} \\

\bottomrule

\end{tabular} 
\end{table}

\newpage
\begin{table}[H] \centering
\ContinuedFloat \singlespacing\scriptsize 
  \caption{Eurozone Creation and Income Smoothing---DID Estimation} 
    \vspace{-0.25cm}
     \begin{minipage}{1\linewidth}\scriptsize
  \singlespacing
  This table reports the results of the DID analysis of the Eurozone creation. Panel A depicts the OLS coefficient estimates of the DID model (\Cref{eq:llpeurodid}), considering (i) the estimation with the full sample of treatment (FEA) and control (NEA) banks (specification 1), (ii) the full sample estimated with a fully interacted model (specification 2), (iii) an analysis with propensity-score-matched subsamples of treatment (FEA) and control (NEA) banks (specifications 3 and 4, respectively  for PSM samples with and without replacement), and (iv) an estimation with a restricted sample of Nordic countries, which yields one FEA (Finland) and three NEA (Denmark, Iceland, Norway, and Sweden) groups (specification 5), also including the pre-event terms as in \citet{bertrand2004much} (specification 6). Bank-type fixed effects are included in all specifications. Robust standard errors clustered at the bank and year level are reported within parentheses.  *, **, and *** indicate significance at the 10\%, 5\%, and 1\% levels, respectively. Panel B presents the effect of the propensity score matching algorithms on the covariate balancing of the baseline loan-loss provisions model. The first two columns represents the t statistics of the differences in means of treatment and control groups pre- and post-matching. The middle six columns report the percent improvement in covariates balance as a result of matching, defined as $(|b|-|a|)/|b|$, where $b$ ($a$) is the statistic for foreign``minus'' the statistic for domestic banks, measured before (after) matching \citep{ho2007matching}. The last two columns represent the Kolmogorov's D statistics comparing the empirical distributions of treatment and control groups pre- and post-matching.   \\
  \end{minipage}
    \subcaption*{Panel B: Matching Properties}
\scriptsize 
\begin{tabular}{@{\extracolsep{-2pt}}L{1.2cm} D{.}{.}{3,4} D{.}{.}{3,4} l D{.}{.}{3,4} D{.}{.}{3,4} D{.}{.}{3,4} D{.}{.}{3,4}l D{.}{.}{3,4} l D{.}{.}{3,4}  D{.}{.}{3,4}} 
\toprule
  \multicolumn{13}{c}{Matching with replacement (PSM 1)} \\ \hline

 & \multicolumn{2}{c}{T-statistic} & &\multicolumn{6}{c}{$\;$} & &\multicolumn{2}{c}{Kolmogorov's}\\
  & \multicolumn{2}{c}{(diff. means)} & &\multicolumn{6}{c}{Percent improvement due to matching} & &\multicolumn{2}{c}{D statistic}\\
  \hline
  & \multicolumn{1}{c}{Pre} & \multicolumn{1}{c}{Post} & &  &  &  &  &  & &  & \multicolumn{1}{c}{Pre} & \multicolumn{1}{c}{Post}\\ 
   
 & \multicolumn{1}{c}{matched} & \multicolumn{1}{c}{matched} & & \multicolumn{1}{c}{Q.10\%} & \multicolumn{1}{c}{Q.25\%} & \multicolumn{1}{c}{Q.50\%} & \multicolumn{1}{c}{Mean} & \multicolumn{1}{c}{Q.75\%} &\multicolumn{1}{c}{Q.90\%} &  & \multicolumn{1}{c}{matched} & \multicolumn{1}{c}{matched}\\ 
  \hline
PSM &50.315 & 2.644& & 91.220 & 95.184 &96.337 & 94.802 & 93.345 & 95.285&&0.570&0.075\\ 
  \ebllp & 13.271 & 1.408& & 69.322 & 85.286 & 85.845 & 97.749 &92.007&98.654&&0.250&0.100 \\ 
  \lagonellp & 13.530 & 1.044 & & -1.417 & 94.381 & 93.826 & 92.318 &95.512&90.001&&0.309&0.059 \\ 
  \lagtwollp & 14.786& 1.005& & 85.858 & 95.598 & 96.776 &99.731 &93.414&89.159&&0.337&0.073 \\ 
  \IInc & 28.780 & 1.760 & & 84.259 &96.880  & 95.011 &97.296 &93.954&99.296&&0.511&0.000 \\ 

\hline \\
   \multicolumn{13}{c}{Matching without replacement (PSM 2)} \\
   \hline

  & \multicolumn{2}{c}{T-statistic} & &\multicolumn{6}{c}{$\;$} & &\multicolumn{2}{c}{Kolmogorov's}\\
  & \multicolumn{2}{c}{(diff. means)} & &\multicolumn{6}{c}{Percent improvement due to matching} & &\multicolumn{2}{c}{D statistic}\\
  \hline
  & \multicolumn{1}{c}{Pre} & \multicolumn{1}{c}{Post} & &  &  &  &  &  & &  & \multicolumn{1}{c}{Pre} & \multicolumn{1}{c}{Post}\\ 
   
 & \multicolumn{1}{c}{matched} & \multicolumn{1}{c}{matched} & & \multicolumn{1}{c}{Q.10\%} & \multicolumn{1}{c}{Q.25\%} & \multicolumn{1}{c}{Q.50\%} & \multicolumn{1}{c}{Mean} & \multicolumn{1}{c}{Q.75\%} &\multicolumn{1}{c}{Q.90\%} &  & \multicolumn{1}{c}{matched} & \multicolumn{1}{c}{matched}\\ 
  \hline
PSM &50.315 & 2.893& & 86.809&	89.792&	88.102&	89.042&	89.576&	92.903&&0.570&0.123\\ 
  \ebllp & 13.271 & 1.252& & 69.471&	85.492&	84.452&	98.039&	88.446&	98.831&&0.250&0.080 \\ 
  \lagonellp & 13.530 & 1.111 & & 36.573&	81.161&	86.756&	93.438&	90.880&	91.916&&0.309&0.071\\ 
  \lagtwollp & 14.786& 0.884& & 85.415&	88.534&	92.055&	96.982&	93.621&	88.615&&0.337&0.072 \\ 
  \IInc & 28.780 & 1.849 & & 75.370&	88.321&	88.680&	95.174&	89.751&	98.869&&0.511&0.096 \\ 
  \bottomrule
\end{tabular}
\end{table}

\newpage

\begin{table}[H]

\centering 
  \caption{Landesbanken Guarantees Removal and Income Smoothing---Sample Description and Summary Statistics} 
  \label{tab.landes.sample}
  
    \vspace{-0.3cm}
   \begin{minipage}{1\linewidth}\scriptsize
   \singlespacing
This table reports the sample description of the Landesbanken (Panel A) and summary statistics of bank-level fundamental variables of the Landesbanken (treatment group), German commercial banks (control-group 1), and French government-owned banks (control-group 2). Correlation coefficients significant at $p<0.05$ are highlighted. \\

    \end{minipage}
    \subcaption*{Panel A: Sample Description---Landesbanken}
\scriptsize

\begin{tabular}{@{\extracolsep{-2pt}}L{6cm}D{.}{.}{3.3}D{.}{.}{3.3}D{.}{.}{3.3}D{.}{.}{-3.3}D{.}{.}{3.3}D{.}{.}{3.3}}
\toprule

&2002& 2003& 2004& 2005& 2006& 2007 \\ \hline

Bayerische Landesbank &x &x &x &x &x &x \\
Landesbank Hessen-Thueringen  - HELABA &x &x &x &x &x &x \\
LRP Landesbank Rheinland-Pfalz &x &x &x &x &x & \\
Norddeutsche Landesbank  NORD/LB &x &x &x &x &x &x \\
Bremer Landesbank &x &x &x &x &x &x  \\
Landesbank Saar-SaarLB &x &x &x &x &x &x \\
Landesbank Berlin Holding AG &x &x &x &x & &\\
Landesbank Baden-Wuerttemberg  &x &x &x &x &x &x \\

\hline
Total Count &8 &8 &8 &8 &7 &6\\
  \bottomrule \\
\end{tabular}

\subcaption*{Panel B: Summary Statistics---Landesbanken and Control Groups 1 and 2}

\begin{tabular}{@{\extracolsep{-2pt}}L{1.2cm}D{.}{.}{2,3}D{.}{.}{2,3}D{.}{.}{2,3}D{.}{.}{2,3}D{.}{.}{2,3}D{.}{.}{2,3}D{.}{.}{2,3}D{.}{.}{2,3}D{.}{.}{2,3}D{.}{.}{2,3}D{.}{.}{2,3}D{.}{.}{2,3}D{.}{.}{2,3}D{.}{.}{2,3}D{.}{.}{2,3}D{.}{.}{2,3}}
  \toprule

& \multicolumn{5}{c}{Summary Statistics: Landesbanken} & \multicolumn{5}{c}{Summary Statistics: Control-Group 1} & \multicolumn{5}{c}{Summary Statistics: Control-Group 2} \\ \hline

 & \multicolumn{1}{c}{N} & \multicolumn{1}{c}{Mean} & \multicolumn{1}{c}{P25} & \multicolumn{1}{c}{P50} & \multicolumn{1}{c}{P75}  &  \multicolumn{1}{c}{N} & \multicolumn{1}{c}{Mean} & \multicolumn{1}{c}{P25} & \multicolumn{1}{c}{P50} & \multicolumn{1}{c}{P75} &  \multicolumn{1}{c}{N} & \multicolumn{1}{c}{Mean} & \multicolumn{1}{c}{P25} & \multicolumn{1}{c}{P50} & \multicolumn{1}{c}{P75}\\ 
  \hline

  \llp &   \multicolumn{1}{c}{45} & 0.005 & 0.002 & 0.004 & 0.007 &    \multicolumn{1}{c}{60} & 0.006 & 0.000 & 0.005 & 0.010 & \multicolumn{1}{c}{25} & -0.005 & -0.004   & 0.001   &   0.004\\ 
  \ebllp &   \multicolumn{1}{c}{45} & 0.007 & 0.005 & 0.008 & 0.010&    \multicolumn{1}{c}{60} & 0.023 & 0.006 & 0.012 & 0.039 & \multicolumn{1}{c}{25}  & 0.121 & 0.028   & 0.051    & 0.081\\ 
  \lagonellp &   \multicolumn{1}{c}{45} & 0.005 & 0.002 & 0.004 & 0.007&     \multicolumn{1}{c}{60} & 0.008 & 0.000 & 0.006 & 0.013 & \multicolumn{1}{c}{25} & -0.001  & -0.006   & 0.001     &  0.007 \\ 
  \lagtwollp &   \multicolumn{1}{c}{45} & 0.005 & 0.002 & 0.004 & 0.007&    \multicolumn{1}{c}{60} & 0.010 & 0.002 & 0.006 & 0.016 & \multicolumn{1}{c}{25} &  0.005 & -0.004   & 0.004      &  0.009 \\ 
  \lagonesize &   \multicolumn{1}{c}{45} & 11.396 & 10.454 & 11.760 & 12.480&    \multicolumn{1}{c}{60} & 9.621 & 8.440 & 9.173 & 10.722 & \multicolumn{1}{c}{25} & 8.261 &  6.510   &  6.788      &  9.753 \\ 
  \lagonecap &   \multicolumn{1}{c}{45} & 0.023 & 0.018 & 0.020 & 0.025&    \multicolumn{1}{c}{60} & 0.061 & 0.023 & 0.037 & 0.046& \multicolumn{1}{c}{25} & 0.125 & 0.090   & 0.117      & 0.168 \\ 
  \changeloans &   \multicolumn{1}{c}{45} & 0.001 & -0.029 & 0.018 & 0.044&    \multicolumn{1}{c}{60} & 0.398 & -0.038 & 0.076 & 0.318& \multicolumn{1}{c}{25} &  0.055 & -0.053   & 0.016      &  0.125 \\ 
  \IInc &   \multicolumn{1}{c}{45} & 0.041 & 0.040 & 0.042 & 0.045&    \multicolumn{1}{c}{60} & 0.052 & 0.040 & 0.048 & 0.060& \multicolumn{1}{c}{25} & 0.048 & 0.044   & 0.054      & 0.058  \\ 
  \IExp &   \multicolumn{1}{c}{45} & 0.037 & 0.036 & 0.037 & 0.040&    \multicolumn{1}{c}{60} & 0.037 & 0.027 & 0.036 & 0.044 & \multicolumn{1}{c}{25} & 0.044 & 0.035   & 0.041      & 0.052\\

   \hline 
\end{tabular}

\begin{tabular}{@{\extracolsep{-2pt}}L{4.2cm}D{.}{.}{-3}D{.}{.}{-3}D{.}{.}{-3}D{.}{.}{-3}D{.}{.}{-3}D{.}{.}{-3}D{.}{.}{-3}D{.}{.}{-3}D{.}{.}{-3}}
\hline
\multicolumn{10}{c}{Correlation Matrix: Landesbanken (shaded values indicate estimates significant with $p<$5\%)} \\

  \hline

 \llp  & 1.000 & \cellcolor{miewblue} 0.586 & 0.129 & 0.141 & -0.046 & 0.237 & -0.243 & -0.182 & -0.187 \\ 
 \ebllp & \cellcolor{miewblue}0.586 & 1.000 & -0.192 & -0.009 & -0.029 & -0.169 & 0.106 & -0.004 & 0.082 \\ 
  \lagonellp& 0.129 & -0.192 & 1.000 & 0.269 & 0.023 & 0.177 & \cellcolor{miewblue}-0.507 & -0.258 & \cellcolor{miewblue}-0.338 \\ 
  \lagtwollp & 0.141 & -0.009 & 0.269 & 1.000 & 0.159 & 0.273 & \cellcolor{miewblue}-0.416 & \cellcolor{miewblue}-0.357 & \cellcolor{miewblue}-0.391 \\ 
  \lagonesize  & -0.046 & -0.029 & 0.023 & 0.159 & 1.000 & 0.144 & 0.066 & 0.139 & 0.198 \\ 
 \lagonecap & 0.237 & -0.169 & 0.177 & 0.273 & 0.144 & 1.000 & \cellcolor{miewblue}-0.388 & -0.229 & \cellcolor{miewblue}-0.319 \\ 
 \changeloans & -0.243 & 0.106 & \cellcolor{miewblue}-0.507 & \cellcolor{miewblue}-0.416 & 0.066 & \cellcolor{miewblue}-0.388 & 1.000 & \cellcolor{miewblue}0.742 & \cellcolor{miewblue}0.782 \\ 
  \IInc & -0.182 & -0.004 & -0.258 & \cellcolor{miewblue}-0.357 & 0.139 & -0.229 & \cellcolor{miewblue}0.742 & 1.000 & \cellcolor{miewblue}0.981 \\ 
  \IExp & -0.187 & 0.082 & \cellcolor{miewblue}-0.338 & \cellcolor{miewblue}-0.391 & 0.198 & \cellcolor{miewblue}-0.319 & \cellcolor{miewblue}0.782 & \cellcolor{miewblue}0.981 & 1.000 \\

   \hline

\multicolumn{10}{c}{Correlation Matrix: Control-Group 1 (shaded values indicate estimates significant with $p<$5\%)} \\

  \hline

  \llp & 1.000 & \cellcolor{miewblue} 0.451 & 0.253 & -0.120 & 0.005 & \cellcolor{miewblue} -0.367 & 0.181 & \cellcolor{miewblue} 0.462 & 0.242 \\ 
  \ebllp & \cellcolor{miewblue} 0.451 & 1.000 & -0.140 & -0.101 & -0.034 & 0.222 & \cellcolor{miewblue} 0.274 & 0.133 & \cellcolor{miewblue} 0.266 \\ 
  \lagonellp & 0.253 & -0.140 & 1.000 & 0.106 & -0.142 & -0.164 & 0.001 & \cellcolor{miewblue} 0.450 & 0.252 \\ 
  \lagtwollp & -0.120 & -0.101 & 0.106 & 1.000 & -0.222 & -0.078 & -0.136 & \cellcolor{miewblue} 0.317 & 0.215 \\ 
  \lagonesize & 0.005 & -0.034 & -0.142 & -0.222 & 1.000 & -0.108 & -0.136 & -0.209 & -0.209 \\ 
  \lagonecap & \cellcolor{miewblue} -0.367 & 0.222 & -0.164 & -0.078 & -0.108 & 1.000 & 0.049 & \cellcolor{miewblue} -0.262 & 0.132 \\ 
  \changeloans & 0.181 & \cellcolor{miewblue} 0.274 & 0.001 & -0.136 & -0.136 & 0.049 & 1.000 & -0.094 & -0.017 \\ 
  \IInc & \cellcolor{miewblue} 0.462 & 0.133 & \cellcolor{miewblue} 0.450 & \cellcolor{miewblue} 0.317 & -0.209 & \cellcolor{miewblue} -0.262 & -0.094 & 1.000 & \cellcolor{miewblue} 0.603 \\ 
  \IExp & 0.242 & \cellcolor{miewblue} 0.266 & 0.252 & 0.215 & -0.209 & 0.132 & -0.017 & \cellcolor{miewblue} 0.603 & 1.000 \\ 
  
   \hline

\multicolumn{10}{c}{Correlation Matrix: Control-Group 2 (shaded values indicate estimates significant with $p<$5\%)} \\

  \hline

  \llp & 1.000 & 0.060 & \cellcolor{miewblue}0.698 & 0.314 & \cellcolor{miewblue}0.435 & \cellcolor{miewblue}-0.506 & 0.373 & -0.097 & 0.214 \\ 
  \ebllp & 0.060 & 1.000 & -0.025 & -0.115 & 0.375 & -0.075 & \cellcolor{miewblue}0.679 & \cellcolor{miewblue}-0.741 & \cellcolor{miewblue}-0.546 \\ 
  \lagonellp & \cellcolor{miewblue}0.698 & -0.025 & 1.000 &\cellcolor{miewblue} 0.464 & 0.304 & -0.281 & 0.302 & 0.090 & 0.296 \\ 
  \lagtwollp & 0.314 & -0.115 & \cellcolor{miewblue}0.464 & 1.000 & 0.005 & -0.184 & 0.176 & 0.174 & 0.272 \\ 
  \lagonesize & \cellcolor{miewblue}0.435 & 0.375 & 0.304 & 0.005 & 1.000 & \cellcolor{miewblue}-0.545 & 0.277 & \cellcolor{miewblue}-0.590 & 0.150 \\ 
  \lagonecap & \cellcolor{miewblue}-0.506 & -0.075 & -0.281 & -0.184 & \cellcolor{miewblue}-0.545 & 1.000 & -0.122 & 0.335 & -0.206 \\ 
  \changeloans & 0.373 & \cellcolor{miewblue}0.679 & 0.302 & 0.176 & 0.277 & -0.122 & 1.000 & -0.342 & -0.180 \\ 
  \IInc & -0.097 & \cellcolor{miewblue}-0.741 & 0.090 & 0.174 & \cellcolor{miewblue}-0.590 & 0.335 & -0.342 & 1.000 & \cellcolor{miewblue}0.663 \\ 
  \IExp & 0.214 & \cellcolor{miewblue}-0.546 & 0.296 & 0.272 & 0.150 & -0.206 & -0.180 & \cellcolor{miewblue}0.663 & 1.000 \\

   \bottomrule
\end{tabular}

\end{table}

\newpage
\begin{table}[H] \centering 
  \caption{Landesbanken Guarantees Removal and Income Smoothing---Differences Estimation}
  \vspace{-0.5cm}
     \begin{minipage}{1\linewidth}\scriptsize
   \singlespacing
This table reports OLS coefficient estimates of the differences model computed for the Landesbanken guarantees removal (\Cref{eq:llplandesdiff}). Column 1 depict the main effect and column 2 present the estimation considering a fully interacted model. Columns 3--10 hold constant the baseline model of column 1 while excluding a specific Landesbanken in each column, with the exclusion following the order they are presented in \Cref{tab.landes.sample}, Panel A. Robust standard errors clustered at the bank and year level are reported within parentheses.  *, **, and *** indicate significance at the 10\%, 5\%, and 1\% levels, respectively. \\
  \end{minipage}
  
  \label{tab.landes.d} 
\scriptsize 
\begin{tabular}{@{\extracolsep{-2pt}}lD{.}{.}{2,5}D{.}{.}{2,5}D{.}{.}{2,5}D{.}{.}{2,5}D{.}{.}{2,5}D{.}{.}{2,5}D{.}{.}{2,5}D{.}{.}{2,5}D{.}{.}{2,5}D{.}{.}{2,5} } 
\toprule
 & \multicolumn{10}{c}{\textit{Dependent variable:} \llp{}} \\ 
\cline{2-11} 

\\[-1.8ex] & \multicolumn{1}{c}{(1)} & \multicolumn{1}{c}{(2)} & \multicolumn{1}{c}{(3)} & \multicolumn{1}{c}{(4)} & \multicolumn{1}{c}{(5)} & \multicolumn{1}{c}{(6)} & \multicolumn{1}{c}{(7)} & \multicolumn{1}{c}{(8)} & \multicolumn{1}{c}{(9)} & \multicolumn{1}{c}{(10)}\\ 
\hline \\[-1.8ex] 
 \PostTussandFive & -0.011^{***} & 0.071^{**} & -0.011^{***} & -0.012^{***} & -0.012^{***} & -0.011^{***} & -0.010^{***} & -0.012^{***} & -0.005^{***} & -0.010^{***} \\ 
  & (0.002) & (0.029) & (0.003) & (0.003) & (0.003) & (0.004) & (0.003) & (0.003) & (0.001) & (0.002) \\ 
  \ebllp & 0.007 & 0.118 & -0.039 & -0.032 & 0.030 & 0.017 & 0.060 & 0.011 & 0.909^{***} & -0.094 \\ 
  & (0.166) & (0.158) & (0.159) & (0.201) & (0.161) & (0.213) & (0.133) & (0.149) & (0.077) & (0.101) \\ 
 \rowcolor{miewblue} \ebllpPostTussandFive & 0.853^{***} & 0.644^{***} & 0.862^{***} & 0.903^{***} & 0.866^{***} & 0.806^{***} & 0.915^{***} & 0.895^{***} & 0.398^{***} & 0.893^{***} \\ 
  & (0.113) & (0.156) & (0.109) & (0.146) & (0.132) & (0.187) & (0.118) & (0.110) & (0.094) & (0.069) \\ 
  \lagonellp & -0.080 & -0.390^{**} & -0.185 & -0.075 & -0.135 & -0.034 & -0.203 & -0.135 & 0.009 & 0.180 \\ 
  & (0.230) & (0.159) & (0.156) & (0.252) & (0.256) & (0.248) & (0.230) & (0.263) & (0.098) & (0.250) \\ 
  \lagtwollp& -0.226 & -0.414^{*} & -0.240 & -0.214 & -0.261 & -0.254 & -0.263 & -0.289 & -0.063 & -0.173 \\ 
  & (0.195) & (0.237) & (0.184) & (0.190) & (0.198) & (0.211) & (0.198) & (0.257) & (0.099) & (0.209) \\ 
  \lagonesize & 0.001 & 0.003^{*} & 0.001 & 0.001 & 0.001 & 0.000 & 0.001 & 0.002 & 0.000 & 0.000 \\ 
  & (0.001) & (0.002) & (0.001) & (0.001) & (0.001) & (0.001) & (0.001) & (0.002) & (0.000) & (0.001) \\ 
  \lagonecap & 0.025 & -0.145 & 0.057 & 0.003 & 0.041 & 0.179 & -0.242^{*} & -0.050 & 0.059^{**} & 0.023 \\ 
  & (0.123) & (0.393) & (0.133) & (0.120) & (0.162) & (0.168) & (0.127) & (0.110) & (0.026) & (0.110) \\ 
  \changeloans & -0.012 & -0.022 & -0.011 & -0.016 & -0.009 & -0.011 & -0.022 & -0.013 & 0.003 & -0.004 \\ 
  & (0.012) & (0.014) & (0.013) & (0.012) & (0.012) & (0.016) & (0.014) & (0.015) & (0.006) & (0.008) \\ 
  \IExp & -0.466 & -3.973 & -0.329 & -0.223 & -1.036 & 0.153 & -2.741 & -1.135 & 0.611 & 0.058 \\ 
  & (1.338) & (2.979) & (1.293) & (1.375) & (1.674) & (0.971) & (1.842) & (1.571) & (0.402) & (1.013) \\ 
  \IInc & 0.787 & 4.416 & 0.621 & 0.622 & 1.362 & 0.246 & 3.197 & 1.392 & -0.555 & 0.172 \\ 
  & (1.395) & (3.042) & (1.323) & (1.449) & (1.724) & (1.080) & (1.935) & (1.624) & (0.398) & (1.007) \\ 

 \hline \\[-1.8ex] 
Observations & \multicolumn{1}{c}{45} & \multicolumn{1}{c}{45} & \multicolumn{1}{c}{39} & \multicolumn{1}{c}{39} & \multicolumn{1}{c}{40} & \multicolumn{1}{c}{39} & \multicolumn{1}{c}{39} & \multicolumn{1}{c}{39} & \multicolumn{1}{c}{41} & \multicolumn{1}{c}{39} \\ 

Adjusted R$^{2}$ & \multicolumn{1}{c}{0.727} & \multicolumn{1}{c}{0.831} & \multicolumn{1}{c}{0.740} & \multicolumn{1}{c}{0.732} & \multicolumn{1}{c}{0.745} & \multicolumn{1}{c}{0.737} & \multicolumn{1}{c}{0.773} & \multicolumn{1}{c}{0.732} & \multicolumn{1}{c}{0.897} & \multicolumn{1}{c}{0.794} \\ 

\bottomrule

\end{tabular} 
\end{table} 

\newpage

\begin{table}[H] \centering\singlespacing\scriptsize  
  \caption{Landesbanken Guarantees Removal and Income Smoothing---DID Estimation} 
  \label{tab.landes.did}
    \vspace{-0.5cm}
     \begin{minipage}{1\linewidth}\scriptsize
   \singlespacing
This table reports the results of the DID analysis of the Landesbanken guarantees removal (\Cref{eq:llplandesdid}). Columns 1 and 2 report the estimates when the sample of large German commercial banks is used as the control sample (control-group 1). Columns 1 and 2 report the estimates when the sample of large French government-owned banks is used as the control sample (control-group 2). Columns 1 and 3 (2 and 4) report the estimates for a standard (fully-interacted) DID model. Robust standard errors clustered at the bank and year level are reported within parentheses.  *, **, and *** indicate significance at the 10\%, 5\%, and 1\% levels, respectively. \\
  \end{minipage}
\scriptsize 
\begin{tabular}{@{\extracolsep{-2pt}}lD{.}{.}{2,5}D{.}{.}{2,5}D{.}{.}{2,5}D{.}{.}{2,5} } 
\toprule 
 & \multicolumn{4}{c}{\textit{Dependent variable:} \llp{}} \\ 
\cline{2-5} 

\\[-1.8ex] & \multicolumn{1}{c}{(1)} & \multicolumn{1}{c}{(2)} & \multicolumn{1}{c}{(3)} & \multicolumn{1}{c}{(4)}\\ 
\hline \\[-1.8ex] 
 \PostTussandFive & 0.006 & -0.043 & 0.007 & -0.529^{***} \\ 
  & (0.005) & (0.030) & (0.012) & (0.191) \\ 
  \Landes & 0.009^{**} & -0.107^{**} & -0.010^{***} & -0.541^{***} \\ 
  & (0.004) & (0.053) & (0.003) & (0.192) \\ 
  \LandesPostTussandFive & -0.013^{***} & 0.114^{**} & -0.016^{*} & 0.599^{***} \\ 
  & (0.005) & (0.051) & (0.009) & (0.193) \\ 
  \ebllp & 0.731^{***} & 0.816^{***} & -0.098 & -0.309^{***} \\ 
  & (0.133) & (0.293) & (0.109) & (0.108) \\ 
  \ebllpPostTussandFive & -0.617^{***} & -0.684^{**} & 0.074 & 0.290^{***} \\ 
  & (0.125) & (0.293) & (0.121) & (0.101) \\ 
  \ebllpLandes & -0.640^{***} & -0.699^{***} & 0.112 & 0.426^{***} \\ 
  & (0.100) & (0.183) & (0.185) & (0.149) \\ 
\rowcolor{miewblue}  \ebllpLandesPostTussandFive & 1.285^{***} & 1.328^{***} & 0.743^{***} & 0.354^{**} \\ 
  & (0.075) & (0.224) & (0.083) & (0.166) \\ 
  \lagonellp& 0.329^{**} & 0.452 & 0.323^{***} & 0.722^{**} \\ 
  & (0.165) & (0.305) & (0.108) & (0.331) \\ 
  \lagtwollp & 0.023 & -0.005 & -0.027 & 0.908^{**} \\ 
  & (0.118) & (0.279) & (0.158) & (0.386) \\ 
  \lagonesize & -0.0001 & -0.003^{*} & 0.002^{***} & -0.027^{**} \\ 
  & (0.001) & (0.002) & (0.001) & (0.011) \\ 
  \lagonecap & -0.056^{***} & -0.212 & -0.113^{**} & 0.266^{**} \\ 
  & (0.012) & (0.149) & (0.051) & (0.113) \\ 
  \changeloans & 0.0003 & 0.001 & 0.007^{**} & 0.020 \\ 
  & (0.003) & (0.002) & (0.003) & (0.012) \\ 
  \IExp & -0.115 & -0.103 & -0.255^{*} & 3.798^{**} \\ 
  & (0.108) & (0.191) & (0.149) & (1.623) \\ 
  \IInc & 0.100 & -0.260 & 0.221 & -9.472^{**} \\ 
  & (0.121) & (0.345) & (0.184) & (3.791) \\ 

 \hline \\[-1.8ex] 
Observations & \multicolumn{1}{c}{105} & \multicolumn{1}{c}{105} & \multicolumn{1}{c}{70} & \multicolumn{1}{c}{70} \\ 

Adjusted R$^{2}$ & \multicolumn{1}{c}{0.767} & \multicolumn{1}{c}{0.804} & \multicolumn{1}{c}{0.680} & \multicolumn{1}{c}{0.857} \\ 

\bottomrule

\end{tabular} 
\end{table} 

\newpage

\appendix

\renewcommand{\thetable}{\thesubsection.\arabic{table}}
\setcounter{table}{0}
\renewcommand{\thefigure}{\thesubsection.\arabic{figure}}
\setcounter{figure}{0}

\section*{Appendix: Theoretical Framework, Data Details, and Additional Results }

\renewcommand{\thesubsection}{\Alph{subsection}}

\subsection{Theoretical Framework---Government Guarantees and Income Smoothing} \label{app:sec:theorymodel}

\renewcommand{\theequation}{\thesubsection.\arabic{equation}}
\setcounter{equation}{0}

We provide a simple model to illustrates how banks' income smoothing decisions relate to government guarantees. We build on the model proposed by \citet{trueman1988explanation}, who examine a firm's decision to smooth earnings by shifting income between periods in an attempt to alter investors' perceptions of the underlying riskiness a firm. Our goal is to describe the theoretical foundations representing the competing forces of (i) government guarantees reducing incentives to smooth earnings (the direct asset pricing effect proposed in this paper) and (ii) such guarantees altering marginal costs of smoothing earnings (e.g., through changes in capital market monitoring incentives).  

There are two stylized players: a bank manager and an outside investor. The investor estimates the bank's value in part based on the bank's reported earnings, which provide a noisy signal of \emph{actual} economic earnings (i.e., earnings excluding the potential income smoothing effect). The bank's economic earnings are defined by the following stochastic process:
\vspace{-0.2cm}
\begin{align*}
\tilde{x}_t = \mu + \epsilon_t 
\end{align*}
where the mean $\mu$ is known to both the manager and the investor but the actual process $\tilde{x}_t$ is only observed by the manager at time $t$. $\epsilon_t$ is distributed normally with a mean of zero and the variance depends on bank type. There are  two possible types of banks: low variance ($Var[\epsilon_t]=\sigma_A^2$) and high variance ($Var[\epsilon_t]=\sigma_B^2>\sigma_A^2$), where variance captures the riskiness of the bank. The bank manager knows her own type but the representative investor does not know this information. Instead, the representative investor forms an expectation about the probability of the bank being of type A (B) defined as  $p_A$ ($p_B=1-p_A$) before observing the bank manager's earnings disclosures. As in \citet{trueman1988explanation}, the assumption that the mean is known serves to simplify the analysis and emphasize the effect of an uncertain variance on managers' income smoothing decisions.

We consider a two period model where the goal of the bank manager is to maximize her proceeds obtained from issuing new debt securities to the representative investor at the end of period 2.\footnote{The fundamental differences between our framework and the one of \citet{trueman1988explanation} are twofold. First, we endogenize the amount of income smoothing chosen by the bank manager given the capital market benefits and monitoring costs she faces. Second, we introduce the asset pricing effects of government guarantees as a censoring parameter to the left tail of the high variance bank. \citet{trueman1988explanation} employ a binary (discrete) smoothing decision and a constant cost, showing that if the accounting system allows managers to shift income from one period to another, the manager will engage in earnings smoothing as long as it is not costly. Last, but not least, our analytical abstracts from important features of the banking sector which naturally do not pertain to \citet{trueman1988explanation}. For example, by building on  \citet{trueman1988explanation} we do not endogenize the role of a bank regulator, which could alter equilibrium costs of monitoring as well.} In our model the bank manager chooses the amount of earnings smoothing to optimize the net benefit of this activity (i.e., capital market benefits minus costs arising from lack of financial transparency, investors' scrutiny, or taxation externalities). 

After  the realization of the economic profit at time 1 ($x_1$ is known to the manager but unobserved by the investor), the bank manager can choose what quantity $s$ of the actual income above (or below) the expected value $E[\tilde{x}_t] = \mu$ to shift to period 2. Since new debt will be issued at time 2, reported income should comprise not only the actual economic performance ($x_2$) but also any delayed income from period 1 (either positive if $x_1>\mu$ or negative if $x_1<\mu$). In other words, reported income at periods 1 and 2 are given by:
\vspace{-0.2cm}
\begin{align*}
x^s_1 = (1-s)x_1+s\mu \\
x^s_2 = x_2 -s(\mu-x_1) 
\end{align*}
where $0\leq s \leq 1$.

Once $x^s_1$ is reported, the representative investor updates her prior probability of the bank being of type A based on the observation of $x^s_1$ and the publicly known properties of $x^s_2$ (still to be reported). As in \citet{trueman1988explanation}, applying Bayes' rule allows us to express the ex-post probability $p'_A(x_1^s,x_2^s)$ as
\vspace{-0.2cm}
\begin{align*}
p'_A(x_1^s,x_2^s) = \displaystyle\frac{\Phi(x^s_1;\sigma_A^2)\Phi(x^s_2;\sigma_A^2)p_A}{\Phi(x^s_1;\sigma_A^2)\Phi(x^s_2;\sigma_A^2)p_A + \Phi(x^s_1;\sigma_B^2)\Phi(x^s_2;\sigma_B^2)p_B} 
\end{align*}
where $\Phi(x^s_1;\sigma_i^2), i=A,B$ represents the probability density function of a normal distribution whose mean is $\mu$ and variance is $\sigma_i^2$.

In an unambiguous setting where the representative investor is certain about the bank type being A (or B), the market value of the debt security to be issued is simply a (decreasing) function of the underlying risk of the banks' income---i.e., $B=V(\sigma)$ with $V_{\sigma}(\sigma)<0$. Since $\sigma_A$ and $\sigma_B$ are static parameters, then $B_A=V(\sigma_A)>V(\sigma_B)>B_B$. As the investor observes the series of reported (managed) earnings and uses such signals to infer the ambiguous underlying volatility of the bank's earnings, the market value of proceeds to be issued from  the bank's debt is equal to 
\vspace{-0.2cm}
\begin{align*}
B\big(p'_A(x_1^s,x_2^s)\big) &= p'_A(x_1^s,x_2^s) B_A + \big(1-p'_A(x_1^s,x_2^s)\big) B_B \\
&= p'_A(x_1^s,x_2^s)(B_A-B_B) + B_B
\end{align*}

In other words, capital market benefits (to the bank manager) can be optimized by choosing a level of smoothing that maximizes the investor's posterior probability of the bank being of type A. Investors update their expected values of $p'_A$ at the end of period 1 when $x^s_1$ is reported and based on  the distribution properties of the $\tilde{x}^s_2$ whose realization is still unknown. Therefore, the manager aims to maximize investors' expectations $E[p'_A(x_1^s,\tilde{x}_2^s)]$ by choosing $s$ after she observes the actual value $x_1$.

We define the costs of income smoothing stemming from investor monitoring and government monitoring respectively as functions $K^{inv}(s)$ and $K^{gov}(s)$ indicating the monitoring costs associated with a smoothing choice of $s$. We assume that both functions are  twice continuously differentiable in the interval $0\leq s \leq 1$.\footnote{It is important to note that the assumption of a cost function that depends solely on the level of smoothing $s$ is inherently simplistic and done for sake of analytical tractability. In particular, one should expect that the same level of smoothing $s$ would be associated with higher costs when the overall effect is income increasing (i.e., under-provisioning when performance is bad) vis-\`{a}-vis income decreasing (over-provisioning when performance is good). Nevertheless, our main result would be inferentially similar if we adopted a cost function that depends on $s$ and $x_1-\mu$.} We assume that the first-order derivatives $K^{inv}_{s}(s)$ and $K^{gov}_{s}(s)$ with respect to the smoothing parameter $s$ are strictly positive, as costs should be increasing in $s$ if larger amounts of income smoothing are associated with greater costs (e.g., higher detection risk, either by the investor or the government). Since the bank manager chooses $s$ to achieve the desirable effect on investors' expectations $E[p'_A(x_1^s,\tilde{x}_2^s)]$ and both $x_1^s$ and $x_2^s$ are a function of the actual earnings she  observes (and her choice of $s$) we can substitute $p'_A(x_1^s,\tilde{x}_2^s) = y(x_1,\tilde{x}_2,s)$.  The bank manager's optimization problem thus is described as
\vspace{-0.2cm}
\begin{equation}
\begin{aligned}
max \;\;\;& E[y(x_1,\tilde{x}_2,s)](B_A-B_B) + B_B - K^{inv}(s) - K^{gov}(s) \\
 subject \; to \;\;\;& 0\leq s \leq 1
\end{aligned}
\end{equation}
where 
\begin{align*}
E[y(x_1,\tilde{x}_2,s)] &=  \int_{-\infty}^{+\infty}  \bigg(\frac{\Phi(x_1^s;\sigma_A^2 )\Phi(\tilde{x}_2^s;\sigma_A^2 ) p_A}{\Phi(x_1^s;\sigma_A^2 )\Phi(\tilde{x}_2^s;\sigma_A^2 ) p_A + \Phi(x_1^s;\sigma_B^2 )\Phi(\tilde{x}_2^s;\sigma_B^2 ) p_B} \bigg) \Phi(\tilde{x}_2;\sigma_i^2) d\tilde{x}_2
\end{align*}
Disregarding corner solutions, the optimal level of smoothing chosen by the bank manager, i.e,  $s^* = \argmax E[y(x_1,\tilde{x}_2,s)](B_A-B_B) + B_B - K(s)$, must satisfy the following first and second order conditions:
\begin{equation}
\begin{aligned}
{F_{s}(x_1,s^{*})}(B_A-B_B) - K^{inv}_s(s^*) - K^{gov}_s(s^*)  =0 \\
{F_{ss}(x_1,s^*)}(B_A-B_B) - K^{inv}_{ss}(s^*) - K^{gov}_{ss}(s^*) <0 
\end{aligned}
\end{equation}
{where $F(x_1,s) \equiv E[y(x_1,\tilde{x}_2,s)]$, $F_{s}(x_1,s) \equiv \frac{\partial}{\partial s} F_{ss}(x_1,s)$, $F_{ss}(x_1,s) \equiv \frac{\partial^2}{\partial s^2} F_{ss}(x_1,s)$}

\paragraph{Introducing Government Guarantees.} Since we are ultimately interested in comparative statics of how the optimal (equilibrium) level of smoothing $s^*$ varies with the introduction of positive government guarantees, we now introduce effects of government guarantees into the model. Since the most direct effect of government guarantees is a left-tail censoring representing potential cash infusions in high-marginal-utility states, such guarantees are represented by the exogenous parameter $g$. In introducing $g$, however, we endogenize a variety of outcome variables to represent the overall effect of government guarantees on banks' choice of optimal smoothing (i.e., $s^*(g)$), as well as other parameters representing the four different channels proposed in this paper.

In our framework, we assume that government guarantees provide an extra layer of protection in states of extreme left tail realizations of economic profits. Consequently, the distribution parameters representing realizations of actual earnings (i.e., $\mu$, $\sigma_A$, and $\sigma_B$) remain unaltered. The presence of government guarantees $g$, however, censors the left tail distribution of the random variable $\tilde{x}_2$, consequently altering the functional form of $F(x_1,s)$ (i.e., investor's subjective belief of the bank being of type A). This effect represents the \emph{tail-risk channel} proposed in this paper.

Regarding the \emph{risk-taking channel}, government guarantees affect the underlying risks of the bank (i.e., $\sigma_A=\sigma_A(g)$ and $\sigma_B=\sigma_B(g)$). We assume that government guarantees will then affect the prices of bonds $B_A$ and $B_B$ through two channels: (i) the direct value of potential cash infusions $g$ and (ii) the indirect value through the effect of $g$ on banks' risk taking $\sigma(g)$. We can then express the value of the debt to be issued as $B=V(\sigma,g)$, where $B_A=V(\sigma_A,g)$ and $B_B=V(\sigma_B,g)$. We assume that function $V(\sigma,g)$ is continuously differentiable with respect to $\sigma$ and $g$. Moreover, we assume that $V_{\sigma}(\sigma,g)<0$ (i.e., $V(\sigma,g)$ is decreasing in $\sigma$ for a given level of government guarantees $g$), $V_{g}(\sigma,g)>0$ (.e., $V(\sigma,g)$ is increasing in $g$ for a given earnings volatility $\sigma$), and $V_{\sigma g}(\sigma,g)>0$ (i.e., $V_{g}(\sigma,g)>0$ is increasing in $\sigma$, meaning that the government guarantees are marginally more valuable for a bank with higher earnings volatility).

The analytical expression of $F(x_1,s)$ is altered to incorporate the censoring effect in the distribution of economic earnings $\tilde{x}_2$ that the bank manager observes, as well as the effect of government guarantees on banks' risk taking.
\vspace{-0.2cm}
\begin{align} \label{eq:model1}
&F(x_1,\sigma_A(g),\sigma_B(g),s(g),g) = E[y(x_1,\tilde{x}_2,\sigma_A(g),\sigma_B(g),s(g),g)] = \nonumber \\
& \phantom{{}={}} \begin{aligned}[t]
=\int_{g}^{+\infty} 
\bigg(\frac{\Phi(x_1^s;\sigma^2_A(g) )\Phi(\tilde{x}_2^s;\sigma^2_A(g) ) p_A}{\Phi(x_1^s;\sigma^2_A(g) )\Phi(\tilde{x}_2^s;\sigma^2_A(g) ) p_A + \Phi(x_1^s;\sigma^2_B(g) )\Phi(\tilde{x}_2^s;\sigma_B^2(g) ) p_B} \bigg) \Phi(\tilde{x}_2;\sigma_i^2) d\tilde{x}_2
 \end{aligned}
\end{align}

In other words, government guarantees should affect $F(x_1,\sigma_A,\sigma_B,s,g)$ through (i) the direct effect on the left tail of the earnings distribution (defined by the integration limit $g$), (ii) the indirect effect on banks' endogenous risk taking ($\sigma_A=\sigma_A(g)$ and $\sigma_B=\sigma_B(g)$), and (iii) the indirect effect on the manager's endogenous choice of $s^*=s^*(g)$.

Last, to represent the effects of government guarantees on the monitoring incentives of investors (\emph{investor-monitoring channel}) and the government (\emph{government-monitoring channel}), we we allow the cost functions $K^{inv}$ and $K^{gov}$ to directly depend on $g$, in addition to its indirect dependence through $s^*=s^*(g)$. We assume that $K_{sg}^{inv}(s(g),g)<0$ (i.e., the investor's marginal incentives to monitor income smoothing, $K_{s}^{inv}$, decreases with $g$) and $K_{sg}^{gov}(s(g),g)>0$ (i.e., the government's marginal incentives to monitor income smoothing, $K_{s}^{inv}$, increases with $g$).

Taken together, the aforementioned effects summarize the four channels proposed in this paper. The first and second order conditions of the manager's optimization problem of choosing $s^*(g)$ are written as
\vspace{-0.2cm}
\begin{equation}\label{eq:model2}
\begin{aligned}
F_s(x_1,\sigma_A(g),\sigma_B(g),s^{*}(g),g)(V(\sigma_A(g),g)-V(\sigma_B(g),g)) - K^{inv}_{s}(s^*(g),g) - K^{gov}_{s}(s^*(g),g) =0 \\
F_{ss}(x_1,\sigma_A(g),\sigma_B(g),s^{*}(g),g)(V(\sigma_A(g),g)-V(\sigma_B(g),g)) - K^{inv}_{ss}(s^*(g),g) - K^{gov}_{ss}(s^*(g),g) <0 
\end{aligned}
\end{equation}
The main result of this framework---i.e., the microfoundations underlying the directional effects of the four channels linking exogenous changes in government guarantees and banks' endogenous choices of income smoothing---is formalized below.
\\ \noindent
\textbf{Proposition 1.} \emph{The sign of the comparative statics $\displaystyle\frac{d}{dg} s^*(g)$ is equivalent to the opposite sign of the sum $\lambda^{tailrisk}+\lambda^{risktaking}+\lambda^{inv} +\lambda^{govv}$, where}
\vspace{-0.2cm}

\begin{equation}
\begin{aligned}
\lambda^{taikrisk}=&-F_{sg} (x_1,\sigma_A(g),\sigma_B (g),s^* (g),g)\bigg(V(\sigma_A(g),g)-V(\sigma_B (g),g)\bigg) \\
&-F_s(x_1,\sigma_{A}(g),\sigma_{B}(g),s^*(g),g)\bigg(V_g (\sigma_{A}(g),g)-V_g(\sigma_{B}(g),g)\bigg)\\
\lambda^{risktaking}=&-\bigg(F_{s\sigma_{A}}(x_1,\sigma_{A}(g),\sigma_{B}(g),s^*(g),g) \frac{\partial \sigma_{A} }{\partial g}+F_{s\sigma_{B}} (x_1,\sigma_{A}(g),\sigma_{B}(g),s^*(g),g)  \frac{\partial \sigma_{B} }{\partial g}\bigg)\bigg(V(\sigma_{A}(g),g) \\
&-V(\sigma_{B}(g),g)\bigg) -F_{s}(x_1,\sigma_{A}(g),\sigma_{B}(g),s^*(g),g)\bigg(V_{\sigma}(\sigma_{A}(g),g)  \frac{\partial \sigma_{A} }{\partial g}-V_{\sigma}(\sigma_{B}(g),g)  \frac{\partial \sigma_{B} }{\partial g}\bigg) \\
\lambda^{inv} =& K_{sg}^{inv}(s^{*}(g),g) \\
\lambda^{gov} =& K_{sg}^{gov}(s^{*}(g),g)
\end{aligned} \nonumber
\end{equation}

\vspace{-0.2cm}
\begin{proof}
Given that functions $F$, $K^{inv}$, and $K^{gov}$ are twice continuously differentiable with respect to their arguments, we can differentiate the first-order condition of the problem with respect to the exogenous parameter $g$. With some algebraic manipulations, the expression below is obtained: 

\begin{equation}\label{eq:model3}
\begin{aligned}
 \Bigl\{F_{ss}(x_1,\sigma_A(g),\sigma_B(g),s^{*}(g),g)(V(\sigma_A(g),g)-V(\sigma_B(g),g))  \\
 - K^{inv}_{ss}(s^*(g),g) - K^{gov}_{ss}(s^*(g),g) \Bigr\} \frac{d}{dg}s^*(g) =  \lambda^{tailrisk}+\lambda^{risktaking}+\lambda^{inv} +\lambda^{govv} 
\end{aligned}
\end{equation}
It is straightforward to note that the term within braces is simply the second-order condition of the manager's problem (hence, it should be negative). As such, the sign of $\frac{d}{dg}s^*(g)$ should be the opposite sign of  $\lambda^{tailrisk}+\lambda^{risktaking}+\lambda^{inv} +\lambda^{govv}$. From the expressions above, $\lambda^{inv}<0$ and $\lambda^{gov}>0$ by construction, implying that the \emph{investor-monitoring channel} and the \emph{government-monitoring channel} should respectively predict a positive and negative association between government guarantees and banks' income smoothing. With some algebraic manipulation, one can show that $\lambda^{risktaking}<0$ and $\lambda^{tailrisk}>0$, meaning that the \emph{risk-taking channel} and the \emph{tail-risk channel} should respectively lead to a positive and negative association between government guarantees and banks' income smoothing.

\end{proof}

\subsection{Variable Definitions} \label{app:sec:variable_definitions}

\begin{table}[H] 
\centering
\caption{Variable Definitions}
\label{tab:variable_definitions}
\vspace{-0.4cm}
{\scriptsize
\begin{longtable}{ p{.15\textwidth}  p{.65\textwidth} p{.15\textwidth}} 
    \toprule
    \textbf{Variable} & \textbf{Description}  & \textbf{Source}\\
    
    \midrule
    \llp{} & Loan loss provisions for bank $i$ at year $t$ normalized by lagged total loans. &Bankscope\\
    
    \hline
    \ebllp{} & Earnings before loan loss provisions and taxes  for bank $i$ at year $t$ normalized by lagged total loans. &Bankscope \\
    \hline
    \lagonesize{} & Natural logarithm of the banks' total assets (measured in millions of USD). &Bankscope\\
        \hline
    \lagonecap{} &  Bank $i$'s total equity by total assets–capital ratio. &Bankscope \\
    
        \hline
    \changeloans{} & Bank $i$'s change in loans scaled by total assets. &Bankscope \\
    
    \hline
    \IInc{} & Interest income of bank $i$ at year $t$ normalized by the average balance of the bank's interest-earning assets. &Bankscope \\
    
    \hline
    \IExp{} & Interest expense of bank $i$ at year $t$ normalized by the average balance of the bank's interest-bearing liabilities. &Bankscope\\
            \hline
                \PercGDPPC{} & Change in GDP per capital of country $c$ between year $t-1$ and $t$. &WorldBank\\
            \hline
    $llr_{i,t-1}$ & Bank $i$'s beginning balance of loan loss reserves normalized by lagged total loans. &Bankscope\\
        \hline
    \BigBath & Indicator variable that takes value of 1 if bank $i$ reports provisions in year $t$ that are at least 50\% greater  than the average provisions of the last 3 reported years conditional on the bank's earnings before loan loss provision being negative. &Bankscope\\
      \hline
    \SmallP & Indicator variable that takes value of 1 if bank $i$ reports in year $t$ return-on-equity greater than zero but less than 1\%. &Bankscope\\   \hline

    $Trade_{c,t}$ & Sum of exports and imports of goods and services measured as a share of gross domestic product for country $c$ and year $t$. & World Bank database \\
                  \hline
                  
                      $Unemp_{c,t}$ &  Share of the labor force that is without work but available for and seeking employment for country $c$ and year $t$. & World Bank database \\
                  \hline
    \RRtussandfour{} & US monetary policy shocks computed by \citet{romer2004new} and aggregated at the year $t$ level. & \citet{romer2004new}, \citet{breitenlechner2018update}\\

        \hline
                \EPUEuro{} &  European version of the Economic Policy Uncertainty index. &\citet{baker2016measuring}\\

    \bottomrule
\end{longtable}
}
\end{table}

\subsection{Supplemental Information} \label{app:sec:eurozoneadd}

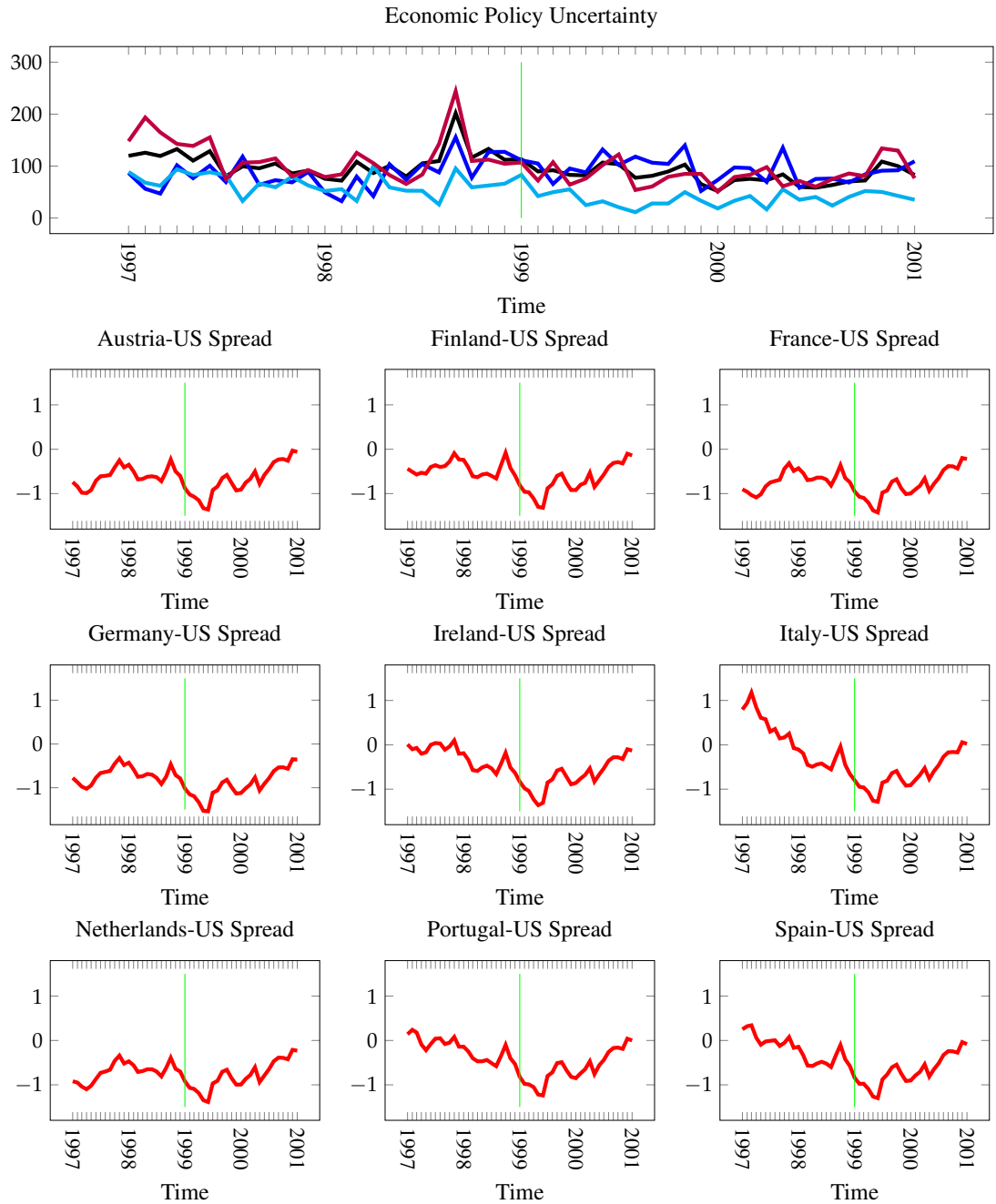
\setcounter{table}{0}

\setcounter{figure}{0}

\begin{table}[H]
  \centering
  \caption{Timeline of Relevant Events Related to the Creation of the Eurozone}
  \label{tab:eurotimeline}
  \scriptsize
  \begin{tabular}{ p{.15\textwidth}  p{.75\textwidth}} 
    \toprule
    \textbf{Year} & \textbf{Event Description} \\

    \hline
    1957 &   Creation of the EEC (predecessor of the EU) by the treaty of Rome. FX fluctuations already seen as a concern for economic stability. \\
    \hline
    1961 & Proposition of a European monetary reserve system---though no actions taken until 1969. \\
    \hline

    1969 & The heads of EEC states agree on the establishment of a economic and monetary union by 1980. \\
    \hline 
    
    1970 & Luxembourg’s prime minister (Pierre Werner) proposes the union to move towards a single economy in 10 years with fixed FX but keeping individual country currencies. Plan failed with the collapse of the Bretton Woods. The EEC proposes actions towards political unity. \\
    \hline 
    
    1972 & European currencies (even of non-EEC countries) are constrained to fluctuations within a 4.5\% limit (called monetary snake). 1974 Oil crisis forces out some currencies later. \\
    \hline 
    
    1979 & The European monetary system (EMS) formally substitutes the monetary snake (including only Germany, Denmark and the Benelux countries by then). \\
    \hline 
    
    1989 & Heads of the European Community states agree on the implementation of the Economic and Monetary Union EMU. \\ \hline

    1990 & Stage 1 of the EMU implementation: capital transactions liberalization and increasing cooperation among national banks. \\ \hline
    
    1992 & The Maastricht Treaty is signed in February. The UK declines during the final state. Denmark rejects by Referendum. On September currency speculation forces the UK to leave the ERM. \\ \hline

    1994 & Stage 2 of the EMU starts, being comprised by the establishment of the European Monetary Institute (EMI) as the predecessor of the European Central Bank (ECB). Member countries commit to pursue currency convergence criteria. \\ \hline
    
    1998 & The European Commission recommends 11 countries to participate in the first wave of monetary union: Austria, Belgium, Finland, France, Germany, Ireland, Italy, Luxembourg, Netherlands, Portugal and Spain. The European Central Bank is established (Frankfurt) and the FX rates between the euro and national currencies are fixed on December 31. \\ \hline
    1999 &On January 1st the Euro is introduced and monetary policy responsibility is transfered from individual countries’ central banks to the ECB. The new currency is mainly used for non-cash transactions as the 11 original currencies from the adopters are simply subdivisions. \\ \hline
    2001 & Greece joins the Euro. \\ \hline
    2002 & Euro notes and coins become legal tender in 12 countries (11 first adopters and Greece). Between January 28 and February 28 all eurozone countries ceased the legal tender aspect of their national currencies. \\
    
    \bottomrule
\end{tabular}

\end{table}

\begin{figure}[H]
    \centering

\scalebox{0.8}{
\pgfkeys{/pgf/number format/.cd,1000 sep={,}}
\begin{tikzpicture}
\begin{axis}[
    width=18.8cm,
   height=5.0cm,
date coordinates in=x,
xtick=data,
xticklabel style={rotate=270},
xticklabel=\month.\year,
title={\large Economic Policy Uncertainty},
xlabel={\large Time},
yticklabel style={
            /pgf/number format/fixed,
            /pgf/number format/precision=0,
            /pgf/number format/fixed zerofill,
	/pgf/number format/1000 sep={,},
        },
ylabel style={yshift=-1pt},
        scaled y ticks=false,
   legend style ={  draw=black, 
        fill=white,align=left},
date ZERO=1997-01-01,
xticklabels={1997,,,,,,,,,,,,1998,,,,,,,,,,,,1999,,,,,,,,,,,,2000,,,,,,,,,,,,2001}
]
\addplot[color=black,line width=2pt] table[x=date,y=value] {EPU_EUR.dat};

\addplot[color=blue,line width=2pt] table[x=date,y=value] {EPU_GER.dat};

\addplot[color=cyan,line width=2pt] table[x=date,y=value] {EPU_FRA.dat};

\addplot[color=purple,line width=2pt] table[x=date,y=value] {EPU_ITA.dat};

\addplot[color=green,line width=0.5pt] coordinates {(1999-01-01,0) (1999-01-01,300)};

\end{axis}

\end{tikzpicture}
}

\scalebox{0.8}{
\pgfkeys{/pgf/number format/.cd,1000 sep={,}}
\begin{tikzpicture}
\begin{axis}[
    width=6.5cm,
   height=4.5cm,
date coordinates in=x,
xtick=data,
xticklabel style={rotate=270},
xticklabel=\month.\year,
title={\large Austria-US Spread},
xlabel={\large Time},
yticklabel style={
            /pgf/number format/fixed,
            /pgf/number format/precision=0,
            /pgf/number format/fixed zerofill,
	/pgf/number format/1000 sep={,},
        },
ylabel style={yshift=-1pt},
        scaled y ticks=false,
   legend style ={  draw=black, 
        fill=white,align=left},
date ZERO=1997-01-01,
xticklabels={1997,,,,,,,,,,,,1998,,,,,,,,,,,,1999,,,,,,,,,,,,2000,,,,,,,,,,,,2001}
]
\addplot[color=red,line width=2pt] table[x=date,y=value] {spread_AUT.dat};

\addplot[color=green,line width=0.5pt] coordinates {(1999-01-01,-1.5) (1999-01-01,1.5)};

\end{axis}

\end{tikzpicture}
} \scalebox{0.8}{
\pgfkeys{/pgf/number format/.cd,1000 sep={,}}
\begin{tikzpicture}
\begin{axis}[
    width=6.5cm,
   height=4.5cm,
date coordinates in=x,
xtick=data,
xticklabel style={rotate=270},
xticklabel=\month.\year,
title={\large Finland-US Spread},
xlabel={\large Time},
yticklabel style={
            /pgf/number format/fixed,
            /pgf/number format/precision=0,
            /pgf/number format/fixed zerofill,
	/pgf/number format/1000 sep={,},
        },
ylabel style={yshift=-1pt},
        scaled y ticks=false,
   legend style ={  draw=black, 
        fill=white,align=left},
date ZERO=1997-01-01,
xticklabels={1997,,,,,,,,,,,,1998,,,,,,,,,,,,1999,,,,,,,,,,,,2000,,,,,,,,,,,,2001}
]
\addplot[color=red,line width=2pt] table[x=date,y=value] {spread_FIN.dat};

\addplot[color=green,line width=0.5pt] coordinates {(1999-01-01,-1.5) (1999-01-01,1.5)};

\end{axis}

\end{tikzpicture}
} \scalebox{0.8}{
\pgfkeys{/pgf/number format/.cd,1000 sep={,}}
\begin{tikzpicture}
\begin{axis}[
    width=6.5cm,
   height=4.5cm,
date coordinates in=x,
xtick=data,
xticklabel style={rotate=270},
xticklabel=\month.\year,
title={\large France-US Spread},
xlabel={\large Time},
yticklabel style={
            /pgf/number format/fixed,
            /pgf/number format/precision=0,
            /pgf/number format/fixed zerofill,
	/pgf/number format/1000 sep={,},
        },
ylabel style={yshift=-1pt},
        scaled y ticks=false,
   legend style ={  draw=black, 
        fill=white,align=left},
date ZERO=1997-01-01,
xticklabels={1997,,,,,,,,,,,,1998,,,,,,,,,,,,1999,,,,,,,,,,,,2000,,,,,,,,,,,,2001}
]
\addplot[color=red,line width=2pt] table[x=date,y=value] {spread_FRA.dat};

\addplot[color=green,line width=0.5pt] coordinates {(1999-01-01,-1.5) (1999-01-01,1.5)};

\end{axis}

\end{tikzpicture}
}

\scalebox{0.8}{
\pgfkeys{/pgf/number format/.cd,1000 sep={,}}
\begin{tikzpicture}
\begin{axis}[
    width=6.5cm,
   height=4.5cm,
date coordinates in=x,
xtick=data,
xticklabel style={rotate=270},
xticklabel=\month.\year,
title={\large Germany-US Spread},
xlabel={\large Time},
yticklabel style={
            /pgf/number format/fixed,
            /pgf/number format/precision=0,
            /pgf/number format/fixed zerofill,
	/pgf/number format/1000 sep={,},
        },
ylabel style={yshift=-1pt},
        scaled y ticks=false,
   legend style ={  draw=black, 
        fill=white,align=left},
date ZERO=1997-01-01,
xticklabels={1997,,,,,,,,,,,,1998,,,,,,,,,,,,1999,,,,,,,,,,,,2000,,,,,,,,,,,,2001}
]
\addplot[color=red,line width=2pt] table[x=date,y=value] {spread_GER.dat};

\addplot[color=green,line width=0.5pt] coordinates {(1999-01-01,-1.5) (1999-01-01,1.5)};

\end{axis}

\end{tikzpicture}
} \scalebox{0.8}{
\pgfkeys{/pgf/number format/.cd,1000 sep={,}}
\begin{tikzpicture}
\begin{axis}[
    width=6.5cm,
   height=4.5cm,
date coordinates in=x,
xtick=data,
xticklabel style={rotate=270},
xticklabel=\month.\year,
title={\large Ireland-US Spread},
xlabel={\large Time},
yticklabel style={
            /pgf/number format/fixed,
            /pgf/number format/precision=0,
            /pgf/number format/fixed zerofill,
	/pgf/number format/1000 sep={,},
        },
ylabel style={yshift=-1pt},
        scaled y ticks=false,
   legend style ={  draw=black, 
        fill=white,align=left},
date ZERO=1997-01-01,
xticklabels={1997,,,,,,,,,,,,1998,,,,,,,,,,,,1999,,,,,,,,,,,,2000,,,,,,,,,,,,2001}
]
\addplot[color=red,line width=2pt] table[x=date,y=value] {spread_IRE.dat};

\addplot[color=green,line width=0.5pt] coordinates {(1999-01-01,-1.5) (1999-01-01,1.5)};

\end{axis}

\end{tikzpicture}
} \scalebox{0.8}{
\pgfkeys{/pgf/number format/.cd,1000 sep={,}}
\begin{tikzpicture}
\begin{axis}[
    width=6.5cm,
   height=4.5cm,
date coordinates in=x,
xtick=data,
xticklabel style={rotate=270},
xticklabel=\month.\year,
title={\large Italy-US Spread},
xlabel={\large Time},
yticklabel style={
            /pgf/number format/fixed,
            /pgf/number format/precision=0,
            /pgf/number format/fixed zerofill,
	/pgf/number format/1000 sep={,},
        },
ylabel style={yshift=-1pt},
        scaled y ticks=false,
   legend style ={  draw=black, 
        fill=white,align=left},
date ZERO=1997-01-01,
xticklabels={1997,,,,,,,,,,,,1998,,,,,,,,,,,,1999,,,,,,,,,,,,2000,,,,,,,,,,,,2001}
]
\addplot[color=red,line width=2pt] table[x=date,y=value] {spread_ITA.dat};

\addplot[color=green,line width=0.5pt] coordinates {(1999-01-01,-1.5) (1999-01-01,1.5)};

\end{axis}

\end{tikzpicture}
}

    \scalebox{0.8}{
\pgfkeys{/pgf/number format/.cd,1000 sep={,}}
\begin{tikzpicture}
\begin{axis}[
    width=6.5cm,
   height=4.5cm,
date coordinates in=x,
xtick=data,
xticklabel style={rotate=270},
xticklabel=\month.\year,
title={\large Netherlands-US Spread},
xlabel={\large Time},
yticklabel style={
            /pgf/number format/fixed,
            /pgf/number format/precision=0,
            /pgf/number format/fixed zerofill,
	/pgf/number format/1000 sep={,},
        },
ylabel style={yshift=-1pt},
        scaled y ticks=false,
   legend style ={  draw=black, 
        fill=white,align=left},
date ZERO=1997-01-01,
xticklabels={1997,,,,,,,,,,,,1998,,,,,,,,,,,,1999,,,,,,,,,,,,2000,,,,,,,,,,,,2001}
]
\addplot[color=red,line width=2pt] table[x=date,y=value] {spread_NED.dat};

\addplot[color=green,line width=0.5pt] coordinates {(1999-01-01,-1.5) (1999-01-01,1.5)};


\end{axis}

\end{tikzpicture}
} \scalebox{0.8}{
\pgfkeys{/pgf/number format/.cd,1000 sep={,}}
\begin{tikzpicture}
\begin{axis}[
    width=6.5cm,
   height=4.5cm,
date coordinates in=x,
xtick=data,
xticklabel style={rotate=270},
xticklabel=\month.\year,
title={\large Portugal-US Spread},
xlabel={\large Time},
yticklabel style={
            /pgf/number format/fixed,
            /pgf/number format/precision=0,
            /pgf/number format/fixed zerofill,
	/pgf/number format/1000 sep={,},
        },
ylabel style={yshift=-1pt},
        scaled y ticks=false,
   legend style ={  draw=black, 
        fill=white,align=left},
date ZERO=1997-01-01,
xticklabels={1997,,,,,,,,,,,,1998,,,,,,,,,,,,1999,,,,,,,,,,,,2000,,,,,,,,,,,,2001}
]
\addplot[color=red,line width=2pt] table[x=date,y=value] {spread_POR.dat};

\addplot[color=green,line width=0.5pt] coordinates {(1999-01-01,-1.5) (1999-01-01,1.5)};
\end{axis}

\end{tikzpicture}
} \scalebox{0.8}{
\pgfkeys{/pgf/number format/.cd,1000 sep={,}}
\begin{tikzpicture}
\begin{axis}[
    width=6.5cm,
   height=4.5cm,
date coordinates in=x,
xtick=data,
xticklabel style={rotate=270},
xticklabel=\month.\year,
title={\large Spain-US Spread},
xlabel={\large Time},
yticklabel style={
            /pgf/number format/fixed,
            /pgf/number format/precision=0,
            /pgf/number format/fixed zerofill,
	/pgf/number format/1000 sep={,},
        },
ylabel style={yshift=-1pt},
        scaled y ticks=false,
   legend style ={  draw=black, 
        fill=white,align=left},
date ZERO=1997-01-01,
xticklabels={1997,,,,,,,,,,,,1998,,,,,,,,,,,,1999,,,,,,,,,,,,2000,,,,,,,,,,,,2001}
]
\addplot[color=red,line width=2pt] table[x=date,y=value] {spread_ESP.dat};

\addplot[color=green,line width=0.5pt] coordinates {(1999-01-01,-1.5) (1999-01-01,1.5)};


\end{axis}

\end{tikzpicture}

}

     \begin{minipage}{0.95\linewidth}\scriptsize
   \singlespacing
   This figure depicts the time series of economic policy uncertainty and sovereign-bond yield spreads of FEA countries. The upper panel shows the evolution of economic policy uncertainty (EPU) indexes computed by \citet{baker2016measuring} for the Eurozone (black), Germany (blue), France (cyan), and Italy (purple). The lower panels show the evolution of the spread between the 10 year sovereign bond yields of different FEA counties and the equivalent 10 year sovereign bond yield of the United States. The vertical green lines represent the date of the creation of the Eurozone (January 1, 1999). \\
  \end{minipage}
\caption{ Eurozone Creation---Macro Series  }
\label{fig:macroseries}
\end{figure}

\begin{table}[!htbp] \centering 
  \caption{Eurozone Creation and Income Smoothing---Interactions with Bank-Performance Variables} 
  \label{tab.euro.interacperf} 
    \vspace{-0.5cm}
     \begin{minipage}{1\linewidth}\scriptsize
   \singlespacing
This table reports OLS coefficient estimates of the differences model augmented to include interaction terms with different bank-level performance measures ($z_{i,t}$). The performance measures considered are changes in loans (\changeloans{}), net-interest margin ($NIM_{i,t}$), interest-income rate (\IInc{}), and interest-expense rate (\IExp{}) Robust standard errors clustered at the bank and year level are reported within parentheses. Country and bank-type fixed effects are included in all specifications.  *, **, and *** indicate significance at the 10\%, 5\%, and 1\% levels, respectively. \\
  \end{minipage}
\scriptsize 
\begin{tabular}{@{\extracolsep{-20pt}}lD{.}{.}{-3} D{.}{.}{-3} D{.}{.}{-3} D{.}{.}{-3} } 
\toprule
 & \multicolumn{4}{c}{\textit{Dependent variable:} \llp{}} \\ 
\cline{2-5} 
\\[-1.8ex] & \multicolumn{1}{c}{(1)} & \multicolumn{1}{c}{(2)} & \multicolumn{1}{c}{(3)} & \multicolumn{1}{c}{(4)}\\ 
\hline \\[-1.8ex]

 \PostNinetyNine{} & 0.001^{*} & 0.002 & 0.004 & 0.004 \\ 
  & (0.001) & (0.002) & (0.003) & (0.002) \\ 
  \ebllp{} & 0.114^{***} & 0.121^{***} & 0.193^{***} & 0.185^{***} \\ 
  & (0.020) & (0.035) & (0.060) & (0.055) \\ 
  \ebllpPostNinetyNine{} & -0.047^{***} & -0.093^{***} & -0.166^{***} & -0.115^{***} \\ 
  & (0.009) & (0.024) & (0.041) & (0.036) \\ 
  \lagonellp{} & 0.279^{***} & 0.274^{***} & 0.276^{***} & 0.279^{***} \\ 
  & (0.059) & (0.059) & (0.058) & (0.057) \\ 
  \lagtwollp{} & 0.188^{***} & 0.184^{***} & 0.186^{***} & 0.187^{***} \\ 
  & (0.030) & (0.034) & (0.032) & (0.030) \\ 
  \lagonecap{} & -0.002 & 0.012 & -0.001 & -0.001 \\ 
  & (0.007) & (0.009) & (0.007) & (0.006) \\ 
  \lagonesize{} & -0.0001 & -0.0001 & -0.0001 & -0.0001 \\ 
  & (0.0002) & (0.0001) & (0.0001) & (0.0002) \\ 
  \changeloans{} & -0.002 & -0.0002 & -0.0002 & -0.0001 \\ 
  & (0.002) & (0.001) & (0.002) & (0.002) \\ 
    \IInc{} & 0.084^{***} & 0.444^{***} & 0.109^{***} & 0.085^{***} \\ 
  & (0.027) & (0.170) & (0.028) & (0.030) \\ 
    \IExp{} & -0.071^{***} & -0.403^{**} & -0.069^{***} & -0.024 \\ 
  & (0.022) & (0.161) & (0.024) & (0.043) \\ 
    \PercGDPPC{} & -0.003 & -0.002 & -0.003 & -0.003 \\ 
  & (0.003) & (0.003) & (0.003) & (0.004) \\ 
  
 $\Delta Loans_{i,t} \times ebllp_{i,t}$ & 0.062 &  &  &  \\ 
  & (0.039) &  &  &  \\ 
  $\Delta Loans_{i,t} \times Post1999_{t}$ & 0.002 &  &  &  \\ 
  & (0.003) &  &  &  \\ 
  $\Delta Loans_{i,t} \times ebllp_{i,t} \times Post1999_{t}$ & -0.080^{*} &  &  &  \\ 
  & (0.043) &  &  &  \\ 
  $NIM_{i,t}$ &  & -0.390^{*} &  &  \\ 
  &  & (0.208) &  &  \\ 
  $NIM_{i,t} \times ebllp_{i,t}$ &  & 0.266 &  &  \\ 
  &  & (1.294) &  &  \\ 
  $NIM_{i,t} \times Post1999_{t}$ &  & -0.043 &  &  \\ 
  &  & (0.061) &  &  \\ 
  $NIM_{i,t} \times ebllp_{i,t} \times Post1999_{t}$ &  & 1.903 &  &  \\ 
  &  & (1.428) &  &  \\ 

  $IInc_{i,t} \times ebllp_{i,t}$  &  &  & -0.927^{*} &  \\ 
  &  &  & (0.560) &  \\ 
  $IInc_{i,t} \times Post1999_{t}$  &  &  & -0.032 &  \\ 
  &  &  & (0.036) &  \\ 
  $IInc_{i,t} \times ebllp_{i,t} \times Post1999_{t}$ &  &  & 1.546^{***} &  \\ 
  &  &  & (0.544) &  \\ 

  $IExp_{i,t} \times ebllp_{i,t}$ &  &  &  & -1.167^{*} \\ 
  &  &  &  & (0.701) \\ 
  $IExp_{i,t} \times Post1999_{t}$  &  &  &  & -0.037 \\ 
  &  &  &  & (0.051) \\ 
  $IExp_{i,t} \times ebllp_{i,t} \times Post1999_{t}$ &  &  &  & 0.987 \\ 
  &  &  &  & (0.614) \\ 

 \hline \\[-1.8ex] 
Observations & \multicolumn{1}{c}{4,425} & \multicolumn{1}{c}{4,425} & \multicolumn{1}{c}{4,425} & \multicolumn{1}{c}{4,425} \\ 

Adjusted R$^{2}$ & \multicolumn{1}{c}{0.267} & \multicolumn{1}{c}{0.271} & \multicolumn{1}{c}{0.268} & \multicolumn{1}{c}{0.268} \\

\bottomrule
 
\end{tabular} 
\end{table}

\begin{table}[H] \centering 
  \caption{Eurozone Creation and Income Smoothing---Additional DID Effects} 
  \label{tab:euro.adddid} 
\scriptsize 
      \vspace{-0.3cm}
   \begin{minipage}{0.95\linewidth}\scriptsize
   \singlespacing
This table reports OLS coefficient estimates of the linear difference-in-differences model (\Cref{eq:yeurodid}) following the Eurozone creation on different bank-level variables ($y_{i,t}$), including (1) earnings before loan loss provisions and taxes (\ebllp), (2) net interest margin ($NIM_{i,t}$), (3) return on assets ($ROA_{i,t}$), (4) return on equity ($ROE_{i,t}$), (5) interbank ratio ($IBank_{i,t}$), (6) indicator of banks' engaging in ``big-bath'' accounting, and (7) indicator of banks' reporting small positive profit.  Robust standard errors clustered at the bank and year level are reported within parentheses. Bank-type fixed effects are included in all specifications.  *, **, and *** indicate significance at the 10\%, 5\%, and 1\% levels, respectively. \\
    \end{minipage}

\begin{tabular}{@{\extracolsep{-2pt}}lD{.}{.}{3,5} D{.}{.}{3,5} D{.}{.}{3,5} D{.}{.}{3,5} D{.}{.}{3,5} D{.}{.}{3,5} D{.}{.}{3,5} } 
\toprule
 & \multicolumn{7}{c}{\textit{Dependent variable:}} \\ 
\cline{2-8} 
\\[-1.8ex] & \multicolumn{1}{c}{\ebllp} & \multicolumn{1}{c}{$NIM_{i,t}$} & \multicolumn{1}{c}{$ROA_{i,t}$} & \multicolumn{1}{c}{$ROE_{i,t}$} & \multicolumn{1}{c}{$IBank_{i,t}$} & \multicolumn{1}{c}{\BigBath} & \multicolumn{1}{c}{\SmallP} \\ 
\\[-1.8ex] & \multicolumn{1}{c}{(1)} & \multicolumn{1}{c}{(2)} & \multicolumn{1}{c}{(3)} & \multicolumn{1}{c}{(4)} & \multicolumn{1}{c}{(5)} & \multicolumn{1}{c}{(6)} & \multicolumn{1}{c}{(7)}\\ 
\hline \\[-1.8ex] 

  \PostNinetyNine{} & -0.002^{***} & -0.001^{***} & -0.0004 & -0.006 & -0.313^{***} & 0.002 & 0.002 \\ 
  & (0.001) & (0.0005) & (0.001) & (0.006) & (0.052) & (0.001) & (0.003) \\ 
  \FEA{} & -0.003 & -0.007^{***} & -0.006^{***} & -0.052^{***} & -0.050 & 0.001 & 0.056^{***} \\ 
  & (0.002) & (0.002) & (0.001) & (0.008) & (0.122) & (0.003) & (0.010) \\ 
  \FEAPostNinetyNine{} & -0.002 & -0.002 & 0.001 & 0.005 & 0.039 & -0.00004 & -0.0001 \\ 
  & (0.001) & (0.001) & (0.001) & (0.007) & (0.143) & (0.003) & (0.006) \\ 
 \hline \\[-1.8ex] 
Observations & \multicolumn{1}{c}{8,110} & \multicolumn{1}{c}{8,110} & \multicolumn{1}{c}{8,110} & \multicolumn{1}{c}{8,110} & \multicolumn{1}{c}{4,968} & \multicolumn{1}{c}{8,110} & \multicolumn{1}{c}{8,110} \\ 
 
Adjusted R$^{2}$ & \multicolumn{1}{c}{0.059} & \multicolumn{1}{c}{0.115} & \multicolumn{1}{c}{0.187} & \multicolumn{1}{c}{0.106} & \multicolumn{1}{c}{0.020} & \multicolumn{1}{c}{0.001} & \multicolumn{1}{c}{0.022} \\ 

\bottomrule
 
\end{tabular} 
\end{table}

\begin{table}[H]
  \centering
  \caption{Eurozone Creation and Income Smoothing---Bounds for Robustness to Proportional Selection on Unobservables}

\scriptsize
      \vspace{-0.3cm}
   \begin{minipage}{0.95\linewidth}\scriptsize
   \singlespacing
This table constructs bounding values for the DID income smoothing coefficient estimate (coefficient of \ebllpFEAPostNinetyNine) following the methodology proposed by \citet{oster2019unobservable}, which builds upon the work by \citet{altonji2005selection}. Following \citet{oster2019unobservable}, we assume that selection on unobservables is proportional to selection on observables. The bounding value of the DID estimate ($\beta^{*}$) is defined as $\beta^{*}=\tilde{\beta} - \frac{(\dot{\beta}-\tilde{\beta})(R^2_{max}-\tilde{R}^2)}{\tilde{R}^2-\dot{R}^2}$, where $\dot{\beta}$ and $\dot{R}^2$ are respectively the point estimate and R-squared for the simplified DID regression and $\tilde{\beta}$ and $\tilde{R}^2$ are the analogue values from the regression with all controls (\Cref{tab:euro.did}, column 1). We consider two different OLS estimations for the simplified model (which gives two values of where $\dot{\beta}$ and $\dot{R}^2$): (i) a model including only the seven key terms of a DID estimation (\FEA{}, \PostNinetyNine{}. \FEAPostNinetyNine, \ebllp, \ebllpPostNinetyNine, \ebllpFEA, and \ebllpFEAPostNinetyNine) without any bank-level or macroeconomic control (Panel A), and (ii) a model including the seven key terms of a DID estimation (\FEA{}, \PostNinetyNine{}. \FEAPostNinetyNine, \ebllp, \ebllpPostNinetyNine, \ebllpFEA, and \ebllpFEAPostNinetyNine) without bank-level controls but including our proxy for business cycles (\PercGDPPC{}) (Panel B). The method assumes that the degree of proportionality between selection on unobservables and selection on observables is one ($\delta=1$), which entails making an assumption about the maximum possible $R^2$ of the regression. We follow the calibration proposed by \citet{oster2019unobservable}, which sets $R^2_{max}=\min {(1,\Pi \times \tilde{R}^2)}$ and $\Pi=1.3$ as our benchmark. As in \citet{verner2020household}, we also consider the conservative value of $\Pi=2.0$ for robustness purposes. \\
    \end{minipage}

\begin{tabular}{@{\extracolsep{-3pt}}llC{1.1cm}C{1.1cm}lC{1.1cm}C{1.1cm}lC{1.1cm}C{1.1cm}lC{1.1cm}C{1.1cm} } 
\toprule
     \multicolumn{13}{c}{Panel A: Simple DID without bank-level and macro controls to DID with all controls}  \\ \hline

 & $\;$& \multicolumn{2}{c}{Simplified model}  & & \multicolumn{2}{c}{All DID controls} && \multicolumn{2}{c}{$R^2_{max}$} && \multicolumn{2}{c}{Bounding values} \\ 
 \cmidrule{3-4} \cmidrule{6-7} \cmidrule{9-10} \cmidrule{12-13}
Outcome && \multicolumn{1}{c}{$\dot{\beta}$} & \multicolumn{1}{c}{$\dot{R}^2$}  & &\multicolumn{1}{c}{$\tilde{\beta}$} & \multicolumn{1}{c}{$\tilde{R}^2$} & & \multicolumn{1}{c}{$\Pi=1.3$} & \multicolumn{1}{c}{$\Pi=2.0$} & & \multicolumn{1}{c}{$\beta^{*}_{\Pi=1.3}$} & $\beta^{*}_{\Pi=2.0}$ \\ \hline
DID smoothing coefficient && $-0.0697$	& $0.0809$	& &	$-0.0602$ &	$0.3519$&	&	$0.4574$ &$0.7037$ & & $-0.0566$ & $-0.0480$ \\

    \\[-1.8ex]\hline 
\hline \\[-1.8ex] 

     \multicolumn{13}{c}{Panel B: Simple DID without bank-level controls to DID with all controls}  \\ \hline

 & $\;$& \multicolumn{2}{c}{Simplified model}  & & \multicolumn{2}{c}{All DID controls} && \multicolumn{2}{c}{$R^2_{max}$} && \multicolumn{2}{c}{Bounding values} \\ 
 \cmidrule{3-4} \cmidrule{6-7} \cmidrule{9-10} \cmidrule{12-13}
Outcome && \multicolumn{1}{c}{$\dot{\beta}$} & \multicolumn{1}{c}{$\dot{R}^2$}  & &\multicolumn{1}{c}{$\tilde{\beta}$} & \multicolumn{1}{c}{$\tilde{R}^2$} & & \multicolumn{1}{c}{$\Pi=1.3$} & \multicolumn{1}{c}{$\Pi=2.0$} & & \multicolumn{1}{c}{$\beta^{*}_{\Pi=1.3}$} & $\beta^{*}_{\Pi=2.0}$ \\ \hline
DID smoothing coefficient && $-0.0706$	& $0.0815$	& &	$-0.0602$ &	$0.3519$&	&	$0.4574$ &$0.7037$ & & $-0.0562$ & $-0.0467$ \\
\bottomrule
    \end{tabular}
  \label{tab:ostereuro}

\end{table}%

\begin{table}[H]
  \centering
  \caption{Landesbanken Guarantees Removal and Income Smoothing---Bounds for Robustness to Proportional Selection on Unobservables}

\scriptsize
      \vspace{-0.3cm}
   \begin{minipage}{0.95\linewidth}\scriptsize
   \singlespacing
This table constructs bounding values for the DID income smoothing coefficient estimate (coefficient of $ebllp_{i,t}\times Landes_{t} \times Post2005_t$) following the methodology proposed by \citet{oster2019unobservable}, which builds upon the work by \citet{altonji2005selection}. Following \citet{oster2019unobservable}, we assume that selection on unobservables is proportional to selection on observables. The bounding value of the DID estimate ($\beta^{*}$) is defined as $\beta^{*}=\tilde{\beta} - \frac{(\dot{\beta}-\tilde{\beta})(R^2_{max}-\tilde{R}^2)}{\tilde{R}^2-\dot{R}^2}$, where $\dot{\beta}$ and $\dot{R}^2$ are respectively the point estimate and R-squared for the simplified DID regression (i.e., only including $Landes_i$, $Post2005_t$, $Landes_i \times Post2005_t$, \ebllp, $ebllp_{i,t} \times Post2005_t$, $ebllp_{i,t} \times Landes_i$, and $ebllp_{i,t}\times Landes_{t} \times Post2005_t$, without any control) and $\tilde{\beta}$ and $\tilde{R}^2$ are the analogue values from the regression with all controls (\Cref{tab.landes.did}, column 1). Panel A depicts the parameter bounds when control-group 1 (large German commercial banks) is used as controls. Panel B depicts the parameter bounds when control-group 2 (French government-owned banks) is used as controls. The method assumes that the degree of proportionality between selection on unobservables and selection on observables is one ($\delta=1$), which entails making an assumption about the maximum possible $R^2$ of the regression. We follow the calibration proposed by \citet{oster2019unobservable}, which sets $R^2_{max}=\min {(1,\Pi \times \tilde{R}^2)}$ and $\Pi=1.3$ as our benchmark. As in \citet{verner2020household}, we also consider the conservative value of $\Pi=2.0$ for robustness purposes. \\
    \end{minipage}

\begin{tabular}{@{\extracolsep{-3pt}}llC{1.1cm}C{1.1cm}lC{1.1cm}C{1.1cm}lC{1.1cm}C{1.1cm}lC{1.1cm}C{1.1cm} } 
\toprule
     \multicolumn{13}{c}{Panel A: DID with Control Group 1 (German Commercial Banks)}  \\ \hline

 & $\;$& \multicolumn{2}{c}{Simplified model}  & & \multicolumn{2}{c}{All DID controls} && \multicolumn{2}{c}{$R^2_{max}$} && \multicolumn{2}{c}{Bounding values} \\ 
 \cmidrule{3-4} \cmidrule{6-7} \cmidrule{9-10} \cmidrule{12-13}
Outcome && \multicolumn{1}{c}{$\dot{\beta}$} & \multicolumn{1}{c}{$\dot{R}^2$}  & &\multicolumn{1}{c}{$\tilde{\beta}$} & \multicolumn{1}{c}{$\tilde{R}^2$} & & \multicolumn{1}{c}{$\Pi=1.3$} & \multicolumn{1}{c}{$\Pi=2.0$} & & \multicolumn{1}{c}{$\beta^{*}_{\Pi=1.3}$} & $\beta^{*}_{\Pi=2.0}$ \\ \hline
DID smoothing coefficient && $1.4518$	& $0.5972$	& &	$1.2849$ &	$0.7983$&	&	$1.0000$ &$1.0000$ & & $1.1176
$ & $1.1176
$ \\

    \\[-1.8ex]\hline 
\hline \\[-1.8ex] 

     \multicolumn{13}{c}{Panel A: DID with Control Group 2 (French Government-Owned Banks)}  \\ \hline
 & $\;$& \multicolumn{2}{c}{Simplified model}  & & \multicolumn{2}{c}{All DID controls} && \multicolumn{2}{c}{$R^2_{max}$} && \multicolumn{2}{c}{Bounding values} \\ 
 \cmidrule{3-4} \cmidrule{6-7} \cmidrule{9-10} \cmidrule{12-13}
Outcome && \multicolumn{1}{c}{$\dot{\beta}$} & \multicolumn{1}{c}{$\dot{R}^2$}  & &\multicolumn{1}{c}{$\tilde{\beta}$} & \multicolumn{1}{c}{$\tilde{R}^2$} & & \multicolumn{1}{c}{$\Pi=1.3$} & \multicolumn{1}{c}{$\Pi=2.0$} & & \multicolumn{1}{c}{$\beta^{*}_{\Pi=1.3}$} & $\beta^{*}_{\Pi=2.0}$ \\ \hline
DID smoothing coefficient && $0.4038$	& $0.4716
$	& &	$0.7427$ &	$0.7451$&	&	$0.9687
$ &$1.0000$ & & $1.0196$ & $1.0584$ \\
\bottomrule 
    \end{tabular}
  \label{tab:osterlandes}

\end{table}%

\begin{table}[H]
\centering
\caption{Serial Properties of Banks' Operating Performance---Eurozone Creation}

    \vspace{-0.25cm}
     \begin{minipage}{0.95\linewidth}\scriptsize
   \singlespacing
This table reports the dynamic properties of banks’ earnings before loan loss provisions (\ebllp{}) for the Eurozone creation sample. The upper panel shows the average values of \ebllp{} for each of the six years in the sample, as well as the p-value of a t-test comparing the average earnings before loan loss provisions for a given year $t$ (\ebllp{}) and the average of the previous year ($ebllp_{i,t-1}$). The lower panels presents the cross-bank difference in the serial volatility of \ebllp{}. For a given FEA bank $i$ and year $t$, the serial volatility (Serial.Var(\ebllp)) is computed as the sample variance of earnings before loan loss provisions for the last three years. The average serial volatilities are shown for the pre-event and post-event years, as well as a t-test comparing the statistical differences between the sample averages.   \\
  \end{minipage}
  \label{tab:landesebllp.appendix} 
\scriptsize
\begin{tabular}{lcccccccc}
  \hline
  & 1996 & 1997 & 1998 & 1999 & 2000 & 2001 \\ 
  \hline
 Average \ebllp & 0.0313 & 0.0287 & 0.0281 & 0.0260 & 0.0255 &  0.0238\\
 t.test(\ebllp,$ebllp_{i,t-1}$), p-values & 0.0844 &  0.02985 & 0.6148 & 0.09994 & 0.7413 & 0.1941\\ 
   \hline
\end{tabular}

\begin{tabular}{L{4.2cm}cccccccc}
  \hline
  & 1996--1998 & 1999--2001 & t-test (difference in means) \\
    \hline

    Serial.Var(\ebllp) & 0.0044 & 0.0039 &  t = 1.2746,  p-value = 0.2025 \\
   \hline
     & 1998 & 2001 & t-test (difference in means) \\
    \hline

\end{tabular}

\end{table}

\begin{table}[H]
\centering
\caption{Serial Properties of Banks' Operating Performance---Landesbanken Guarantees Removal}

    \vspace{-0.25cm}
     \begin{minipage}{0.95\linewidth}\scriptsize
   \singlespacing
This table reports the dynamic properties of banks’ earnings before loan loss provisions (\ebllp{}) for the Landesbanken guarantees removal  sample. The upper panel shows the average values of \ebllp{} for each of the six years in the sample, as well as the p-value of a t-test comparing the average earnings before loan loss provisions for a given year $t$ (\ebllp{}) and the average of the previous year ($ebllp_{i,t-1}$). The lower panels presents the cross-bank difference in the serial volatility of \ebllp{}. For a given Landesbank $i$ and year $t$, the serial volatility (Serial.Var(\ebllp)) is computed as the sample variance of earnings before loan loss provisions for the last three years. The average serial volatilities are shown for the pre-event and post-event years, as well as a t-test comparing the statistical differences between the sample averages.    \\
  \end{minipage}
  \label{tab:landesebllp.appendix} 
\scriptsize
\begin{tabular}{lcccccccc}
  \hline
  & 2002 & 2003 & 2004 & 2005 & 2007 & 2007 \\ 
  \hline
 Average \ebllp & 0.0089 & 0.0040 & 0.0069 & 0.0129 & 0.0034 & 0.0071\\
 t.test(\ebllp,$ebllp_{i,t-1}$), p-values & 0.6228 &  0.4729 & 0.5362 & 0.3722 & 0.1801 & 0.08587\\ 
   \hline
\end{tabular}

\begin{tabular}{L{4.2cm}cccccccc}
  \hline
  & 2002--2004 & 2005--2007 & t-test (difference in means) \\
    \hline

    Serial.Var(\ebllp) & 0.0005 & 0.0003 &  t = 0.9634, p-value = 0.3414 \\
   \hline
   
     & 2004 & 2007 & t-test (difference in means) \\
    \hline

\end{tabular}

\end{table}

\end{document}